\documentclass[11pt]{article}\pdfoutput=1
\usepackage{amsmath,amsfonts,amssymb,amscd}
\usepackage{cite}
\usepackage{pstricks,pst-node}
\usepackage[small,bf,hang]{caption}
\usepackage{tensor}
\usepackage{slashed}
\usepackage{amsmath}
\usepackage{latexsym,epsfig}
\usepackage{arydshln}
\usepackage{extarrows}
\usepackage[utf8]{inputenc}
\usepackage[linktoc=all,hidelinks]{hyperref}
\usepackage{amsthm,tikz-cd,accents}

\newcommand\starred[1]{\accentset{\star}{#1}} 
\newcommand\lf{$L_\infty${}}
\DeclareMathOperator{\im}{im}

\newcommand\prone{\pi_1}

\usepackage[makeroom]{cancel}

\usepackage{color}

\def\hybrid{
        \topmargin -20pt
        \oddsidemargin 0pt
        \headheight 0pt \headsep 0pt
        \textwidth 6.25in 
        \textheight 9.5in 
        \marginparwidth .875in
        \parskip 5pt plus 1pt \jot = 1.5ex}

\hybrid

 \csname
@addtoreset\endcsname{equation}{section}

\def\moth{\mathsurround=0pt}
\newdimen\zo \zo=0pt

\def\tick{\leaders\hrule height 0.5ex depth 0pt \hskip 0.5pt}
\def\upboxfill{$\moth \setbox\zo\hbox{\tick}%
  \hskip 3pt\hbox to 0pt{$\tick$\hss}\hrulefill \hbox to 7.5pt{$\tick$\hss}$}

\def\dtick{\leaders\hrule height .34pt depth 0.5ex \hskip 0.5pt}
\def\downboxfill{$\moth \setbox\zo\hbox{\dtick}%
  \hskip 2pt\hbox to 0pt{$\dtick$\hss}\hrulefill \hbox to 2pt{$\dtick$\hss}$}

\def\bS{{\bf S}}
\def\pd{\partial}

\def\bec{\begin{center}}
\def\ec{\end{center}}

\def\cF{{\cal F}}

 \def\det{{\rm det\,}}
\def\be{\begin{equation}}
\def\ee{\end{equation}}
\def\bea{\begin{eqnarray}}
\def\eea{\end{eqnarray}}
\def\ba{\begin{array}}
\def\ea{\end{array}}

\begin{document}

\begin{titlepage}
\rightline{}
\rightline{Imperial-TP-2020-CH-02}
\begin{center}
\vskip 2cm
{\Large \bf{Homotopy Transfer and Effective Field Theory I: Tree-level}
}\\
\vskip 1.2cm


  \vskip 1.5cm
 {\large {Alex S.~Arvanitakis$^{a}$, Olaf Hohm$^{b}$, Chris Hull$^{c}$ and Victor Lekeu$^{c}$ }}
\vskip 1cm

  {\it  
      $^a$Theoretische Natuurkunde, Vrije Universiteit Brussel, \\ and the International Solvay Institutes, \\ Pleinlaan 2, B-1050 Brussels, Belgium \\ \ \\}

{\it  $^b$Institute for Physics, Humboldt University Berlin,\\
 Zum Gro\ss en Windkanal 6, D-12489 Berlin, Germany}\\
\vskip .5cm

{\it $^c$The Blackett Laboratory, Imperial College London, \\
Prince Consort Road, 
London
SW7 2AZ, 
U.K.}

\vskip .3cm
alex.s.arvanitakis@vub.be, ohohm@physik.hu-berlin.de, c.hull@imperial.ac.uk, v.lekeu@imperial.ac.uk

\vskip 1.5cm
{\bf Abstract}

\end{center}


\noindent
\begin{narrower}

We use the dictionary between general  field theories and strongly homotopy algebras 
to provide an algebraic formulation of the  procedure of integrating out of degrees of freedom 
in terms of homotopy transfer. This includes more general effective theories in which some massive modes are kept 
while other modes of a comparable mass scale are integrated out, as first explored by Sen in the context of closed string field theory. 
We treat  \lf-algebras both in terms of a nilpotent coderivation and, on the dual space, 
in terms of a nilpotent derivation (corresponding to the BRST charge of the field theory) and provide explicit formulas for homotopy transfer. 
These are then 
shown to govern the integrating out of degrees of freedom at tree level, 
while the generalization to loop level will be explored in a sequel to this paper.

\end{narrower}

\end{titlepage}

\tableofcontents


\section{Introduction}

Classical and quantum field theories are traditionally formulated by specifying a set of fields and providing an action functional  which yields the  classical field equations as Euler-Lagrange equations, determines the measure of the quantum functional integral and 
  sets up the quantum perturbation theory in terms of propagators, vertices and Feynman diagrams. 
The gauge symmetries (if any) of the action are parameterised by
 a set of gauge parameters, and 
 for reducible theories there can be higher order gauge symmetries (symmetries of the symmetries). The symmetries of the theory then satisfy a symmetry algebra, which in the simplest cases is a Lie algebra.
 There is, however,  a more algebraic formulation of classical and quantum field theories in terms of strongly homotopy algebras that 
will be the focus of this paper. Remarkably, this combines the symmetry algebra with the field equations into a single overarching structure.
 The usual field theory data can be encoded in an \lf-algebra, which is a generalization 
of a differential graded Lie algebra in which the Jacobi identity need not hold exactly but only `up  to homotopy', i.e.~up to terms which are topologically trivial 
in a certain sense. 
Specifically,  the failure of the `Jacobiator' to vanish is controlled by an (abstract) differential and higher order brackets. 
These higher order brackets 
 encode higher order interactions and non-linear contributions to gauge transformations,  
and they are subject to higher Jacobi identities, many of which translate to familiar consistency conditions in field theory (such as gauge invariance). 
Strongly homotopy Lie algebras or \lf-algebras were first discovered in the context of closed string field theory  \cite{Zwiebach:1992ie}, 
and it is only  more recently that it has become clear that any field theory can be encoded in an \lf-algebra, see \cite{Hohm:2017pnh} for the dictionary, 
which will be briefly reviewed below. 
The importance of \lf-algebras   was already recognized in early work by mathematicians  \cite{Lada:1992wc,Lada:1994mn,Stasheff:1997iz,Barnich:1997ij}.

One major motivation for our investigation was a seminal paper by Sen \cite{Sen:2016qap} in which 
he analyzed, in the context of closed string field theory, 
the 
integrating out
of a set of   fields or states to arrive at a (generalised)  
effective action for the remaining fields. 
In the conventional Wilsonian approach,  one  integrates out all degrees of freedom whose energy scale is above a certain 
threshold  
in order to arrive at a Wilsonian  effective theory valid for all processes below that scale. After discussing the Wilsonian  effective action for string field theory, Sen went on to 
discuss  integrating out much more general subsets of fields.
 For instance, on toroidal backgrounds one may in principle integrate out all states except the massless 
fields together with their massive Kaluza-Klein and winding modes, thereby arriving at a consistent theory carrying some massive string modes 
while truncating other string modes of comparable mass scales.\footnote{This would be a true double field theory as envisioned in \cite{Hull:2009mi}, with only the so-called weak constraint.}
Sen's analysis shows in particular that the \lf-algebra that governs the full closed string field theory gives rise to a `smaller'  \lf-algebra
for the generalised  effective theory that governs the remaining states.

In this paper and its sequel we will analyze, for general field theories, the integrating out of a general subset of fields
in the language of \lf-algebras.
Specifically, we interpret the  procedure of integrating out degrees of freedom 
in terms of `homotopy transfer', which transports a given algebraic structure from one vector space to another. 
This second vector space  is generally `smaller'  than the original one, 
but still homotopy equivalent to it in a certain sense that we will explain in later sections.  It seems to be known to some that homotopy transfer in principle corresponds to `integrating out' degrees of freedom in field theory at tree level and this has been discussed for certain theories.\footnote{ {In physics literature this has been discussed very recently in  string field theory  \cite{Masuda:2020tfa}. In the mathematical literature this is an idea advocated by Andrey S.~Losev since around 2004, and has been realised in explicit examples involving not just topological (BF or Chern-Simons) theories \cite{Mnev:2006ch,losev2007berezin,Mnev:2008sa,Cattaneo:2008ph,Alekseev:2010ud,Cattaneo:2015vsa,Cattaneo:2017tef,losev2019tqft} but also in the (non-topological) 10-dimensional super Yang-Mills theory in the pure spinor formulation \cite{Krotov:2006th,Alexandrov:2007pd}.}}
Here we show that this is the case for general field theories formulated in terms of  \lf-algebras, 
provide explicit formulas and illustrate them with a number of examples.
One of the prime objectives of this paper is  to extend Sen's analysis to the algebraic formulation of  general field theories, and in particular to the understanding of integrating out general sets of fields in terms of homotopy transfer.
Here we discuss this at tree level and in a second paper we   will 
 extend this  to provide  new results on  the algebraic structure of the corresponding  quantum theories.

The $L_{\infty}$ formulation of field theories is closely related to the Batalin-Vilkovisky (BV) field-antifield formalism. An \lf-algebra encodes 
the data of a classical field theory in terms of multi-linear maps on a graded vector space that includes the space of fields, the space of gauge parameters and if needed 
higher spaces.  In contrast, the BV formalism is defined in terms of a
BRST operator $Q$ satisfying $Q^2=0$, which is an odd vector field defined on a graded (super) manifold with odd symplectic structure \cite{Batalin:1981jr,Batalin:1984jr}. This vector field has vanishing Lie bracket with itself, $[Q,Q]=0$; such a vector field is called a \emph{homological vector field}. Coordinates on this manifold are given by fields of the theory, ghosts (corresponding to gauge parameters), ghosts-for-ghosts etc., along with their `antifields'; the condition $Q^2 = 0$ then fully encodes the consistency of the gauge theory.
The grading is according to the ghost number.\footnote{Here we will restrict ourselves to bosonic field theories whose classical fields are all commuting, so that ghost number is the only grading. The generalisation to include fermionic fields, resulting in a double grading according to both ghost number and fermion number, is straightforward but will not be considered here.}
Roughly speaking, the \lf-algebra arises from choosing local coordinates $\Phi$ on the BV manifold, and the Taylor coefficients obtained by expanding $Q$ in $\Phi$ can be thought of as the structure constants of the \lf-algebra. In particular, the condition $Q^2=0$ then translates into the higher Jacobi identities of the \lf-algebra, and the tangent space to the BV manifold can then be identified with the \emph{dual} of the vector space on which the $L_\infty$ brackets are defined. In this way, the BV formalism gives rise to an $L_\infty$ algebra for any field theory.

While a finite dimensional vector space is isomorphic to its dual, 
it is important to note that for   infinite dimensional vector spaces, such as those  arising in field theory, the relationship between dual spaces is less straightforward.
As well as the formulation of \lf-algebras arising from a homological vector field $Q$ acting on a vector space, there is another formulation  based on a \textit{coderivation} (the dual 
map to the derivation given by  $Q$) acting on a suitable coalgebra, which presents technical advantages in the infinite-dimensional case. While the algebra formulation is more geometric 
(and perhaps more intuitive for most physicists), the coalgebra formulation is more algebraic 
(and perhaps less intuitive for most physicists). 
Both formulations have their advantages, and we will present our results using both the algebra and coalgebra pictures, 
providing explicit formulae that we hope will
make 
the  abstract formalism more generally accessible.

Before turning to the detailed outline of this paper and a summary of our results, 
let us briefly  review the dictionary between field theories  and \lf-algebras following \cite{Hohm:2017pnh}. 
An \lf-algebra is an algebraic structure defined on a graded vector space  $X=\dots  \oplus  X_{-1}\oplus  X_{0} \oplus X_{1} \oplus \cdots$, 
where the dictionary relates $X_1$ to the space of gauge parameters, collectively denoted by $\Lambda$, $X_0$ to the space of fields,  collectively denoted by $\Psi$, 
and $X_{-1}$ to the space of field equations (or antifields, in BV language): 
\be\label{Hextendedvector}
  \begin{split}
  \cdots \  \longrightarrow\; & X_{1} 
\longrightarrow X_{0} \longrightarrow  X_{-1} \longrightarrow  \  \cdots \, \\
   &\,  \Lambda\ \   \qquad  \Psi  
 \qquad \     {\rm EoM}  \end{split}
 \ee
In general there may be further spaces in both directions, corresponding to gauge symmetries for gauge 
symmetries and (higher) Bianchi identities. The algebraic structures defined on this vector space are given by an abstract differential $\partial$, 
acting along the  arrows indicated in
(\ref{Hextendedvector})
and squaring to zero, $\partial^2=0$, and by graded symmetric  brackets $\big[\,\cdot\,,\ldots,\,\cdot\,\big]$ (of intrinsic degree $-1$) with 
a potentially unlimited number of inputs.\footnote{It is important to note that $\partial$ is an abstract map and thus in general \textit{not} a first-order differential operator.}    
 The differential $\partial$ encodes the linearised gauge transformations and the   linearised field equations, while the brackets encode non-linear interactions and the non-linear parts of the gauge transformations.
 
 The action of the field theory for $\Psi$ can be  completely formulated using this data plus an inner product $\langle \cdot\,,\,\cdot\rangle$ 
 (referred to as $\kappa(\cdot\,,\,\cdot)$ later): 
 \be
  S = \frac{1}{2} \langle \Psi, \partial \Psi\rangle + \frac{1}{3!}\langle \Psi, \big[\Psi,\Psi\big]\rangle +\frac{1}{4!}\langle \Psi,\big[\Psi,\Psi, \Psi\big]\rangle+ \cdots  \;. 
 \ee
Similarly, the field equations obtained from this action (assuming that the inner product $\langle\cdot\,,\,\cdot\rangle$ obeys suitable cyclicity properties) read 
 \be\label{EoM}
   0 = \partial \Psi  + \tfrac{1}{2} \big[\Psi, \Psi\big] + \tfrac{1}{3!} \big[\Psi, \Psi, \Psi\big]
  +\tfrac{1}{4!}\big[ \Psi, \Psi, \Psi, \Psi\big]+\cdots\;,
 \ee
while any gauge symmetries take the form  
\be\label{equall}
\begin{split}
 \delta_{\Lambda}\Psi = \partial\Lambda + \big[\Lambda,\Psi\big] + \tfrac{1}{2} 
 \big[\Lambda, \Psi,\Psi\big]
  + \tfrac{1}{3!}  \big[\Lambda,\Psi,\Psi,\Psi\big]  + \cdots \;. 
\end{split}
\ee
Note that thanks to the grading of the underlying vector space, the graded symmetry of the bracket may yield a symmetric bracket, as when evaluated on fields,  $\big[\Psi,\Psi\big]$,  
or an antisymmetric bracket, as when evaluated on gauge parameters, $\big[\Lambda_1,\Lambda_2\big]$, where it encodes the gauge algebra. 

The claim is that any perturbative field theory (for which one has a definite notion of terms in the action being quadratic, cubic, etc.) 
can be encoded in an \lf-algebra. Perhaps surprising at first sight, one can check  this fact with examples, 
as explained in \cite{Hohm:2017pnh}. For instance, for a massless scalar field theory without gauge symmetries the space $X_1$ is trivial, 
and the differential is only non-trivial as a map $\partial: X_0\rightarrow X_{-1}$, given by the d'Alembert operator, $\partial=\square$. The higher order brackets 
in (\ref{EoM}) then encode the Taylor expansion of any scalar potential, and thus (\ref{EoM}) yields a familiar equation of motion. 
More interesting are genuine gauge theories for which $X_1$ (and possibly higher spaces) need to be included so that $\partial^2=0$ 
encodes, in particular, linearised gauge invariance. 
The consistency conditions of any (classical) field theory (such as gauge covariance of the field equations and closure of the gauge algebra) 
are equivalent to the (higher) Jacobi identities of \lf-algebras. At low level the Jacobi identities read  
 \be
  \begin{split}
    0&= \partial^2\,, \\
    0&= \partial\big[x_1,x_2\big]+\big[\partial x_1, x_2\big]+(-1)^{x_1}\big[x_1, \partial x_2\big]\,,\\
    0 &=  \big[\big[x_1,x_2\big],x_3\big]+(-1)^{x_2x_3}\big[\big[x_1,x_3\big],x_2\big] +(-1)^{x_1(x_2+x_3)}\big[\big[x_2,x_3\big],x_1\big] \\
  &\;\;\;+\partial\big[x_1,x_2,x_3\big]\\
  &\;\;\; +\big[\partial x_1 ,x_2,x_3\big]+(-1)^{x_1}\big[x_1, \partial x_2 ,x_3\big] +(-1)^{x_1+x_2}\big[x_1,x_2, \partial x_3\big]  \;. 
  \end{split}
 \ee  
The first two relations state that $\partial$ is a nilpotent operator that acts as a derivation on $\big[\cdot,\cdot\big]$. The third  relation states that the Jacobiator 
need not vanish but is rather governed by the failure of $\partial$ to act as a derivation on $\big[\cdot,\cdot,\cdot\big]$. 
The signs here are somewhat unconventional since we take the `$b$-picture' conventions, which are more convenient for our 
subsequent applications.\footnote{See e.g. \cite{Hohm:2017pnh} for a discussion of the 
$b$ and $\ell$ pictures and  the relation between their sign conventions.} In the main text we will give a closed-form characterization of the infinite tower of Jacobi identities. 
The general result that any field theory can be formulated this way also follows from the BV formalism, and the fact that this yields an \lf-algebra, as we shall discuss in section \ref{sec:fieldtheoriesandlfalgebras}.

There is also an $A_\infty$ or ``strongly homotopy associative'' algebra
\cite{stasheff1963homotopy} formulation
for certain theories, such as open string field theory
\cite{Kajiura:2001ng} and gauge theories like Chern-Simons or Yang-Mills
theory \cite{Hohm:2017pnh}.
It must be emphasized, however,  that currently there is no such
formulation for general theories (including gravity or closed string
field theory).
Even when there is both an $L_{\infty}$ and an $A_\infty$ formulation,
such as for Chern-Simons theory,
the $A_\infty$ formulation may require additional data, such as a matrix
representation for
the Lie algebra. The $L_{\infty}$ formulations are thus both more
general and more natural, and for that reason we focus on \lf-algebras here, even though the $L_{\infty}$ case is technically more involved with regard
to homotopy transfer \cite{manetti,huebschmann,berglund}. 

We now give an outline of this paper and briefly summarize our   results.
Part of this paper brings together and reviews
 material that is scattered throughout the mathematics 
literature and/or known to experts, providing explicit formulae and examples. In addition, we  have a number of new results, which we highlight in the following for the different sections. 
(Some of the relevant earlier literature is contained in \cite{Kajiura:2001ng,kontsevich2003deformation,kontsevich2000homological,fukaya}.)

\textit{Section 2:} We present the coalgebra formulation of \lf-algebras, which has the advantage 
of giving the $L_{\infty}$ structure directly on its underlying  vector space $X$. 
The maps of the \lf-algebra are encoded in a coderivation acting on the coalgebra of all graded symmetric monomials 
of vectors in $X$, and the $L_{\infty}$ relations are encoded in the coderivation squaring to zero. 
We introduce the notion of homology for the differential $\partial$ and define quasi-isomorphisms, which leave the homology invariant. 
One can transport $L_{\infty}$ structures by  such quasi-isomorphisms, and we provide explicit formulas in terms of a homotopy map $h$. 
There are well established explicit homotopy transfer formulas for the closely related $A_{\infty}$ algebras, but the $L_{\infty}$ case is significantly more involved 
due to the graded symmetry of the multi-linear maps, and 
for that reason explicit treatments are much scarcer in the literature. 
We hope that our presentation condenses, and makes accessible for physicists, results that can be found in the mathematics literature, e.g. \cite{Kajiura:2001ng,kontsevich2003deformation,kontsevich2000homological,fukaya,Lazarev,berglund,crainic2004perturbation,manetti,huebschmann,huebschmann2011lie}.

\textit{Section 3:} Here we introduce \lf-algebras from the dual viewpoint of a 
nilpotent derivation   $Q$
 acting on 
the dual space $X^{\star}$. While this formulation operates on the `wrong' vector space it does have a number of advantages, notably 
that it is more closely related to the familiar BV formalism of quantum field theory. In addition, this formulation is more geometric, 
which in turn allows us to give an alternative proof of homotopy transfer { via a morphism ``going the other way'' with respect to the one of Section 2.} 
On a technical level, one of our new results is to provide novel and more efficient formulas for the homotopy transfer 
that, in particular, do not require so-called side conditions.  We also set the stage for our field theory applications by explaining homotopy transfer 
for cyclic \lf-algebras that feature an invariant inner product. {Here there is some overlap with the mathematical physics literature on the path integral in the BV formalism, whose results should agree with ours in a tree-level limit \cite{braun2018minimal,Mnev:2006ch,Mnev:2008sa,Krotov:2006th,Alexandrov:2007pd,Doubek:2017naz}.} 

{The results of Section 2 and Section 3 together establish a \emph{strongly homotopy equivalence} between the two \lf-algebras in the sense of Huebschmann \cite{huebschmann}, under different assumptions.}

\textit{Section 4:}
In this section we discuss in more detail how the formulation of field theories in terms of \lf-algebras is related to the BV formalism. 
Notably, while \lf-algebras are defined  on a vector space, the BV formalism is based upon a more general symplectic (super-)manifold, 
carrying an odd symplectic structure, 
and so their relation is not entirely straightforward. 
The  \lf-algebra emerges from a  suitable ``linearisation" of  the BV manifold.
Some approaches to this have  been discussed in  the literature, e.g. \cite{Jurco:2018sby,Sachs:2019gue}, but we
provide an alternate viewpoint that we
 believe  clarifies a number of issues concerning the relation 
between \lf-algebras and the BV formalism.

\textit{Section 5:} In this section, we start by giving a general prescription of the homotopy map $h$ in terms of the propagator projected to the subspace of 
fields that are integrated out. This establishes the relation to Sen's formulation in the context of closed string field theory.
Then, we give explicit examples of field theories to illustrate that integrating out degrees of freedom in field theory 
(at tree level) corresponds to homotopy transfer in the language of \lf-algebras. We start with the simplest example of a 0-dimensional 
field theory with two sets of fields, one of which is integrated out by eliminating it according to its own equation of motion, and proceed to genuine 
field theory examples, including a homotopy-transfer realisation of a sharp momentum cutoff in scalar field theory.

\textit{Section 6:}
We conclude  with a discussion and summary of our results. A considerable part of this paper is review, but we
 we believe that a number  of our formulas and some of our treatment and detailed physical interpretation 
are new. Moreoever, they lay the groundwork for a second paper that will extend our results to the quantum theory.

\textit{Note added:} While finalizing the present paper the two preprints \cite{Erbin:2020eyc,Koyama:2020qfb} appeared which also 
investigate homotopy transfer and tree-level effective field theory, although with an emphasis on $A_{\infty}$-algebras and string field theory, while our 
results are applicable to arbitrary field theories. {In the preprint \cite{Erbin:2020eyc} in particular there is also a proposal for homotopy transfer of \lf-algebras in an appendix.}

\section{Homotopy Transfer and Coderivations }
\label{section:coderivation}

We introduce \lf-algebras and the homotopy transfer theorem from the viewpoint of coderivations 
acting on a tensor algebra, which arguably is the most efficient formulation. 
In the first subsection we introduce the notion of coderivations obeying the  co-Leibniz property 
with respect to a coassociative coproduct. 
Then, in the second subsection, we discuss homology and define quasi-isomorphisms. 
Finally, in the third subsection, we explain and prove the homotopy transfer theorem according to which 
an \lf-algebra is transported under quasi-isomorphisms to another \lf-algebra, and we provide 
explicit formulas for the resulting coderivation. 

\subsection{\lf-algebras and Coderivations}
\subsubsection{Lie Algebras}

Let us begin by explaining the notion of coderivations for the example of Lie algebras. A Lie algebra is a vector space $X$ 
equipped with an antisymmetric bracket $[\cdot,\cdot]$ satisfying the Jacobi identity. This data 
can be equivalently encoded in a nilpotent coderivation acting on the space of antisymmetric tensors over $X$.
This space is  given by the tensor algebra over $X$ consisting of  all formal sums of tensor powers of vectors in $X$, 
modded out by even permutations to leave only the antisymmetric tensors:
 \be
   {\rm Alt}(X) \ \equiv \ \bigoplus_{k=0}^{\infty} \Lambda^k X\;, 
 \ee
where we set $\Lambda^0X\equiv \mathbb{R}$.  
Concretely, this means that the elements are formal sums of vectors $v$,  $v\wedge w$, $v\wedge w\wedge x$, $\ldots$, with the (anti)symmetry 
properties $v\wedge w = -w\wedge v$,  etc. Now, the coderivation encoding a Lie algebra is a linear map
$D: {\rm Alt}(X)\to  {\rm Alt}(X)$
that acts  as 
 \be
  { D}\;:\quad \Lambda^kX \; \rightarrow\;  \Lambda^{k-1}X\;. 
 \ee
On a single vector $v$, i.e. an element of $\Lambda^1X=X$, it acts trivially, $D(v)=0$. 
Its action on an element $v\wedge w\in \Lambda^2X$ is identified with the Lie bracket: 
  \be\label{Liebracket}
  [v,w] \equiv  {D}(v\wedge w)\;.  
 \ee 
Next, we have to define how $D$ acts on higher tensors.  This is uniquely determined by linearity and  the action of $D$ 
on homogeneous elements $v_1\wedge \ldots \wedge v_n\in \Lambda^nX$, which is given by 
 \be\label{extendedLieAction}
  {D}(v_1\wedge \ldots \wedge v_n) \  = \ \sum_{1\leq i<j\leq n}(-1)^{i+j+1} { D}(v_i\wedge v_j)\wedge v_1
  \wedge \ldots\wedge \hat{v}_i\wedge \ldots \wedge \hat{v}_j\wedge \ldots \wedge v_n\;, 
 \ee
where the hat indicates that these elements are to be omitted.  
 More intuitively, this action can be described as follows. By (\ref{Liebracket}) we know how $D$ 
acts on a tensor product with two factors. The action of $D$ on $n$ factors is 
obtained by picking out two factors and then moving them to the front, keeping their order unchanged and including the signs obtained by doing so. 
Then one acts with $D$ on the first two factors and, finally, sums over all possibilities of picking out these two factors. 
We illustrate this for 
an element in $\Lambda^3X$: 
 \be\label{threeelementaction}
 \begin{split}
  {D}(v_1\wedge v_2\wedge v_3) \ &= \ {D}(v_1\wedge  v_2)\wedge v_3
  + {D}(v_2\wedge  v_3)\wedge v_1-D(v_1\wedge  v_3)\wedge v_2 \\
  \ &= \  [v_1,  v_2]\wedge v_3
  +[v_2,  v_3]\wedge v_1-[v_1, v_3]\wedge v_2\;. 
 \end{split} 
 \ee
Note that the relative order of the two elements out of the three has been kept, leading to three terms. 
Permutations with this property are called unshuffles. 
We can now understand that the Jacobi identity is equivalent to the coderivation squaring to zero:  
 \be
  D^2  =  0\;. 
 \ee 
Indeed, with (\ref{threeelementaction}) we have 
 \be
  {D}^2(v_1\wedge v_2\wedge v_3)  =   [[v_1,  v_2], v_3]
  +[[v_2,  v_3], v_1] +[ [v_3, v_1], v_2]  =  0 \;,  
 \ee
whose vanishing is equivalent to the  Jacobi identity. The condition $D^2=0$ is trivially satisfied on elements in $X$ and $\Lambda^2X$. 
 
\subsubsection{Coderivations and Coproducts on Graded Tensor Algebras}

In order to show that $D^2=0$ is satisfied on the entire tensor algebra and also to set the stage 
for the subsequent generalizations to \lf-algebras
we will now discuss the properties of coderivations and coalgebras more generally. 
First, we allow the vector space $X$ to be integer graded, 
 \be\label{gradedVectorspace}
  X=\bigoplus_{i\in \mathbb{Z}}X_i\;, 
 \ee
so that each homogenous element $x$ has an 
integer degree denoted by $|x|\in \mathbb{Z}$ (sometimes also $\deg x$ or simply $x$ if there is no ambiguity). 
We then take the graded symmetric tensor algebra  
  \be
  \label{gradedsymmetricten}
  {\bf S}(X) \equiv \bigoplus_{n=1}^{\infty} S^nX
 \ee 
to be given by tensor products that are \textit{graded symmetric} according to this grading. 
(Often we just write ${\bf S}$ if it is clear what the underlying vector space is.) 
Concretely, for any homogeneous elements $x_1, x_2\in X$ we have 
 \be\label{gradedsymmetric}
  x_1\wedge x_2 = (-1)^{x_1x_2}x_2\wedge x_1\;, 
 \ee 
where here and in the following we employ  the short-hand notation $(-1)^{x_1x_2}\equiv (-1)^{|x_1||x_2|}$.  
(Note that the  example above is included here if all vector spaces except  $X\equiv X_1$ are trivial, implying that (\ref{gradedsymmetric}) is antisymmetric.)  
More generally, for any tensor power the exchange of two adjacent vectors gives a sign whenever both vectors have odd degree. 
With this one defines the Koszul sign $\epsilon(\sigma;x)$ for any permutation $\sigma$ of $n$ elements and a choice 
of such elements $x=(x_1,\ldots, x_n)$ by 
 \be
  x_1\wedge \cdots \wedge x_n \ = \ \epsilon(\sigma;x)\,x_{\sigma(1)}\wedge \cdots \wedge x_{\sigma(n)}\;. 
 \ee
In words, the  Koszul sign is the sign one picks up when permuting vectors according to (\ref{gradedsymmetric}). 
 
Note that, as is usual in the discussion of infinite direct sums,
we take the graded symmetric tensor algebra   (\ref{gradedsymmetricten}) to consist of sums of \emph{finite} numbers of terms.
If we take a basis $T_a$ of $X$, then ${\bf S}$ can be viewed as the space of polynomials in the $T_a$. In some contexts, it is interesting to consider generalisations to infinite power series in the 
$T_a$ or to continuous functions of the $T_a$; these will be discussed further in the next section.

In order to state the properties of coderivations on ${\bf S}$ we first note that there is a natural  
coproduct, i.e. a map 
 \be
   \Delta: \;{\bf S}\;\rightarrow\; {\bf S}\otimes {\bf S}\;, 
 \ee  
that given a vector in ${\bf S}$ as an input produces two vectors in ${\bf S}$ as an output.   
It is  defined by 
 \be\label{tensorCoproduct}
  \Delta(x_1\wedge\ldots \wedge x_n) \ = \ \sum_{i=1}^{n-1} \sum_{\sigma\in (i,n-i)}
  \epsilon(\sigma; x)\, (x_{\sigma(1)}\wedge\ldots \wedge x_{\sigma(i)})\,\otimes \, 
  (x_{\sigma(i+1)}\wedge \ldots \wedge x_{\sigma(n)})\;, 
 \ee
where  the second sum extends over all permutations $\sigma$  for which 
 \be
 \sigma(1)\leq \cdots \leq \sigma(i)\;, \qquad \sigma(i+1)\leq \cdots \leq \sigma(n)\;, 
 \ee
which defines the set of `unshuffles', which we  denote by $(i,n-i)$.  
Let us describe this operation in words. 
The coproduct of  $x_1\wedge\ldots \wedge x_n\in {\bf S}$ is the sum of all possible splittings into 
two tensor factors, the first with $i$ factors, the second with $n-i$ factors and summed over $i=1,\ldots, n-1$, 
where the order of vectors  in the two sets are left unchanged. Thus, the sum only extends over 
unshuffles $\sigma\in (i,n-i)$. 
It is instructive to display this for low tensor powers: 
 \be\label{lowcoproducts}
  \begin{split}
    \Delta(x) &=  0\;, \\
    \Delta(x_1\wedge x_2) &=  x_1\otimes x_2 + (-1)^{x_1x_2}x_2\otimes x_1\;, \\
     \Delta(x_1\wedge x_2\wedge x_3)  &=   x_1\otimes (x_2\wedge x_3)
   +(-1)^{x_1x_2}x_2\otimes (x_1\wedge x_3)+(-1)^{x_3(x_1+x_2)}x_3\otimes (x_1\wedge x_2) \\
    &+(x_1\wedge x_2)\otimes x_3
   +(-1)^{x_2x_3}(x_1\wedge x_3)\otimes x_2+(-1)^{(x_2+x_3)x_1}(x_2\wedge x_3)\otimes x_1. 
  \end{split}
 \ee   
Note that the coproduct is well-defined in the sense that its action on, say, $x_1\wedge x_2$ is the same 
as on $(-1)^{x_1x_2}x_2\wedge x_1$.

This coproduct has the important property of being  \textit{coassociative}, 
which means that the two ways of mapping  
${\bf S}\rightarrow {\bf S}\otimes {\bf S}\rightarrow {\bf S}\otimes {\bf S}\otimes {\bf S}$ give 
the same result, i.e.  the following diagram  commutes: 
\be\label{DIAGRAM}
\begin{array}{cccccccc}{\bf S}&\xlongrightarrow{\Delta} &{\bf S} \otimes {\bf S} &
\\[1.5ex]
\Big{\downarrow}{\Delta}&&\Big{\downarrow}{\Delta \otimes {\bf 1}}&
\\[1.5ex]
{\bf S}\otimes {\bf S}&\xlongrightarrow{{\bf 1}\otimes \Delta }&{\bf S}\otimes {\bf S}\otimes {\bf S} &
\\[1.5ex]
\end{array}
\ee
This condition can be written as 
 \be
  (\Delta\otimes {\bf 1})\circ \Delta  \ = \ ({\bf 1}\otimes \Delta)\circ \Delta \;. \label{eq:coassoc}
 \ee  
As is common we denote by $\circ$ the composition of maps, but in the following we will often leave it out 
when the meaning is clear from the context. 
Note that the diagram (\ref{DIAGRAM}) is the natural `dual' (or mirror image) to the diagram expressing associativity 
of a product $m:{\bf S}\otimes {\bf S}\rightarrow {\bf S}$. 
With the above relations one may convince oneself that the coproduct (\ref{tensorCoproduct}) is coassociative. 
As it is a vector space with coassociative coproduct,
${\bf S}$ is a \emph{coalgebra} (in this case, without a co-unit).

We now define a  coderivation to be a map $D: {\bf S}\rightarrow {\bf S}$ that satisfies the co-Leibniz property 
with respect to the coproduct (\ref{tensorCoproduct}). This means that the following diagram commutes:  
 \be\label{DIAGRAM2}
\begin{array}{cccccccc}{\bf S}&\xlongrightarrow{D} &{\bf S} &
\\[1.5ex]
\Big{\downarrow}{\Delta}&&\Big{\downarrow}{\Delta}&
\\[1.5ex]
{\bf S}\otimes {\bf S}&\xlongrightarrow{{\bf 1}\otimes D+D\otimes {\bf 1} }&{\bf S}\otimes {\bf S} &
\\[1.5ex]
\end{array}
\ee 
This condition can be written as 
\be \label{eq:coder}
  \Delta  D \ = \ ({\bf 1}\otimes D+ D\otimes {\bf 1}) \Delta\;,  
 \ee
where we have now left the composition $\circ$ implicit.  
Again, the diagram (\ref{DIAGRAM2}) is naturally the `dual' to the diagram expressing the Leibniz property of a 
derivation with respect to a product. 
In the following we will work with conventions in which $D$ has an intrinsic degree of $-1$, and is a \emph{left} coderivation in the sense that the action of, say,  
${\bf 1}\otimes D$ is such that 
  $({\bf 1}\otimes D)(x_1\otimes x_2) \equiv  (-1)^{x_1}x_1\otimes D(x_2)$, 
i.e.~one picks up a  sign when moving $D$ past $x_1$ if $x_1$ is odd. 
More generally, one has 
 \be
  ({\bf 1}\otimes f)(g\otimes {\bf 1})=(-1)^{|f||g|}g\otimes f\;, 
 \ee 
for maps $f$, $g$ of intrinsic degrees $|f|$, $|g|$.

With these conventions and the formulae (\ref{tensorCoproduct}), (\ref{lowcoproducts}) 
one may then verify that the   map (\ref{extendedLieAction})  defining a Lie algebra 
does satisfy the co-Leibniz property and is hence a coderivation. Also note that the sum of two coderivations also satisfies the co-Leibniz property and therefore is also a 
coderivation. 
Furthermore, for any coderivation $D$ of odd degree, the square $D^2$ is also a coderivation (of even degree): 
 \be
 \begin{split}
  \Delta D^2=\Delta DD&=({\bf 1}\otimes D+D\otimes {\bf 1})\Delta D\\
  &=({\bf 1}\otimes D+D\otimes {\bf 1})({\bf 1}\otimes D+D\otimes {\bf 1})\Delta \\
  &=({\bf 1}\otimes D^2-D\otimes D+D\otimes D+ D^2\otimes {\bf 1})\Delta  \\
  &=({\bf 1}\otimes D^2+ D^2\otimes {\bf 1})\Delta\;, 
 \end{split}
 \ee
where we used the co-Leibniz rule twice together with $({\bf 1}\otimes D)(D\otimes {\bf 1})=-D\otimes D$, 
which follows since $D$ had odd degree. This proves that $D^2$ obeys the co-Leibniz rule and is hence a coderivation. 
[Note that, on the contrary, the composition of arbitrary coderivations in general is not a coderivation.] 
Since a coderivation such as $D^2$ acts by decomposing a monomial into sub-monomials and then acting on them, 
and since for Lie algebras $D^2=0$ holds on monomials up to and including cubic ones 
it follows that $D^2=0$ holds on arbitrary monomials. 

\subsubsection{\lf-algebras formulated
in terms of coderivations }

After this introduction into coproducts and  coderivations, we can now efficiently define \lf-algebras 
in terms of coderivations that square to zero and that generalise (\ref{Liebracket}) by having arbitrary higher brackets. 
Specifically, we consider coderivations of the form 
 \be\label{Dwithb's}
  D  \equiv  \sum_{i=1}^{\infty} \, b_i\;,  
 \ee
arising from graded symmetric multilinear maps $b_i: X^{\otimes i}\rightarrow X$, which we assume to have intrinsic degree $-1$ (so that $D$ has intrinsic degree $-1$). 
In particular, $b_2$ corresponds to the original Lie bracket, 
but we also allow for a map $\partial\equiv b_1:X\rightarrow X$ that equips the vector space with a differential. 
The notation in the introduction is related to the $b_i$ via $b_{i}(\,\cdot\,, \ldots, \,\cdot\,)\equiv \big[\,\cdot\,, \ldots, \,\cdot\,\big]$.
The $b_i$ act on the full tensor algebra 
as coderivations. Specifically, for the graded symmetric algebra the $b_i$ are  extendable to maps 
 \be
  b_i: \; S^jX \; \rightarrow \; S^{j-i+1}X\;. 
 \ee
For $j<i$  these maps act trivially, and for $j\geq i$ we define, in analogy to (\ref{extendedLieAction}), 
 \be
  b_i({x}_1\wedge \ldots \wedge {x}_j) =  \sum_{\sigma\in (i,j-i)}\epsilon(\sigma;{x})\;
  b_i\big({x}_{\sigma(1)}, \ldots,{x}_{\sigma(i)}\big)\wedge {x}_{\sigma(i+1)}\wedge \ldots 
  \wedge {x}_{\sigma(j)}\;, 
 \ee
where the sum extends over all un-shuffles in $(i,j-i)$. 
We can now define an \lf-algebra: \\[1ex]
{\bf Definition:}\\
An \lf-algebra is  a $\mathbb{Z}$-graded vector space 
equipped with multilinear graded symmetric maps $b_i: X^{\otimes i}\rightarrow X$ of intrinsic degree $-1$ so that the 
coderivation $D =  \sum_{i=1}^{\infty}  b_i$ is nilpotent: $D^2=0$. 

\medskip

Let us work out explicitly the lowest-order relations for \lf-algebras. 
The condition $D^2=0$ can be written as 
 \be\label{Dsquaredb}
  D^2 = \sum_{n=1}^{\infty}\sum_{i+j=n+1} b_i b_j =0\;, 
 \ee
and implies infinitely many relations, one for each integer $n$. 
For $n=1,2,3$ they become 
 \be\label{firstHtree}
 \begin{split}
  b_1^2 &= 0\;, \\
  b_1b_2+b_2b_1 &= 0\;, \\
  b_2^2+b_1b_3+b_3b_1 &=0\;. 
 \end{split}
 \ee
The first two relations state that $b_1$ squares to zero and is a derivation of the 2-bracket $b_2$.
Let us write out the second relation by acting on $x_1\wedge x_2$: 
 \be
 \begin{split}
  0 = (b_1b_2+b_2b_1)(x_1\wedge x_2)&=b_1(b_2(x_1,x_2))+b_2(b_1(x_1)\wedge x_2+(-1)^{x_1}x_1\wedge b_1(x_2))\\
  &=b_1(b_2(x_1,x_2))+b_2(b_1(x_1), x_2)+(-1)^{x_1}b_2(x_1, b_1(x_2))\;. 
 \end{split}
 \ee
 We will sometimes write $b_1$ as $\partial$ and $b_2$ as $[\cdot,\cdot]  $. Then the first two relations in (\ref{firstHtree}) become $\partial ^2=0$ and
 \be  
 \partial [x_1,x_2] =-  [ \partial x_1,x_2] - (-1)^{x_1}   [x_1, \partial x_2] 
 \ee
 so that  $\partial=b_1$ indeed  acts as a derivation on $[\cdot,\cdot]\equiv b_2$. Similarly, upon evaluating the third relation of (\ref{firstHtree}) on $x_1\wedge x_2\wedge x_3$ 
one obtains 
 \be\label{n=3Relation}
 \begin{split}
  0 = &\;b_2(b_2(x_1,x_2),x_3)+(-1)^{x_2x_3}b_2(b_2(x_1,x_3), x_2)+(-1)^{x_1(x_2+x_3)}b_2(b_2(x_2,x_3),x_1) \\
  &+b_1(b_3(x_1,x_2,x_3))\\
  &+b_3(b_1(x_1),x_2,x_3)+(-1)^{x_1}b_3(x_1,b_1(x_2),x_3)+(-1)^{x_1+x_2}b_3(x_1,x_2,b_1(x_3))\;. 
 \end{split}
 \ee
In the first line we recognize the (graded) Jacobiator, and so this relation tells us that the Jacobi identity need not be satisfied. 
Its failure is then governed by the three-bracket $b_3$. Analogously, one may write out (\ref{Dsquaredb}) for any $n$ and thereby 
obtain the $L_{\infty}$ relations in terms of brackets. 

Next, we introduce the notion of \emph{cyclic} $L_\infty$ algebra. This is an $L_\infty$ algebra equipped with an inner product $\kappa : X\times X \rightarrow \mathbb{R}$ such that
\begin{equation}
     \kappa(x_1, b_n(x_2, x_3, \dots, x_{n+1}) ) = (-1)^{x_1 x_2} \kappa( x_{2}, b_n(x_1, x_3, \dots x_{n+1}) ) \, ,
\end{equation}
so that $\kappa(x_1, b_n(x_2, x_3, \dots, x_{n+1}) )$ defines a graded symmetric map from 
$S^{n+1}X$
to $\mathbb{R}$.
Then this gives the inner product used in the introduction, with
$ \langle \Psi,  \Phi \rangle= 
 \kappa( \Psi,  \Phi)$.

We now define the notion of $L_{\infty}$ morphisms. It can be characterized by a collection of graded symmetric maps $F=(f_1,f_2,f_3,\ldots)$ of intrinsic degree zero from ${\bf S}(X)$ 
to ${\bf S}(X')$ for some vector space $X'$ (which may be the same as $X$), so that $f_{n}:S^n(X)\rightarrow X'$. One first demands this to be a morphism of coalgebras, which means 
that the coproduct is preserved in that the coproduct $ \Delta' $ on ${\bf S}(X')$ satisfies
\be\label{coalgebramorphism}
 \Delta' F = (F\otimes F) \Delta \;. 
\ee
This relation can be taken to determine how the individual maps are extended to a map ${\bf S}(X)\rightarrow {\bf S}(X')$, i.e. how $f_1,f_2,f_3,\ldots$ should act 
on monomials of arbitrary power. 
For instance, the action on $S^1(X)=X$, $S^2(X)$ and $S^3(X)$ is given by 
 \be\label{coalgebramorphismaction}
  \begin{split}
   F(x)  \equiv&\, f_1(x)\;, \\
   F(x_1\wedge x_2) \equiv&\, f_2(x_1,x_2)+f_1(x_1)\wedge f_1(x_2)\;, \\
   F(x_1\wedge x_2\wedge x_3) \equiv &\, f_3(x_1,x_2,x_3) +f_2(x_1,x_2)\wedge f_1(x_3)+(-1)^{x_2x_3}f_2(x_1,x_3)\wedge f_1(x_2) \\
   &+(-1)^{(x_2+x_3)x_1}f_2(x_2,x_3)\wedge f_1(x_1)+ f_1(x_1)\wedge f_1(x_2)\wedge f_1(x_3)\;. 
  \end{split}
 \ee 
It is straightforward to verify with (\ref{lowcoproducts}) that (\ref{coalgebramorphism}) is satisfied. 
A coalgebra morphism is then a morphism of \lf-algebras if both $X$ and $X'$ carry an $L_{\infty}$ structure, encoded in 
coderivations $D$  acting on ${\bf S}(X)$ and $D'$  on ${\bf S}(X')$ that commute with $F$, 
 \be\label{Dcommut}
  D'F = FD\;. 
 \ee

\subsection{Homology and Homotopy Equivalence }
\label{sec:homologyhomotopy}

Any \lf-algebra is, in particular,  a \textit{chain complex}: an integer-graded vector space (\ref{gradedVectorspace}) 
equipped with a differential $\partial: X_i\rightarrow X_{i-1}$ squaring to zero, $\partial^2=0$. 
There is an associated notion of \textit{homology}, encoded in the space of $\partial$-closed vectors modulo 
$\partial$-exact vectors.  We will show that for a projection from $X$ to the vector subspace $\bar{X}$ that preserves the 
homology (quasi-isomorphism) the $L_{\infty}$ structure on $X$ is transported to an $L_{\infty}$ structure on $\bar{X}$.

We begin by considering  two chain complexes, 
  \be\label{Hextendedvector2}
  \begin{split}
  \cdots \ \ \longrightarrow\; & X_{i+1} 
\xlongrightarrow{\partial} \; X_{i} \xlongrightarrow{\partial} \; X_{i-1}  \longrightarrow  \ \ \ldots \ \\
  \cdots \ \ \longrightarrow\; & \bar{X}_{i+1} 
\xlongrightarrow{\bar{\partial}} \; \bar{X}_{i} \xlongrightarrow{\bar{\partial}} \; \bar{X}_{i-1}  \longrightarrow \ \ \cdots \
  \end{split}
 \ee
and assume that $\bar{X}$ is obtained from $X$ by means of a projection operator $p$: 
 \be
  \bar{X}_i  =  p(X_i)\;. 
 \ee
Since we view $\bar{X}$ as the subspace of $X$ obtained by projection, we can define  
the  inclusion map $\iota : \bar{X}\rightarrow X$ that regards a vector  of $\bar{X}$ as a vector in $X$. 
By definition, it satisfies 
\be
p\circ \iota={\rm id}_{\bar{X}}\;,
 \ee
 where $id$ is the identity map.
Furthermore, we assume that $p$ is a \textit{chain map}, meaning that it commutes with the differentials: 
 \be\label{derproj}
  p\circ \partial  = \bar{\partial}\circ p\;. 
 \ee
(This  equation reads more precisely  $p_{i-1}\circ \partial_i = \bar{\partial}_i\circ p_i$, with  
the subscript indicating the space on which the operator acts;  similar comments apply to many of the following equations.) 
We will often use the notation that for $\bar{x}\in\bar{X}$: 
 \be\label{genNot}
  x  :=  \iota(\bar{x}) \in  X \quad \Rightarrow \quad p(x)  =  (p\circ \iota)(\bar{x}) =  \bar{x}\;, 
 \ee
i.e. we distinguish vectors in $\bar{X}$ (and operators on $\bar{X}$) from those in $X$ by a bar.  
The differential projects naturally according to 
 \be\label{derREL}
  \bar{\partial}\bar{x} \ = \ (\bar{\partial} \circ p)(\iota(\bar{x})) \ = \ (p\circ \partial)(x) \ = \ p(\partial x)\;. 
 \ee

Let us now turn to the notion  of homology. The $i$-th homology for the chain complex $X$ is the vector space 
 \be
  H_i  \equiv  \frac{{\rm ker}(\partial_i)}{{\rm im}(\partial_{i+1})}  \equiv  \big\{\, [x]\,|\, x\in X_i\,,\; \partial x=0\,\big\}\;,  
 \ee
where we denote by $[x]$ the equivalence class containing $x\in {\rm ker}(\partial_i)$, with the 
equivalence relation $x \sim x+\partial u$ that identifies two vectors that differ by a $\partial$ exact vector. 
The analogous definition holds for the $i$-th homology space 
$\bar{H}_i$ of $\bar{X}$. We will now show that the projection $p:X_i\rightarrow \bar{X}_i$ induces a well-defined map 
of homologies: 
 \be
  H_i  \ni [x] \; \mapsto \; [p(x)]  \in  \bar{H}_i\;. 
 \ee
In order to show that this map is well-defined, 
we have to verify that $[p(x)] \in\bar{H}_i$ and that the map does not depend on the chosen representative $x$. 
The first follows since $\bar{\partial}(p(x))=p(\partial x)=0$ by (\ref{derproj}) and  $x\in {\rm ker}(\partial_i)$. 
To see that the map does not depend on the representative $x$, we note that $x+\partial u$ is mapped 
to $[p(x+\partial u)]=[p(x)+\bar{\partial}(p(u))]=[p(x)]$,  again using  (\ref{derproj}), and thus is mapped to the same equivalence class. 
Thus, the projection $p$ induces a well-defined map of homologies, but in general this map is not invertible 
and so in general $H_i$ and $\bar{H}_i$ are not isomorphic. 
In the special case that $H_i$ and $\bar{H}_i$ 
are isomorphic, the chain map $p$ is said to be  a \textit{quasi-isomorphism}.

In the following we will study conditions under which the projection is a quasi-isomorphism. 
To this end we need a definition: 
Two chain maps $f, g: X\rightarrow{X}$ (of intrinsic degree zero) are \textit{chain homotopic}, 
in symbols 
$f \sim g$,  if there is a  map $h:X\to X$ consisting of a collection of
maps $h_i: X_i\rightarrow {X}_{i+1}$ such that 
 \be
  f_i - g_i \ = \  {\partial}_{i+1}\circ h_i + h_{i-1}\circ \partial_i\;. 
 \ee
Chain homotopic maps are useful because if $f \sim g$ then
the map that $f$ induces on homologies is the same as the
  map that $g$ induces on homologies: 
 \be
  H_i \ \ni \ [x] \; \mapsto \; [f_i(x)] \ = \ [g_i(x)+{\partial}_{i+1}(h_ix)+h_{i-1}(\partial_ix)] \ = \ [g_i(x)]\;, 
 \ee
using $\partial_ix=0$ and the equivalence relation ${x}\sim {x}+{\partial}{u}$.

We now assume that the composition of inclusion and projection is homotopic to 
the identity on ${X}$, $\iota\circ p\sim {\rm id}_{X}$, so that  there is a map $h:X\to X$ such that
 \be\label{homotopyID}
  p\circ \iota \ = \ {\rm id}_{\bar{X}}\;, \qquad \iota\circ p \ = \ {\rm id}_{X} + \partial\circ h + h\circ \partial\;. 
 \ee
 The projection $p$ then induces as isomorphism on homologies, as we now discuss.
First, let $\bar{x}\in \bar{X}$ and 
set, as in (\ref{genNot}), $x=\iota(\bar{x})$, so that $p(x)=\bar{x}$. 
We then obtain with (\ref{homotopyID})
 \be
  \partial x \ = \ \partial(\iota(\bar{x})) \ = \ \partial((\iota\circ p)(x)) \ = \ \partial (x+\partial(hx)+h(\partial x))
  \ = \ \partial x + \partial(h(\partial x))\;, 
 \ee
using $\partial^2=0$. From this we infer $\partial(h(\partial x))=0$ or, more precisely, 
 \be\label{strangeRELLL}
  \partial(h\partial(\iota(\bar{x}))) = 0\;. 
 \ee
Next, using this relation and acting with $\iota$ on (\ref{derREL}) we obtain 
 \be\label{usefulRELLL}
  \iota(\bar{\partial}\bar{x})  =  (\iota\circ p)(\partial x)  = \partial x+ \partial(h(\partial x))  =  \partial x\;. 
 \ee
This  means that 
 \be\label{iotadelcomm}
  \partial\circ \iota  =  \iota\circ \bar{\partial}\;, 
 \ee
as a map on $\bar{X}$, which is  analogous to (\ref{derproj}).   
Recalling (\ref{derREL}), we have learned that
 \be\label{derunimap}
  \bar{\partial}\bar{x}  =  p(\partial x)\;, \quad   \partial x =  \iota(\bar{\partial}\bar{x})\;, \quad {\rm where} \;\;
  x =  \iota(\bar{x})\;. 
 \ee
Thus, exact vectors and closed vectors in $X$ and $\bar{X}$ are mapped into each other under $p$ and $\iota$, 
confirming that the homologies are isomorphic. 
In particular, there is an inverse map $\bar{H}_i\rightarrow H_i$ defined as follows. 
Let 
 $[\bar{x}]\in \bar{H}_i$  be a class represented by $\bar{x}$, 
with $\bar{\partial}\bar{x}=0$, then $x:=\iota(\bar{x})$ satisfies $\partial x=0$ and 
 we can take  $[x]\in H_i$ as the image of $ [\bar{x}]\in \bar{H}_i$.  Thus the map is $[\bar{x}] \mapsto [\iota(\bar{x}) ]$.

We close this subsection by establishing some useful relations. 
Acting with $p$ on the second equality of (\ref{homotopyID}) 
and using $p\circ \iota={\rm id}_{\bar{X}}$ one obtains 
\be
 p = (p\circ \iota)\circ p  = p\circ (\iota\circ p) =  p\circ ({\rm id}_X+\partial\circ  h+ h\circ \partial)
 \ = \ p + p\circ (\partial \circ h+ h\circ \partial)\;, 
\ee
from which we infer 
 \be\label{anotherprojrel}
 p\circ (\partial \circ h+ h\circ \partial)  = 0\;, 
 \ee 
i.e. for all $x\in X$: $p((\partial h+ h\partial)(x))  =  0$. Similarly, acting with $\iota$ on $p\circ \iota={\rm id}_{\bar{X}}$ one obtains 
\be
 \iota  =   \iota\circ (p\circ \iota)  =   (\iota\circ p)\circ \iota =  ({\rm id}+\partial \circ h+h\circ \partial)\circ \iota
 =  \iota + (\partial\circ  h+h\circ \partial)\circ \iota\;, 
 \ee
from which we infer  
 \be\label{iotaprojectorrel}
  (\partial h+h\partial)\iota(\bar{x}) =  0\;, 
 \ee 
for all $\bar{x}\in \bar{X}$.

\subsection{Homotopy Transfer}
\label{sec:homotopytransfercoalgebra}

We are now ready to state and prove the homotopy transfer theorem for \lf-algebras in the following form: 
If the projection $p:X\rightarrow \bar{X}$ is a quasi-isomorphism, then an \lf-algebra on $X$
can be transported to an \lf-algebra on $\bar{X}$. More specifically,  we will show that a coderivation $D$ on 
${\bf S}(X)$ that squares to zero and hence defines an \lf-algebra gives rise a to coderivation $\bar{D}$
on ${\bf S}(\bar{X})$ that squares to zero and hence defines an \lf-algebra on $\bar{X}$, 
thereby providing explicit formulas for the homotopy transfer.

To this end, we will have to extend the definition of the projection $p$, inclusion $\iota$ and homotopy map $h$ to 
the tensor algebra ${\bf S}$. 
The inclusion and projection $\iota$ and $p$ have natural extensions to ${\bf S}(\bar{X})$ and ${\bf S}(X)$, respectively,  
acting as morphisms: 
 \be\label{iotaPext}
 \begin{split}
  \iota(\bar{x}_1\wedge\bar{x}_2\wedge\cdots) \ &= \ \iota(\bar{x}_1)\wedge \iota({\bar{x}_2})\wedge\cdots\;, \\
  p(x_1\wedge x_2\wedge \cdots ) \ &= \ p(x_1)\wedge p(x_{2})\wedge \cdots \;. 
 \end{split}
 \ee
It is more subtle to extend the action of $h$, which can be done, following \cite{Lazarev,berglund}, by the so-called 
$(\iota p, {\bf 1})$-Leibniz rule. It is defined recursively by 
 \be\label{strangeLeibniz2}
 \begin{split}
  h(x_1\wedge \ldots\wedge x_n) \equiv \frac{1}{n!}\sum_{\sigma\in S_n}&\epsilon(\sigma;x)\Big(
  h(x_{\sigma(1)}\wedge\ldots\wedge x_{\sigma(n-1)})\wedge  x_{\sigma(n)}\\
  &\;\;\;\quad +(-1)^{x_{\sigma(1)}+\cdots x_{\sigma(n-1)}}\iota p(x_{\sigma(1)}\wedge\ldots \wedge x_{\sigma(n-1)})\wedge   h x_{\sigma(n)}\Big)\;. 
  \end{split} 
 \ee
Note that, according to (\ref{iotaPext}),  the action of $\iota p$ in here is given by 
$\iota p(x_{\sigma(1)}\wedge\ldots \wedge x_{\sigma(n-1)})=(\iota p)(x_1)\wedge \ldots \wedge (\iota p)(x_{\sigma(n-1)})$.
As usual, it is instructive to describe the action (\ref{strangeLeibniz2}) in words: $h$ acts like a derivation, with the important addition 
that the factors `over which $h$ has jumped' are acted upon by $\iota p$. Moreover, at the end one symmetrizes in the  arguments by 
summing over all permutations $\sigma \in S_n$ with the appropriate Koszul signs. This step is necessary so that the action 
of $h$ on graded symmetric arguments is well-defined. 
For instance,  the $h$ action on quadratic monomials reads 
 \be\label{quadraticH}
  h(x_1\wedge x_2) = \frac{1}{2}\big(hx_1\wedge x_2+(-1)^{x_1}\iota p(x_1)\wedge hx_2+(-1)^{x_1x_2}hx_2\wedge x_1
  +(-1)^{(x_1+1)x_2}\iota p(x_2)\wedge h x_1\big)\;. 
 \ee

The claim is that the action of  $h$ so defined  has the property of lifting the homotopy relation (\ref{homotopyID}) 
to the entire symmetric algebra ${\bf S}(X)$: 
 \be\label{iotaPRELLL}
  \iota p =  {\bf 1} + \partial h +h\partial\;:\;\; {\bf S}(X) \,\rightarrow\, {\bf S}(X)\;. 
 \ee
A useful  corollary is that then $\partial$ and $\iota p$ commute as operators on the entire {\bf S}(X): 
 \be\label{commutationnn}
  [\partial, \iota p]  =  \partial(\partial h+h\partial) - (\partial h+ h\partial)\partial  =  0\;. 
 \ee
The proof of (\ref{iotaPRELLL}) proceeds by induction on the power $n$ of the monomials 
on which both sides act. The details are presented in an appendix.

We are now ready to prove the homotopy transfer theorem. 
We assume that $X$ carries a general \lf-algebra, encoded in 
the coderivation 
 \be
  D= \partial+B \;, \qquad B\equiv b_2+b_3+b_4+\cdots\;, 
 \ee
satisfying $D^2=0$. Therefore, 
 \be\label{Dsquaredwritten}
  \partial B+B\partial+B^2=0\;. 
 \ee
The claim is that the homotopy-transported $L_{\infty}$ structure on $\bar{X}$ is encoded in the coderivation 
 \be
  {\bar D} =  \bar{\partial}+\bar{B}\;, 
 \ee
where $\bar{B}$ is given by 
 \be\label{BFormula}
 \begin{split}
  \bar{B}  &=   p\left(B +B hB+B hB hB+\cdots\right)\iota\\
  &=  p\left(\sum_{k=0}^{\infty}B (hB)^k\right)\iota
 =p\left(\sum_{k=0}^{\infty} (B h)^kB \right)\iota\,.
 \end{split}
 \ee
We have to verify that $\bar{D}$ is a coderivation and that it squares to zero, 
i.e. satisfies the analogue of (\ref{Dsquaredwritten}), 
  \be
  \bar{\partial}\bar{B}+\bar{B}\bar{\partial}+\bar{B}^2=0\;. 
 \ee
This is verified  by a direct computation, using  the homotopy relation (\ref{iotaPRELLL}) and recalling 
 \be
  \begin{split}
   p\partial&=\bar{\partial}p\;, \quad \partial\iota = \iota\bar{\partial}\;. 
  \end{split}
 \ee
 We then compute (remembering that we are manipulating maps, which each operate on everything to the right of them)
 \be\label{firsttwo}
 \begin{split}
  \bar{\partial}\bar{B}=p\left(\sum_{k=0}^{\infty}\partial B (hB)^k\right)\iota\;, \qquad 
  \bar{B}\bar{\partial}=p\left(\sum_{k=0}^{\infty}(B h)^kB \partial\right)\iota\;, 
 \end{split}
 \ee
and 
 \be\label{secondTerm}
 \begin{split}
  \bar{B}^2 &= p\left(\sum_{k=0}^{\infty}\sum_{l=0}^{\infty}(B h)^kB\,\iota p\, (B h)^l B \right)\iota \\
  &= p\left(\sum_{k=0}^{\infty}\sum_{l=0}^{\infty}(B h)^kB({\bf 1}+\partial h+h\partial) (B h)^l B \right)\iota \\
  &\equiv p(U+V+W)\iota\;, 
 \end{split}
 \ee
where we defined 
 \be
  \begin{split}
   U&\equiv \sum_{k,l=0}^{\infty}(B h)^kB(B h)^lB\;,  \\
   V&\equiv  \sum_{k,l=0}^{\infty}(B h)^kB \partial h(B h)^lB\;,  \\
   W&\equiv  \sum_{k,l=0}^{\infty}(B h)^k B h\partial (B h)^lB\;. 
  \end{split}
 \ee
We will now manipulate these three structures so that (\ref{Dsquaredwritten}) can be used to show that 
the sum of (\ref{firsttwo}) and (\ref{secondTerm}) vanishes. For $U$ we find 
 \be
 \begin{split}
  U&=\sum_{k,l=0}^{\infty}(B h)^kB^2 (hB )^l =\sum_{k=1}^{\infty}\sum_{l=0}^{\infty}(B h)^kB^2 (hB )^l 
  +\sum_{l=0}^{\infty}B^2 (hB )^l  \\
  &= \sum_{k=1}^{\infty}\sum_{l=1}^{\infty}(B h)^kB^2 (hB )^l +\sum_{k=1}^{\infty}\left((B h)^kB^2+B^2(hB)^k\right)+B^2\;. 
 \end{split}
 \ee
For $V$ we find, upon employing an index shift,   
 \be
  \begin{split}
   V&= \sum_{k,l=0}^{\infty}(B h)^kB\partial(hB)^{l+1}=\sum_{k=0}^{\infty}\sum_{l=1}^{\infty}(B h)^kB\partial(hB)^{l}\\
   &=\sum_{k=1}^{\infty}\sum_{l=1}^{\infty}(B h)^kB\partial(hB)^{l} +\sum_{l=1}^{\infty} B\partial(hB)^{l}\;. 
  \end{split}
 \ee 
Finally, for $W$ we find  
 \be
 \begin{split}
  W&= \sum_{k,l=0}^{\infty}(B h)^{k+1}\partial B (hB)^l = \sum_{k=1}^{\infty}\sum_{l=0}^{\infty}(B h)^k\partial B (hB)^l\\
  &=\sum_{k=1}^{\infty}\sum_{l=1}^{\infty}(B h)^{k}\partial B (hB)^l+\sum_{k=1}^{\infty}(Bh)^{k}\partial B\;. 
 \end{split}
 \ee 
Therefore, 
 \be
 \begin{split}
  U+V+W&= \sum_{k=1}^{\infty}\sum_{l=1}^{\infty}(B h)^k(B^2+B\partial +\partial B)(hB )^l \\
 &\;\;\;\;+\sum_{k=1}^{\infty}\left((B h)^k(\partial B +B^2)+(B \partial+B^2)(hB)^k\right)+B^2\\
 &=\sum_{k=1}^{\infty}\left((B h)^k(\partial B +B^2)+(B\partial+B^2)(hB)^k\right)+B^2\;, 
 \end{split}
 \ee
where we used  (\ref{Dsquaredwritten}). 
Finally, adding up 
(\ref{firsttwo}) and (\ref{secondTerm}), 
and using (\ref{Dsquaredwritten}) once more, one obtains 
 \be
  \begin{split}
    &\bar{\partial}\bar{B}+\bar{B}\bar{\partial}+\bar{B}^2\\
    &\quad =p\left(\sum_{k=1}^{\infty}\left((B h)^k(\partial B+B\partial+B^2)+(\partial B+B \partial+B^2)(hB)^k\right)
    +\partial B +B \partial+B^2\right)\iota\\
    &\quad =0\;. 
  \end{split}
 \ee 
This completes the proof that $\bar{D}$ squares to zero.

In order to complete the proof that $\bar{D}$ defines an \lf-algebra on $\bar{X}$ we have to show that 
$\bar{D}$ acts as a coderivation. This is actually not true in general, but only provided $h$ satisfies
certain constraints that are usually referred to as
\emph{side conditions}. 
More abstractly, we want to establish that there is an `$L_{\infty}$-morphism' from $X$ to $\bar{X}$, i.e. a map $F=(f_1, f_2,\ldots):{\bf S}(X)\rightarrow {\bf S}(\bar{X})$ 
that is a coalgebra morphism and respects the coderivation $D$ in that 
 \be\label{morphism}
  F D = \bar{D} F 
 \ee
holds, where both sides of the equation are regarded as maps ${\bf S}(X)\rightarrow {\bf S}(\bar{X})$. Here the map $F$ is explicitly given by\footnote{Note that,
 while $F$ is defined as an infinite series,  its action on any polynomial in the basis elements $T_a$ of $X$ yields a polynomial  in the basis elements   of $\bar X$  since the sum terminates. Thus $F$ is indeed a map ${\bf S}(X)\rightarrow {\bf S}(\bar{X})$. }
 \be
 \label{Fformula}
  F = p({\bf 1}-Bh)^{-1} \equiv p({\bf 1}+Bh+BhBh+\cdots)\;. 
 \ee
The compatibility condition (\ref{morphism}) can be verified using (\ref{iotaPRELLL}) and (\ref{Dsquaredwritten}), with a calculation very similar 
to the derivation above.

While $F$ is not invertible it does have a `right-inverse' $\widetilde{F}:
   {\bf S}(\bar{X})  \rightarrow {\bf S}(X) $, defined by 
 \be
  \widetilde{F}=({\bf 1}-Bh)\iota\;, 
 \ee
satisfying 
 \be\label{weirdinverses}
  F\widetilde{F}={\rm id}_{\bar{X}}
  \;, 
 \ee
as follows immediately with $p\iota={\rm id}_{\bar{X}}$. With (\ref{morphism}) 
one may then express $\bar{D}$ in terms of $D$: 
 \be
  \bar{D}=FD\widetilde{F}
   \;. 
 \ee
Note that since $F$ has only  a right-inverse 
it is not possible to 
reconstruct $D$ from $\bar{D}$. In this sense the \lf-algebra on $X$ is larger than the $L_{\infty}$
algebra on $\bar{X}$.

Next, we would like to verify that $F$ and $\widetilde{F}$
are coalgebra morphisms, 
so that they actually define $L_{\infty}$ morphisms. While everything established so far is valid for any homotopy $h$ 
satisfying the basic homotopy relation (\ref{iotaPRELLL}), it turns out that the morphism properties only hold provided 
$h$ obeys the so-called side conditions 
 \be\label{ipsides}
  ph = h\iota = h^2=0\;. 
 \ee
This can be assumed without loss of generality since for any $p$, $\iota$ and $h$ satisfying the homotopy relation (\ref{iotaPRELLL})
one can first find a new homotopy, defined by 
 \be
  h' = ({\bf 1}-\iota p)h({\bf 1}-\iota p)\;, 
 \ee
that satisfies  the same homotopy relation  and $ph' = h'\iota=0$, as follows by a quick computation using $p\iota={\rm id}_{\bar{X}}$ and (\ref{commutationnn}). 
One may then define another homotopy 
 \be
  h''=h'\partial h'\;, 
 \ee
that satisfies also the third condition in (\ref{ipsides})\footnote{To prove this, first notice that $\partial h' h' \partial$ is zero using the homotopy relations, $\partial^2 = 0$ and the side conditions for $h'$:
\be 
\label{dfgdd}
\partial h' h' \partial = (\iota p - {\bf 1} - h'\partial) h' \partial = - h' \partial - h'(\iota p - {\bf 1} - h' \partial) \partial = - h'\partial + h' \partial = 0 \, .
\ee
Then, $(h'')^2 = 0$ follows immediately, and so does the homotopy relation
\begin{equation}
\partial h'' + h''\partial = \partial h' \partial h' + h' \partial h'\partial = (\partial h' + h' \partial)^2 = (\iota p - {\bf 1})^2 = \iota p - {\bf 1}
\end{equation}
since the cross-terms in the third expression vanish
using (\ref{dfgdd}).}.
We will thus assume that $h$ obeys (\ref{ipsides}) (and drop the primes). 
Under these conditions we have the following relations \cite{berglund}
\be\label{module}
\begin{split}
(h\otimes {\bf 1} - {\bf 1} \otimes h) \Delta h &= (h\otimes h) \Delta\, , \\
(p\otimes {\bf 1}) \Delta h &= (p\otimes h) \Delta\,,  \\
({\bf 1}\otimes p) \Delta h &= (h\otimes p) \Delta \,, 
\end{split}
\ee
which follow by an explicit computation using the $(\iota p,{\bf 1})$-Leibniz rule whose details can be found in the appendix.\footnote{\label{footnotehb}Furthermore, defining as before 
$x_1=\iota(\bar{x}_1)$, etc., a crucial simplification is that 
\be
 h(B(x_1\wedge x_2)\wedge x_3)=h(b_2(x_1,x_2))\wedge x_3\;, 
\ee
as follows with (\ref{quadraticH}), using that $hx_3=h\iota(\bar{x}_3)=0$ by the side condition and $\iota p(x_3)=\iota p\iota(\bar{x}_3)=\iota(\bar{x}_3)=x_3$.}  

We will now show that $F$ and $\widetilde{F}$ are coalgebra morphisms. For $\widetilde{F}$ this is trivially the case since under the side 
conditions it reduces to the inclusion $\iota$. The proof for $F$ proceeds as follows. We first note that $F$ obeys the recursive relation 
\be
\label{eq:bigFrecursion}
F =  p + F Bh\;. 
\ee
We then compute 
\be
\begin{split}
(F \otimes F) \Delta Bh &= (F \otimes F)(B\otimes {\bf 1} + {\bf 1} \otimes B) \Delta h \\ 
&= \left[ FB \otimes (p + FBh) + (p + FBh)\otimes FB  \right] \Delta h \\ 
&= \left[ (FB\otimes p) + (p \otimes FB) + (FBh\otimes FB) + (FB\otimes FBh) \right] \Delta h\, , \label{eq:FBline}
\end{split}
\ee
where we used the fact that $B$ is a coderivation and the recursive definition of $F$. The first term in here gives
\be
\begin{split}
(FB\otimes p) \Delta h &= (FB \otimes {\bf 1})({\bf 1}\otimes p) \Delta h = (FB\otimes {\bf 1}) (h\otimes p) \Delta = (FBh \otimes p) \Delta \\
&= (F \otimes p) \Delta - (p\otimes p) \Delta\, ,
\end{split}
\ee
using (\ref{module}) and the recursive property of $F$ again. Similarly, the second term in (\ref{eq:FBline}) is
\begin{equation}
(p \otimes FB) \Delta h = (p\otimes F) \Delta - (p \otimes p) \Delta\, .
\end{equation}
Finally, the last two terms in (\ref{eq:FBline}) can be written as 
\be
\begin{split}
(FB\otimes FB) (-h\otimes {\bf 1} + {\bf 1}\otimes h) \Delta h &= - (FB \otimes FB) (h\otimes h) \Delta = + (FBh \otimes FBh) \Delta \\
&= (F-p)\otimes (F-p) \Delta \\
&= (F \otimes F) \Delta - (p \otimes F) \Delta - (F \otimes p) \Delta + (p\otimes p) \Delta \, ,
\end{split}
\ee
where signs come from $(-1)^{|FB|\,|h|} = -1$ when $h$ `jumps' over $FB$, and we used the property of $h$ in (\ref{module}). Putting things together, we have
\begin{equation}
(F\otimes F) \Delta Bh = (F\otimes F) \Delta - (p\otimes p) \Delta\, ,
\end{equation}
which implies
\begin{align}
(F\otimes F) \Delta ({\bf 1} - Bh) &= (p\otimes p) \Delta = \bar{\Delta} p 
= \bar{\Delta} F ({\bf 1} - Bh)\, ,
\end{align}
where we used the fact that $p$ is a morphism of coalgebras and the recursive definition of $F$. Since $({\bf 1}-Bh)$ is invertible, this gives the desired result
\begin{equation}
(F\otimes F) \Delta = \bar{\Delta} F\, , 
\end{equation}
completing the proof that $F$ is a coalgebra morphism.

Using the morphism property and (\ref{morphism}) we can finally prove that $\bar{D}$ obeys the co-Leibniz rule and is hence a coderivation, 
thereby completing the proof that $\bar{D}$ defines an \lf-algebra on $\bar{X}$. To this end we compute
 \be
 \begin{split}
  ({\bf 1}\otimes \bar{D}+\bar{D}\otimes {\bf 1})\bar{\Delta}  
  &= ({\bf 1}\otimes \bar{D}+\bar{D}\otimes {\bf 1})(F\otimes F)(\iota\otimes \iota)\bar{\Delta}\\
  &= (F \otimes \bar{D}F+\bar{D}F\otimes F){\Delta}\iota\\
  &= (F \otimes F{D}+F{D}\otimes F){\Delta}\iota\\
  &= (F\otimes F)({\bf 1} \otimes {D}+{D}\otimes {\bf 1}){\Delta}\iota\\
  &= (F\otimes F){\Delta}D\iota\\
  &= \bar{\Delta}FD\iota\\
   &= \bar{\Delta}\bar{D}F\iota\\
  &= \bar{\Delta}\bar{D}\;. 
 \end{split}
 \ee
Here we inserted in the first line the identity ${\bf 1}\otimes {\bf 1}=F\iota\otimes F\iota$, using (\ref{weirdinverses}); 
we used (\ref{morphism}) in the third and seventh line;  we used in the second line that $\iota$ is a coalgebra morphism
and in the fifths line that $F$ is a coalgebra morphism; 
and we used in the fifth line that $D$ obeys the co-Leibniz rule. 
This completes the proof that $\bar{D}$ obeys the co-Leibniz rule.

As a result, we have 
the formulae for the new coderivation $\bar D$ defining the new \lf-algebra structure and for the morphism $F:\bS(X)\to\bS(\bar X)$ of \lf-algebras:
\be
\bar D=FD\iota\,,\qquad F=p+FBh\,.
\ee
These are valid whenever the homotopy $h:X\to X$ obeys the side conditions \eqref{sideconditions} and its lift to $\bS(X)$ (also denoted $h$) is defined by \eqref{strangeLeibniz2}.

We close this section by displaying explicitly the homotopy-transported brackets to low orders. 
To simplify the formulas let us assume that $X$ carries a \textit{differential graded Lie algebra} structure: an \lf-algebra 
for which the three-bracket and all higher brackets vanish. 
Thus, the coderivation on ${\bf S}(X)$ is given by $D=\partial+b_2$, where $b_2=[\cdot,\cdot]$ is the (graded) Lie bracket. 
From (\ref{BFormula}) we infer that the coderivation on ${\bf S}(\bar{X})$ is given by $\bar{D}=\bar{\partial}+\bar{B}$, where
 \be\label{barBBB}
  \bar{B}=\bar{b}_2+\bar{b}_3+\cdots = p(b_2+b_2hb_2+\cdots )\iota\;. 
 \ee
In order to determine $\bar{b}_2$ we act on $\bar{x}_1\wedge \bar{x}_2$: 
 \be
  \bar{b}_2(\bar{x}_1\wedge \bar{x}_2) = pb_2\iota(\bar{x}_1\wedge \bar{x}_2) = p(b_2(\iota(\bar{x}_1), \iota(\bar{x}_2)))\;, 
 \ee 
where we used that all terms of higher order in $b_2$ act trivially on $\bar{x}_1\wedge \bar{x}_2$. 
Thus, the transported 2-bracket is given by 
 \be\label{transported2bracket}
  \bar{b}_2(\bar{x}_1,\bar{x}_2) = p([\iota(\bar{x}_1), \iota(\bar{x}_2)])\;. 
 \ee
In words, one maps the two input vectors $\bar{x}_1, \bar{x}_2\in \bar{X}$ via the inclusion map to $X$, then takes them as inputs for 
the bracket of $X$ and finally projects back to $\bar{X}$. 
This is the canonical way to transport the bracket from $X$ to $\bar{X}$, but due to the projection the Jacobi identity for $b_2=[\cdot,\cdot]$ 
does not translate to a Jacobi identity for $\bar{b}_2$. Rather, a non-trivial 3-bracket is needed, which can be computed by acting with (\ref{barBBB}) 
on $\bar{x}_1\wedge \bar{x}_2\wedge \bar{x}_3$:
 \be\label{transported3bracket}
 \begin{split}
  &\bar{b}_3(\bar{x}_1,\bar{x}_2,\bar{x}_3) = p\,b_2\,h\,b_2\,\iota(\bar{x}_1\wedge \bar{x}_2\wedge \bar{x}_3)
  =p \,b_2\,h\,b_2({x}_1\wedge {x}_2\wedge {x}_3)\\
   & = p\,b_2\,h\left(b_2(x_1,x_2)\wedge x_3 + (-1)^{x_2x_3} b_2(x_1,x_3)\wedge x_2+(-1)^{x_1(x_2+x_3)} b_2(x_2,x_3)\wedge x_1\right)\\
   & = p\,b_2 \left(h(b_2(x_1,x_2))\wedge x_3 + (-1)^{x_2x_3} h(b_2(x_1,x_3))\wedge x_2+(-1)^{x_1(x_2+x_3)} h(b_2(x_2,x_3))\wedge x_1\right)\\
   & = p\Big(\big[h\big[x_1,x_2\big],x_3\big]+ (-1)^{x_2x_3} \big[h\big[x_1,x_3\big], x_2\big]+(-1)^{x_1(x_2+x_3)} \big[h\big[x_2,x_3\big], x_1\big]\Big)
  \;, 
 \end{split}
 \ee
where we used the simplification of the $h$ action noted in footnote \ref{footnotehb} above.
Similarly, the higher brackets can be algorithmically determined from (\ref{barBBB}) to any desired order. 
This concludes our discussion of homotopy transfer in the coalgebra formulation.

\section{Homotopy Transfer in the Dual Picture}
\label{homotopytransferviahomologicalvectorfields}

\subsection{Dual picture of $L_\infty$ algebras}\label{sec:dualpicture}

We now turn to the dual picture of $L_\infty$ algebras, which makes contact with the BRST-BV formalism.
Our purpose in this section is to study homotopy transfer in the dual picture and in particular to
provide a different proof of the homotopy transfer theorem, while in the next section we will apply this to field theory. {This picture is originally due to Kontsevich \cite{kontsevich2003deformation}, and the $A_\infty$ version due to Kajiura \cite{Kajiura:2001ng}.}

In the last section, we formulated $L_\infty$ algebras in terms of a set of brackets $b_n$, which were then encoded in terms of a coderivation $D$.
We introduce a basis $T_a$ for $X$,
with each basis vector $T_a$ having  definite degree $|a|=\deg (T_a)$, so that a general vector in $X$ is a linear combination $\lambda^aT_a$ with real coefficients $\lambda^a$. Then
 the brackets are specified by a set of structure constants $C^a_{c_1\dots c_n}$, 
\begin{equation}
\label{bnis}
    b_n(T_{c_1}, \dots,  T_{c_n}) = C^a_{c_1\dots c_n} T_a\, .
\end{equation}
These satisfy a set of generalised Jacobi identities which are elegantly formulated as $D^2=0$.
In the dual picture, the $T_a$ are replaced by dual variables $z^a$ and one introduces a (left) \emph{derivation} $Q$ defined by
\begin{equation}
\label{Qis}
    Q = \sum_{n=1}^\infty \frac{1}{n!} \, C^a_{b_1\dots b_n} z^{b_1} \, \cdots z^{b_n} \,\frac{\partial}{\partial z^a}\, ,
\end{equation}
where the coefficients $C^a_{b_1\dots b_n}$ are   the structure constants of the bracket $b_n$ appearing in (\ref{bnis}).
In this picture, the generalised Jacobi identities are now expressed in the nilpotency of $Q$:
\begin{equation}
\label{q2is0}
    Q^2 = 0\, .
\end{equation}

We will discuss two closely related realisations of this dual picture of  $L_\infty$ algebras.
In this section, we will consider a dual of the formulation of the last section, while in the next section we will take the $z^a$ to be fields 
in the   BRST-BV formalism  in which $Q$ becomes the BRST operator. Both cases have the same algebraic structure governed by an operator $Q$  of the form (\ref{Qis}), but in this section it is a derivation acting on a vector space while in the next section it is a vector field acting on a manifold.

The relation between the   picture of $L_\infty$ algebras in the last section and the dual formulation we will develop in this section is straightforward in the case in which $X$ is 
 finite dimensional. Then the $z^a$ are a basis for the dual space $X^*$
and  the derivation $Q$ is the dual map to the coderivation $D$,  $Q = D^{\star}$.
The isomorphisms
\be
\label{dualdual}
(X^\star)^\star\cong X\,,\quad (X\otimes Y)^\star\cong (X^\star\otimes Y^\star)\,
\ee
can then be used to show the equivalence of the two dual formulations.
Just as $D$ acts on ${\bf S}(X)$, $Q$ can be taken to act on the dual of this, ${\bf S}(X)^{\star} $.
However, in many of the applications to field theory and string theory, the spaces involved are infinite dimensional. If $X$ is infinite dimensional, the $z^a$ do not necessarily form a complete basis for $X^{\star}$ in general and \eqref{dualdual} need not hold, so the relation between the two formulations can be more involved.
We  discuss the finite and infinite dimensional cases in turn below.

We shall henceforth  take $\deg x\to -\deg x$ for all $x\in X$. This ensures that the differential defining the \lf-structure is of degree $+1$ (cohomological convention), in accordance with the BRST-BV literature. This introduces no relative signs (since $-1\equiv 1\mod 2$).

\subsubsection{The Finite Dimensional Case}

In this subsection we  consider the case in which $X$ is finite dimensional,
so that only a finite number of the $X_i$ are non-trivial, and all are finite dimensional.
 Choosing a basis $\{ T_a \}$ for $X$, we introduce a dual basis $\{ z^a \}$ for the dual space $X^{\star}$, the space of linear functions on $X$.
The $z^a$'s have degree opposite to the $T_a \in X$, and the natural pairing $\langle \cdot , \cdot \rangle :X\times X^{\star} \rightarrow \mathbb{R}$ is
\begin{equation}
\label{dualbas}
    \langle z^a, T_b \rangle 
    =z^a(T_b)= \delta^a_b\, .
\end{equation}
The space ${\bf S}(X)$ has a basis given by the graded symmetric monomials $T_{b_1} \dots T_{b_m}$ (where we suppress the $\wedge$ in the following) and can be viewed as the space of graded symmetric polynomial functions in the $T_a$.
Note that, as ${\bf S}(X)$  is infinite dimensional, its dual ${\bf S}(X)^{\star}$ is \emph{not} the infinite direct sum
    $\bS(X^{\star}) = \bigoplus_{n=1}^\infty S^n(X^{\star})$, but rather the infinite direct product
    \begin{equation}
        {\bf S}(X)^{\star} = \prod_{n=1}^\infty S^n (X^\star)\, .
    \end{equation}
    (A direct sum of vector spaces dualises into a direct \emph{product} of the duals, see e.g. \cite[exercise 2.6]{topologicalvectorspaces}, and for finite-dimensional $X$ we have indeed $(S^n(X))^{\star} = S^n(X^{\star})$.) Therefore, $\bS(X)^{\star}$ consists of 
 \emph{formal power series} in $z^a$, to be contrasted with ${\bf S}(X^\star)$ which is a  subspace of $\bS(X)^\star$ that includes only \emph{terminating} power series, which is to say polynomials in $z^a$.
 It will sometimes be convenient to slightly enlarge this to
\begin{equation}
\bS^\star\equiv \mathbb{R} \times {\bf S}(X)^{\star} = \prod_{n=0}^\infty S^n(X^{\star}),
\end{equation}
so the power series are allowed to include constant terms. In particular, this enlargement makes the derivative operator $\pd / \pd z^a$ well-defined.

Conversely, the algebraic dual of  the space of polynomials in $z^a$, ${\bf S}(X^\star)^\star$,
 will include infinite power series in $T_a$, not just  polynomials, and so is larger than ${\bf S}(X)$.
Although graded-symmetric monomials $z^{a_1} \dots z^{a_n}$ are not a basis of ${\bf S}^\star$ --- because ${\bf S}^\star$ contains non-terminating power series --- 
the pairing of $X$ and $X^\star$ extends to $\bS \times \bS^*$ as
\begin{equation}\label{eq:pairingS}
    \langle z^{a_1} \dots z^{a_n}, T_{b_1} \dots T_{b_m} \rangle = \begin{cases} \;0 \text{ if } n \neq m \\
    \; n!\,  \delta^{(a_1}_{b_1} \dots \delta^{a_n)}_{b_n} \text{ otherwise,} \end{cases}
\end{equation}
(with weight one graded symmetrization).

There is a natural product operation $m: \bS^* \otimes \bS^* \rightarrow \bS^*$, given by
\begin{equation}
    m(f_1 \otimes f_2) = f_1f_2\, , 
\end{equation}
where $f_1f_2$ is the product of the two functions $f_1(z^a)$ and $f_2(z^a)$.
It is dual to the coproduct $\Delta$ of last section, in the sense that\footnote{On the right-hand side of this equation is the pairing between the tensor products $\bS^* \otimes \bS^*$ and $\bS \otimes \bS$, defined in the usual way by $\langle f_1 \otimes f_2 , x_1 \otimes x_2 \rangle = \langle f_1, x_1\rangle \langle f_2, x_2 \rangle$. We will use the same notation for all those pairings.}
\begin{equation}
    \langle m(f_1 \otimes f_2), x \rangle = \langle f_1 \otimes f_2, \Delta x \rangle
\end{equation}
for all $f_1$, $f_2 \in \bS^*$ and $x \in \bS$. This can be checked on basis elements using definition \eqref{tensorCoproduct} of the coproduct. The associativity of $m$, i.e. $(f_1 f_2) f_3 = f_1(f_2f_3)$, reads 
\begin{equation}
m\circ (m\otimes {\bf 1}) = m\circ ({\bf 1} \otimes m)
\end{equation}
in this more abstract notation. Taking the dual, this is equivalent to the coassociativity \eqref{eq:coassoc} of $\Delta$.

If $f:V\to W $ is a map between two vector spaces $V,W$, the dual map $f^{\star}: W^{\star}\to V^{\star}$ is defined by 
\be 
\label{dualmap}
f^{\star}(u)= u \circ
 f
 \ee
  for all $u\in W^{\star}$. If $f$ is a linear map, then the matrix representing $f^{\star}$ is the transpose of the matrix representing $f$. In terms of the pairing above, the dual of a map $f:\bS \rightarrow \bS$ satisfies $\langle f^{\star}(u), x \rangle = \langle u, f(x) \rangle$ with $u \in \bS^{\star}$ and $x \in \bS$.

A (left) derivation is a linear map $Q: \bS^* \to \bS^*$ of degree $+1$ which satisfies the (left) Leibniz rule, i.e. for any two monomials $p_1, p_2 \in \bS^*$, we have
\begin{equation}
    Q(p_1 p_2) = Q(p_1) p_2 + (-1)^{\deg p_1} p_1 (Qp_2)\, .
\end{equation}
(We will sometimes write $(-1)^{\deg p_1}=(-1)^{p_1}$.) Linearity then yields the Leibniz rule for non-monomial elements of $\bS^*$.
This can be rewritten as
\begin{equation}
    Q \circ m = m \circ \left(Q \otimes {\bf 1} +{\bf 1} \otimes Q\right)\, .
\end{equation}
The dual of a derivation is then clearly a coderivation, satisfying \eqref{eq:coder}.

In this picture, an $L_\infty$ algebra is  encoded in a \emph{derivation} $Q = D^*$ that squares to zero\footnote{Derivations $Q$ on ${\bf S}^\star$ that arise as the duals of coderivations $D$ on ${\bf S}(X)=\bigoplus_{n=1}^\infty S^n(X)$ also have ``vanishing constant term'': $Q(z^a)|_{z=0}=0$. This is reflected in \eqref{Qis}. One needs the extra condition because ${\bf S}(X)$ is the \emph{reduced} symmetric coalgebra, while the extended $\bf S^\star$ is the dual to the non-reduced symmetric coalgebra which includes $S^0(X)\cong\mathbb R$. Relaxing this condition leads to ``curved'' \lf-algebras, which we shall not consider.},
\begin{equation}
    Q^2 = 0\, ,
\end{equation}
instead of a coderivation on $\bS$.
Here  $Q$ is given by  (\ref{Qis}) where
 the coefficients $C^a_{b_1\dots b_n}$ are   the structure constants defining  the $n$-bracket (\ref{bnis}).
 That $Q$  given by  (\ref{Qis}) is  the  dual of $D = \sum_{n=1}^\infty b_n$ can be seen by
 comparing
\begin{align}
    \langle Qz^a, T_{c_1} \cdots T_{c_m} \rangle &= \langle \sum_{n=1}^\infty \frac{1}{n!} \, C^a_{b_1\dots b_n} z^{b_1} \, \cdots z^{b_n}, T_{c_1} \cdots T_{c_m} \rangle \\
    &= \frac{1}{m!} \, C^a_{b_1\dots b_m} \langle z^{b_1} \cdots z^{b_m}, T_{c_1} \cdots T_{c_m} \rangle  \\
    &= C^a_{c_1\dots c_m}
\end{align}
with
\begin{align}
    \langle D^{\star} z^a, T_{c_1} \cdots T_{c_m} \rangle &= \langle z^a, \sum_{n=1}^\infty b_n(T_{c_1} \cdots T_{c_m}) \rangle \\
    &= \langle z^a, \sum_{n=1}^m \sum_{\sigma \in (n, m-n)} b_n(T_{c_{\sigma(1)}}, \dots, T_{c_{\sigma(n)}}) T_{c_{\sigma(n+1)}} \cdots T_{c_{\sigma(m)}} \rangle \\
    &= \langle z^a, \sum_{\sigma \in (m, 0)} b_m(T_{c_{\sigma(1)}}, \dots, T_{c_{\sigma(m)}}) \rangle \\
    &= \langle z^a, b_m(T_{c_1}, \dots, T_{c_m}) \rangle
    \\
     &= C^a_{c_1\dots c_m} \, .
\end{align}
The relation $Q^2 = 0$ then implies an infinite number of relations among the structure constants; those are just the $L_\infty$ relations written out in a basis.

In this language, the dual of a $L_\infty$ morphism $F : \bS \to \bar{\bS}$ (in the sense of the  last section) is a degree-preserving linear map
\be
F^{\star} : \bar\bS^{\star} \to {\bS}^{\star}
\ee
that preserves both the multiplication and the derivation structure,\footnote{Morphisms of algebras $F^\star:\bar{\bf S}^\star\to{\bf S}^\star$ that arise as the duals of morphisms $F$ of coalgebras also have $F^\star(z^a)|_{z=0}=0$ for the same reason as in the previous footnote. This is reflected in \eqref{algebramorphismincoordinates}.}
\begin{align}
    \begin{cases}F^{\star}(a_1 a_2) = F^{\star}(a_1) F^{\star}(a_2) \quad \forall a_1, a_2 \in \bar{\bS}^{\star}\,, \quad (\Leftrightarrow F^{\star} \circ \bar{m} = m \circ (F^{\star} \otimes F^{\star})) \\
    F^{\star} \circ \bar Q = {Q} \circ F^{\star}\, .
    \end{cases}
\end{align}
The first requirement implies that a morphism is specified by its action on $S^1(\bar X^\star)\cong \bar X^\star$, i.e. by the value of all $F^\star(\bar z^{\bar a})$ with  $\{ \bar z^{\bar a} \}$ a basis for $\bar{X}^\star$. Expanding those as
\be
\label{algebramorphismincoordinates}
F^\star(\bar z^{\bar a})=\sum_{n=1}^{\infty}\frac{1}{n!} F^{\bar a}_{a_1\dots a_n} z^{a_1}\cdots z^{a_n}\,,
\ee
one sees that $F$ is completely specified by the collection of degree-zero linear maps $S^n(X) \to \bar{X}$ defined by the coefficients $F^{\bar a}_{a_1\dots a_n}$.

\subsubsection{The Infinite Dimensional Case}
\label{sec:infinited}

Before entering into the details, it is useful to outline some of the key points that arise in the infinite dimensional case. We start from the coderivation $D$ on a space $X$ with basis $T_a$ and construct the duals $z^a$ of the $T_a$. In the finite dimensional case, the $z^a$ are a basis for the dual space $X^\star$, we define $Q$ as $Q=D^\star$ and, as $(\star)^2=1$, there is a complete equivalence between the two dual formulations. 
However if $X$ is infinite dimensional the $z^a$ are not a complete basis for $X^\star$,  but we can formally restrict ourselves to the subspace $\tilde X \subset X^\star$ spanned by the $z^a$ and our construction then gives a derivation $Q$ acting on the space $\tilde{\bS}$ of  formal (graded symmetric) power series in the $z^a$.
Then there is instead a formal \lq duality' between $X$ and $\tilde{X}$ and relating $D$ acting on functions of $T_a$ to $Q$ acting on functions of $z^a$.

The link between the homotopy transfer formulas which are independently developed in sections \ref{sec:homotopytransfercoalgebra} and \ref{sec:homotopytransferalgebra} can then always be done \emph{in this formal sense}. Since we mostly aim at purely algebraic questions in this paper, this is sufficient for our purposes and looks exactly like the finite-dimensional case; however, it is still useful to know that this can be made precise in some special infinite-dimensional cases. We discuss this in the remainder of this section.

In physics applications, the relevant vector space $X$ is an infinite-dimensional \emph{topological} vector space. The study of \lf-algebras of topological vector spaces appears to have been initiated only very recently \cite[section 2.3]{baarsmathesis}.  We will provide here a sketch of the key features that are relevant for our applications
 along with pointers to relevant literature, to highlight the subtleties in this case.
 In particular,  for topological vector spaces one can consider the continuous dual (defined below)  and restrict to the spaces for  which the double (continuous)  dual of a space is the space itself.
Of particular interest are the special spaces for which many of the properties  we have discussed of the finite dimensional case also apply.

Let us first give more details on the case in which $X$ is any infinite dimensional vector space; we will use standard results on infinite dimensional vector spaces, see e.g. \cite{topologicalvectorspaces}.
 The \emph{algebraic dual} $X^\star$ of $X$ is defined as the vector space of linear functions $X \to \mathbb{R}$.
The algebraic dual does not satisfy (\ref{dualdual}) and moreover the cardinality of a basis of $X^\star$ is greater than that of $X$. For example, if $X$ has a countable basis, then any basis for $X^*$ will be uncountable. Given a basis $T_a$ of $X$, we can define a set of linear functions $z^a:X \to \mathbb{R}$
by
\begin{equation}
 z^a(T_b)= \delta^a_b\, .
\end{equation}
However, the $z^a$ do not form a complete basis for $X^\star$ in the infinite dimensional case.
The dual of a map is again defined as in (\ref{dualmap}). In particular, the dual $D^\star$ of $D$ is now no longer given by $Q$.
However, as discussed above, there is a vector subspace $\tilde X \subset X^\star$ which is spanned by the $z^a$ and then $Q$ can be taken to act on $\tilde{\bS}$.
 Then $Q$ gives the action of $D^\star$ restricted to $\tilde X$, and $Q$ encodes all the structure constants and Jacobi identities that characterise the $L_\infty$ algebra.

 If $X$ is a topological vector space, one can  instead define the \emph{continuous dual} $X'$ of $X$ which is defined as the vector space of \emph{continuous} linear functions $X \to \mathbb{R}$. If $f:V\to W $ is a continuous map between two topological vector spaces $V,W$, the dual map $f': W' \to V'$ is defined by 
\be 
\label{dualmap'}
f '(u)= u \circ f
\ee
for all $u\in W'$.
 
If $X$ is a Hilbert space, then the continuous dual
satisfies $(X')' \cong X$.
In fact,  (\ref{dualdual}) 
 is satisfied if $X$ and $Y$ are separable Hilbert spaces, $X^\star, Y^\star$ are \emph{continuous} duals, and the tensor product is a completed tensor product.   We will be particularly interested in Hilbert spaces for which the $z^a$ define a basis for $X'$ and the dual ${\bf S}(X)'$ of ${\bf S}(X)$ includes  the space of continuous functions of the $z^a$, with the dual $D'$ of $D$ is given by $Q$, as defined in (\ref{Qis}).

Hilbert spaces  behave well under duality and tensor products. However, for \lf-algebras associated with field theories we need a different kind of space: for scalar field theory, for example, 
a vertex such as $\int d^Dx\; (\varphi(x))^D$ is not well-defined if $\varphi$ lies in the Hilbert space of square-integrable functions as the integral may diverge. For that reason we could instead take $X$ to be e.g.~a $D$-dimensional Schwartz space $\mathcal S^D$: the space of smooth functions on $\mathbb R^D$ decaying faster than any polynomial at infinity 
(or, depending on the field content, a sum of Schwartz spaces).
 The continuous dual $(\mathcal S^D)'$ of $\mathcal S^D$ is known as the space of tempered distributions on $\mathbb{R}^D$. Both can be equivalently characterised concretely as spaces of sequences \cite{barrysimon}:
\be
\begin{split}
{\mathcal S}^1\cong \Big\{\{a_n\} \quad \big| \quad a_n\in\mathbb R\,,n\in\mathbb N\,, \sum_{n\geq 0}|a_n|^2 (n+1)^m  \text{ converges for all }  m\in\mathbb N\Big\},\\
({\mathcal S}^1)'\cong \Big\{\{b_n\}\quad \big| \quad b_n\in\mathbb R\,,n\in\mathbb N \,, |b_n|\leq C(1+n)^m  \text{ for some } C\in\mathbb R\,,m\in\mathbb N\Big\}\,,
\end{split}
\ee
with the natural $\ell_2$ pairing $\langle b,a\rangle\equiv \sum_{n\in\mathbb N} a_n b_n$. The coefficients $a_n$ arise in the expansion of a function $\varphi \in \mathcal S^1$ in Hermite functions $\varphi_n$: $\varphi\equiv\sum_{n\geq 0} a_n\varphi_n$; see \cite{barrysimon}. (The Hilbert space $\ell_2$ of square-summable sequences --- corresponding to square-integrable functions --- lies in between these two spaces: $\mathcal S^1\subset \ell_2\subset (\mathcal S^1)'$.) If $X=\mathcal S^1$ we can then define
\be
T_a\equiv\{\delta_{a,n}\}\in\mathcal S^1\,,\quad z^a\equiv \{\delta_{a,n}\}\in (\mathcal S^1)'\,,\qquad a\in\mathbb N\,,
\ee
i.e. sequences $\{a_n\}$, $\{b_n\}$ that have zeros everywhere except a $1$ in position $a$. 
These provide bases of the two spaces (which are in fact topological (Schauder) bases), and in this infinite-dimensional setting all formulas should work as in the finite-dimensional case. The generalisation to $\mathcal{S}^D$ then involves multisequences \cite{barrysimon}.

Schwartz spaces are closed with respect to not just pointwise multiplication of functions but also with respect to the action of differential operators such as the Klein-Gordon operator $(\square+m^2)$. Therefore they can accommodate the kinds of \lf-algebra brackets that will appear for field theories on Minkowski or Euclidean space.
However, $\mathcal S^D$ does not contain nonzero solutions of $(\square+m^2)\phi=0$, because those fail to decay. For the \lf-algebras associated to scalar theories this implies $H(X)=0$ \cite{Arvanitakis:2019ald,Macrelli:2019afx}, and this will be true generally due to the existence of inverses --- for the differential operators defining the linearised equations of motion --- as linear maps on $\mathcal S^D$. The restriction to $H(X) = 0$ is not very relevant for our purposes (see the discussions section); still, these questions involving both $L_\infty$ algebras and functional analysis certainly deserve further study.

To finish this discussion, we can be slightly more general and locate $\mathcal S^D$ and $(\mathcal S^D)'$ in a nice class of topological vector spaces which is closed under (completed) tensor products and behaves manageably under duality just as Hilbert spaces do. In particular, we are interested in spaces for which the duality properties \eqref{dualdual} hold. One such category is \emph{nuclear Fr\'echet} spaces or $\mathcal{ NF}$-spaces, and their duals known as $\mathcal{NDF}$-spaces.
(Properties and definitions of $\mathcal{NF}$ and $\mathcal{NDF}$ spaces are conveniently collected in appendix 2 of Costello's book \cite{costello2011renormalization}.)
In particular,  { $\mathcal S^D$ is $\mathcal{NF}$, and its strong dual $(\mathcal S^D)'$ is $\mathcal{NDF}$}.
 If $V_1,V_2$ are $\mathcal{NF}$, then e.g.~$V_1\otimes V_2$ is also $\mathcal{NF}$, and its strong (continuous) dual is $(V_1\otimes V_2)'\cong V_1'\otimes V_2'$ of type $\mathcal{NDF}$. Both types are closed in particular under finite tensor products, direct products, and direct sums, and are reflexive ($V\cong (V')'$). If $X$ is $\mathcal{NF}$, we can form $\bS(X)$; as it is a countable direct sum of $\mathcal{NF}$ spaces, it is not $\mathcal{NF}$. It is, however, still nuclear, and therefore locally convex. Therefore the strong dual is the direct product \cite[IV.13 Proposition 14]{bourbaki2013topological}
\be
\bS(X)'=\prod_{n=1}^\infty S^n(X')
\ee
just as in the finite-dimensional case. We can also dualise this direct product again to arrive at the direct sum of the duals \cite[IV.14 Proposition 15]{bourbaki2013topological}. Therefore, given a continuous coproduct and continuous coderivations on $\bS(X)$ and $\bS(\bar X)$, one could dualise to arrive at the dual picture of products and derivations, and vice versa.

\subsubsection{Cyclic $L_\infty$ algebras }
\label{section:cyclicLinftyBRSTBV}
As we will discuss in the next section, the dual  picture of $L_\infty$ algebras is closely related to the BRST-BV formalism, which provides   a nilpotent  $Q$ (and therefore an $L_\infty$ algebra) for any field theory, with or without gauge symmetry. 
Recall that  a \emph{cyclic} $L_\infty$ algebra  is an $L_\infty$ algebra equipped with a degree $-1$ inner product $\kappa : X\times X \rightarrow \mathbb{R}$ such that
\begin{equation}\label{eq:kappacyclic}
     \kappa(x_1, b_n(x_2, x_3, \dots, x_{n+1}) ) = (-1)^{x_1 x_2} \kappa( x_{2}, b_n(x_1, x_3, \dots x_{n+1}) ) \, .
\end{equation}
The degree condition means that non-zero products appear only between $X_n$ and $X_{-n+1}$. In components, this corresponds to a matrix $\kappa_{ab} = \kappa( T_a, T_b )$
satisfying $\kappa_{ab}= (-1)^{(a+1)(b+1)}\kappa_{ba}$. Note however that since non-zero components only appear between spaces $X_n$ and $X_{-n+1}$ with degrees of different parity, this sign is always positive and $\kappa$ is actually symmetric, $\kappa_{ab} = \kappa_{ba}$.
In components, equation \eqref{eq:kappacyclic} means that the index-down structure constants $C_{a_1 \dots a_n} \equiv \kappa_{a_1 b}C^b_{a_2\dots a_n}$ have the expected graded symmetry,
\begin{equation}
C_{a_1 a_2\dots a_n}=(-1)^{a_1 a_2}C_{a_2a_1\dots a_n}\, . \label{cyclicity}
\end{equation}

We will assume in the following that this product is \emph{non-degenerate}, and this is the crucial assumption that makes the link with the BV formalism. 
We  write $\kappa^{ab}$ for the inverse matrix. This allows us to define the antibracket $(\cdot, \cdot): \bS^{\star} \times \bS^{\star} \rightarrow \bS^{\star}$ by
\begin{equation}
\label{antibracketdef}
(f, g) = (-1)^{(\deg f)(\deg z^a)} \frac{\partial f}{\partial z^a} \kappa^{ab} \frac{\partial g}{\partial z^b}
\end{equation}
(we always use left derivatives unless explicitly stated otherwise).
This is graded symmetric and satisfies the graded Jacobi identity. 
We  define
\begin{equation}
\label{thetadefinition}
    \Theta = \sum_{n=1}^\infty \frac{1}{(n+1)!} \, C_{a b_1\dots b_n} z^a z^{b_1} \, \cdots z^{b_n}\, .
\end{equation}
The differential $Q$ can now  be written in terms of the antibracket as
\begin{equation}\label{eq:Qtheta}
Q = ( \Theta\, , \, \cdot)\, ,
\end{equation}
and the fact that $Q^2 = 0$ (i.e. all $L_\infty$ relations) is equivalent to 
\begin{equation}
(\Theta, \Theta) = 0\, .
\end{equation}
In mapping to the BV formalism $ \Theta$
becomes the BV master action and \eqref{eq:Qtheta} is the classical master equation.

The non-degeneracy of $\kappa$ implies in particular that the spaces $X_n$ and $X_{-n+1}$ always have the same dimension.
Moreover, (\ref{dualbas}) implies that a given $z^a$ has the opposite degree to the corresponding $T_a$. Let $\tilde X_n$ be the  vector space with basis given by the $z^a$ of degree $n$.
For the finite dimensional case, $\tilde X_n = (X_{-n})^{\star}$ while in general $\tilde X_n \subseteq (X_{-n})^{\star}$. Then the spaces $\tilde X_n$ and $\tilde X_{-n-1}$ always have the same dimension.

We  remark that cyclic \lf-algebras in the dual picture admit a nice geometric interpretation \cite{kontsevich2003deformation} which parallels the geometric interpretation of the BRST-BV formalism \cite{Schwarz:1992nx}: if $\bS^\star$ is interpreted as the ring of functions over a $\mathbb Z$-graded supermanifold\footnote{For this to be true, one needs the enlargement discussed above to include constant functions.}, $Q$ is a homological vector field on it. A morphism $F$ of \lf-algebras is then a map of supermanifolds, defined as the \emph{pullback} map $F^\star$; \emph{invertible} morphisms are identified as diffeomorphisms preserving the homological vector field $Q$:
\be
\bar Q=(F^{-1})^\star\circ Q\circ F^\star\,;
\ee
and a cyclic structure $\kappa$ is equivalent to a degree $-1$, \emph{constant} symplectic form
\be
\kappa\equiv \frac{1}{2}\,dz^a \wedge \kappa_{ab}\, dz^b\,,
\ee
annihilated by $Q$ acting by Lie derivative:
\be
L_Q \kappa=0\,.
\ee
The antibracket $(\,\cdot\, , \,\cdot\,)$ is the graded Poisson bracket of degree $+1$ canonically associated to the degree $-1$ symplectic form $\kappa$; we refer to \cite{Arvanitakis:2018cyo} appendix A for details {on the convention used here. In the context of this geometric interpretation, $Q$ along with the underyling supermanifold are collectively known as ``Q-manifolds'' since \cite{Schwarz:1992gs}.}

\subsection{Homotopy Transfer in the Dual Picture}
\label{sec:homotopytransferalgebra}
  
In the previous section we displayed explicit formulae for homotopy transfer from an \lf-algebra on $X$ --- as defined by a coderivation $D$ on ${\bf S}(X)$ --- into a quasi-isomorphic chain complex $(\bar X,\bar \partial)$, along with an \lf-morphism $F:\bS(X)\to \bS(\bar X)$ extending the projection $p:X\to \bar X$. 
In this section, we shall construct a new \lf-morphism $E:\bS(\bar X)\to \bS(X)$ extending the \emph{inclusion} $\iota: \bar X\to X$, using intuition from the supergeometric intepretation as well as effective field theory. This   provides an explicit construction of a \emph{quasi-inverse} to the map $F$ of the previous section. (A quasi-inverse to a morphism of \lf-algebras is a morphism going ``the other way'' whose linear part inverts the linear part of the other in cohomology.) {In fact $E$ will be shown to be a \emph{strict} partial inverse to $F$ in the sense of \eqref{eq:FE=1}.}

Our chief aim in the rest of this section is to construct the dual map $E^{\star}: \bS^{\star} \to \bar{\bS}^{\star}$, and re-express the results for the new \lf-algebra structure in terms of $E^{\star}$. The map $E^{\star}$ will then give homotopy transfer in the dual picture, complementing the original construction in the coalgebra picture involving $F$.
This is most straightforward in the case in which $X$ is finite-dimensional so that $Q=D^{\star}$ and the $z^a$ provide a dual basis; we use the notation appropriate for this case. Whereas in the last section we considered a map $p$ from a vector space $X$ with basis $T_a$ to a vector space $\bar X$ with basis $\bar T_{\bar a}$ and extended it to a morphism of coalgebras sending a coderivation $D$ acting on functions of $T_a$ to a coderivation $\bar D$ acting on functions of
$\bar T_{\bar a}$, we will here instead consider a map $\iota^\star$ from a vector space with basis $z^a$
to a space with basis $\bar z ^{\bar a}$
and extend this to a morphism of algebras sending a derivation $Q$ acting on functions of $z^a$ to a derivation $\bar Q$ acting on functions of $\bar z ^{\bar a}$.

In the case of general infinite dimensional $X$, we can formally construct  algebraic dual maps $D^{\star},E^{\star},F^{\star}$ acting on the algebraic dual $\bS(X) ^{\star}$
and then restrict these to {$\tilde{\bS}$}, while if $X$ is a topological vector space we can do the same with continuous duals. In both cases, many of the results for the  finite dimensional
case  apply, and in particular we have the same results for homotopy transfer. However, in  the case in which $X$ is
one of the  special infinite dimensional spaces discussed in subsection \ref{sec:infinited} (with $\star$ denoting the appropriate topological dual) that share many of the properties of the finite dimensional case,
there will be a complete equivalence between the two dual formulations, just as there is in the finite dimensional case.
This then looks very similar to the finite dimensional construction.

\subsubsection{Homotopy transfer}

We now turn to the dual picture of homotopy transfer. We shall first consider the case in which $X$ is finite dimensional, and later discuss the infinite dimensional case.
In the previous section, we introduced 
the projection $p:X\to \bar X$ and  the {inclusion} $\iota: \bar X\to X$, and so we also have $P=\iota p: X\to X$.
The dual maps are
$p^{\star}:  \bar X ^{\star}\to X^{\star} $ 
which is an \emph{inclusion} and
$\iota ^{\star}:  X^{\star} \to \bar X^{\star}$ which is a \emph{projection}, together with $P^{\star}=p^{\star}\iota ^{\star} : X^{\star}\to X^{\star}$.
These are then extended to maps of algebras, so e.g.
\be
P^\star:{\bf S}(X^\star)\to {\bf S}(X^\star)\,,\quad P^\star z^a=P^a_b z^b\,,\quad P^\star(z^a z^b)=(P^\star z^a)(P^\star z^b)\,,
\ee
etc.,~similarly to \eqref{iotaPext}. 

 In the previous section we found an extension of  the projection $p:X\to \bar X$ to  an \lf-morphism $F:\bS(X)\to \bS(\bar X)$ that
 takes the coderivation $D$ on ${\bf S}(X)$ to a coderivation $\bar D$ on ${\bf S}(\bar X)$.
  Our aim here  is to find a dual construction in which we extend the projection
 $\iota ^{\star}:  X^{\star} \to \bar X^{\star}$ to a morphism
 $E^{\star}: \bS(X) ^{\star} \to \bS(\bar X)^{\star}$ taking the derivation $Q$ on $ \bS(X) ^{\star}$ to a derivation $\bar Q$ on $\bS(\bar X)^{\star}$.

If  $\{ \bar z^{\bar a} \} $ are a basis for $\bar X^{\star}$, the morphism of algebras $E^\star:{\bf S}(X)^\star\to {\bf S}(\bar X)^\star$ 
takes the general form
\be
\label{Estarexpansion}
E^\star(z^a)=\sum_{n=1}^{\infty}\frac{1}{n!}E^a_{\bar a_1\bar a_2 \dots \bar a_n} \bar z^{\bar a_1} \bar z^{\bar a_2}\cdots \bar z^{\bar a_n}\,
\ee
and so 
is determined by the coefficients $E^a_{\bar a_1\bar a_2 \dots \bar a_n}\,, n\geq 1$.
Each $E^a_{\bar a_1\bar a_2 \dots \bar a_n}$ defines a linear degree-0 map $S^n(\bar X ^\star)\to X^\star$. We will set $E^a_{\bar a}$ equal to the matrix representation of the projection map  $\iota ^*:  X^* \to \bar X^*$, and the higher maps will simply serve to make $E$ an \lf-morphism.

From a geometric perspective the problem is: given a vector field $Q$ with $Q^2=0$ on a linear space $X$ and a linear subspace $\bar X\cong PX$, can we construct a vector field $\bar Q$ on $\bar X$ also with $\bar Q^2=0$? Of course vector fields do not pull back so there is no canonical construction in general. In this particular context where $X$ and $\bar X$ are quasi-isomorphic chain complexes viewed as linear formal supermanifolds, homotopy transfer can be viewed as such a construction.

Consider  first the simple ansatz
\be
\bar Q=\iota^\star Q p^\star \nonumber
\ee
 This satisfies the correct Leibniz rule, due to $\iota^\star p^\star={\rm id}_{{\bf S}(\bar X)^\star}$.\footnote{If we were instead trying to lift some vector field $\bar V$ on $\bar X$ to $X$ using the obvious formula $V=p^\star \bar V \iota^\star$, the Leibniz rule would fail; $V$ would satisfy what should probably be called a $(p^\star \iota^\star,p^\star \iota^\star)$-Leibniz rule.} 
 We then calculate
\be
\bar Q^2=\iota^\star Q p^\star\iota^\star Q p^\star=\iota^\star Q P^\star Q p^\star\;,  \nonumber
\ee
which fails to vanish in general. However, it will vanish if
 $Q$ is in fact tangent to $\bar X\cong PX$ up to terms vanishing under the $\iota^\star$ above:
\be
\iota^\star Q^b P^a_b=\iota^\star Q^a\,.\nonumber
\ee
In that case $\bar Q^2=0$ and $\iota^\star$ in fact defines a morphism of \lf-algebras:
\be
\bar Q\iota^\star=\iota^\star Qp^\star\iota^\star=\iota^\star Q \nonumber\,,
\ee
so we can simply set $E^\star=\iota^\star$. Algebraically speaking, this scenario occurs when the projector $P$ defines a subalgebra in the strict sense in which all \lf-brackets close on $PX$.

We can  solve the problem in the general case if we can find a $Q'$ defining an isomorphic \lf-algebra structure that is a vector field  tangent to $PX$ (up to terms that vanish under $\iota^\star$).
We then seek a $Q' =Q'{}^a\partial _a$  that
satisfies the
the ``tangentiality'' condition:
\be
\label{tancon}
\iota^\star Q'{}^b P^a_b=\iota^\star Q'{}^a\,
\ee
and take
\be
\label{ansatzQ}
 \bar Q=\iota^\star Q' p^\star\, .
\ee
The condition (\ref{tancon}) is equivalent to the more convenient
\be
\label{tangencycondition_derivation}
P^\star Q'P^\star=P^\star Q' \,.
\ee

One approach is to seek a
morphism
$R$  that transforms $Q$ to a  $Q'$ with these properties.
 This point of view was advocated in Kajiura's work \cite{Kajiura:2001ng,kajiura} on homotopy transfer into the \emph{homology} $H(X)$ of an $A_\infty$-algebra; we treat the generalisation to an arbitrary quasi-isomorphic subcomplex $\bar X$ of $X$, but for an \lf-algebra over $X$ instead. 
We then
write the ansatz
\be
\label{ansatzstar}
E^\star=\iota^\star R^{-1}\,,\quad  Q'= R^{-1} Q R  \, ,
\ee
where $R$ is an \emph{invertible} (degree 0) morphism of algebras ${\bf S}^\star\to {\bf S}^\star$ of the form
\be
R:z^a \to r^a(z)\,,
\ee
where the function 
\be 
r^a(z)= z^a + \mathcal O(z^2)\,
\ee
is chosen so that the  higher terms ensure the ``tangentiality'' requirement  (\ref{tancon}).
The new derivation $Q'=R^{-1}Q R$ defines an isomorphic \lf-algebra structure on $X$, whose brackets all close on $PX$ when \eqref{tangencycondition_derivation} is true.

An alternative geometric viewpoint is as follows. We regard the $z^a$ as coordinates of a manifold $M\simeq \mathbb{R}^{d|d'}$ where $d + d'$ is the dimension of $X$, and regard
$\bS^\star$ as a function space on $M$. (Here $\bS^\star$ is the space of formal power series in $z$, but much of the following discussion works just as well if we replace this with the space of polynomials in $z$ or the space of continuous functions on $M$.)
Then $Q=Q^a(z)\partial/\partial z^a$ can be regarded as a vector field on  $M$. We then regard the transformation of $z$ as  a diffeomorphism $\rho$ of $M $
\be
\rho :z^a \to 
r^a(z)
\ee
with the same functions $r^a(z)$ as above 
and  take the vector field $Q'$ to be the push-forward of the vector field $Q$, 
\be
Q'=\rho _\star Q\,. 
\ee
As we shall see, both the algebraic and geometric approaches give the same results.

We now turn to the choice of function $r^a(z)$. We
split the  $Q$ given by (\ref{Qis}) into its linear  part $\partial^\star$ and nonlinear part  $B^\star \equiv W$, as we did for its dual $D=Q^\star$ in section \ref{section:coderivation}:
 \be
 \label{Qsplit}
 Q=\partial^\star + W\,. 
 \ee
 From (\ref{Qis}) we have 
 \be 
 \label{dsta}
 \partial^\star =C^a_b z^b \partial_a\,,
 \ee
 and $W=W^a\partial_a$, where
 \be
 W^a(z)= \sum_{n=2}^\infty \frac{1}{n!} \, C^a_{b_1\dots b_n} z^{b_1} \, \cdots z^{b_n} \, .
 \ee
 We choose the functions $r^a(z)$
defined by
\be
\label{rmorphism}
r^a(z)=z^a-h^a_b W^b(z)\, .
\ee
 This form for $R$ is motivated below, with $h$  the same map as that arising in (\ref{homotopyID}).
We use the notation $s^a(z)$ for the inverse function
to $r^a$, so that $s^a(r(z))=z^a$.
This function is defined recursively by
  \be
  \label{srecur}
 s^a(z)=z^a+ h^a_b W^b(s(z))\, .
\ee

Before proceeding, we note some properties of the function $r^a$ that will be useful later.
Its derivative is
\be
\label{derrr}
\partial _b r^a=\delta _b^a-h^a_c \partial _b W^c\, , 
\ee    
so that 
\be
\label{Qderrr}
Q^b\partial _b r^a=Q^a-h^a_c Q^b \partial _b W^c\, .
\ee    
The original \lf-algebra Jacobi identities in the form
\be
Q^2=0\iff 0=Q^a\partial_a Q^b=Q^a\partial_a(C^b_c z^c + W^b)
\ee
give
\be
Q^a\partial_aW^b=- Q^a C^b_a\,.
\ee
Using  (\ref{Qis}) and $C^a_bC^b_c=0$, which follows from  $(\partial^\star)^2=0$ and
(\ref{dsta}), this can be rewritten as
\be
Q^a\partial_aW^b=- W^a C^b_a\,.
\ee
Then  using this, (\ref{Qderrr}) becomes
\be
\label{Qderrra}
Q^b\partial _b r^a=Q^a+    h^a_c C^c_b W^b\,.
\ee

We first consider the geometric derivation of $Q'$.
The diffeomorphism $\rho$ is realised as  the active coordinate transformation
\be \rho:z^a \to {z'} ^a(z) = r^a(z)\,.
\ee
The vector field
\be 
Q= Q^a(z)  \frac{\partial}{\partial z^a}  
\ee
transforms to
\be 
Q'= {Q'}^a(z')  \frac{\partial}{\partial  {z'}  ^a} \,,
\ee
where
\be
{Q'}^a({z'}) = Q^b(z)  \frac{\partial  {z'}^a} {\partial z^b}(z)
= Q^b(z)  \frac{\partial  r^a} {\partial z^b}(z)\,.
\ee
(This is the push-forward by the diffeomorphism $\rho$.)

The LHS can be written as a function of $z$ using $z'{} ^a(z) = r^a(z)$ to give a relation between functions of $z$:
\be
{Q'}^a(r(z)) 
= Q^b(z)  \frac{\partial  r^a} {\partial z^b}(z)\,.
\ee
Alternatively, we can write this as a relation between functions of $z'$. The inverse function of $ {z'} ^a(z)$ is
\be
{z} ^a(z') = s^a(z')
\ee
and can be used in the RHS to give
\be
{Q'}^a({z'}) 
= Q^b(z)  \frac{\partial  r^a} {\partial z^b}(z) \bigg|_{z=s(z')}
=Q^b(s(z')) \,  \left[  \frac{\partial  r^a} {\partial z^b}(z) \right] \bigg|_{z=s(z')}\;. 
\ee
Then
\be 
Q'= {Q'}^a(z')  \frac{\partial}{\partial  {z'}  ^a} 
\ee
with
\be
{Q'}^a({z'}) 
= Q^b(s({z'}))  \frac{\partial  r^a} {\partial z^b}(s({z'}))\,. 
\ee
Now we can drop the prime on $z$ (i.e.~changing the name of the coordinate) to get
\be 
Q'= {Q'}^a(z)  \frac{\partial}{\partial  {z}  ^a} \,,
\ee
with
\be
\label{qprimis}
{Q'}^a({z}) 
= Q^b(s({z}))  \frac{\partial  r^a} {\partial z^b}(s({z}))\,.
\ee
The RHS can be written as 
\be
\left(   \Big[Q^b  \frac{\partial  r^a} {\partial z^b}\Big]\circ \rho^{-1} \right)(z)\,,
\ee
or as 
\be
\label{morphrsi}
 R^{-1}
\left[ Q^b  \frac{\partial  r^a} {\partial z^b} \right]\,,
\ee
where $ R^{-1}$ is the morphism that takes $z$ to $s(z)$.

We now turn to the algebraic picture.
We take the morphism $R$ to act on $X^{\star}$ as
\be
\label{Rmorphism}
R(z^a)=z^a-h^a_b W^b\, .
\ee
That is, the morphism acts to take $z^a $ to the functions $r^a(z)$
defined by
\be
r^a(z)=z^a-h^a_b W^b(z)\, ,
\ee
so that $R:z^a\to r^a(z)$.
This is then
 extended to the whole of $\bS^\star$ by the morphism property $R(z^a z^b) = R(z^a) R(z^b)$ so that
 a function $f(z)\in \bS^\star$ is mapped as 
 \be
 \label{rmap}
 R:f(z)\to f(r(z))\, .
 \ee
  The inverse morphism  $R^{-1}:z^a\to s^a(z)$ is given by the recursive formula
  \be
 R^{-1}(z^a)=z^a+ h^a_b R^{-1}(W^b)\, .
\ee
$R^{-1}$ is to be interpreted as the map $R^{-1}:z^a\to s^a(z)$ where $s^a$ is the inverse function to $r^a$, defined recursively by (\ref{srecur}).
 These recursive definitions are 
    well-defined since $W^a(z)$ is of quadratic order and higher in $z$. In particular the coefficients $E^a_{\bar a_1\bar a_2\dots \bar a_n}$ defining the morphism $E:\bar X\to X$ through \eqref{Estarexpansion} are finite sums. A non-recursive formula can be found by inverting $R$ as a geometric series.

We regard $  R^{-1} Q R  $ as acting on functions of $z$.
We use the chain rule, \eqref{Rmorphism}, (\ref{rmap}) and the split \eqref{Qsplit} to write
\be
\label{RQR1}
R^{-1}QR=R^{-1}\left[Q^a\partial_a(r^b(z))\right]\partial_b=R^{-1}(Q^b-h^b_c Q^a\partial_a W^c)\partial_b\,.
\ee
Thus we recover  the formula (\ref{morphrsi}) for $Q'^a$.

We will now prove that 
$Q'$ given by (\ref{qprimis}) with the 
$R$ of \eqref{Rmorphism} and the
ansatz (\ref{ansatzQ}), (\ref{ansatzstar}) 
 define an \lf-structure $\bar Q$ and an \lf-morphism $E:\bar X\to X$, by proving $Q'$ satisfies \eqref{tangencycondition_derivation}. This is  analogous to the calculation in \cite{Arvanitakis:2019ald} for the minimal model theorem (homotopy transfer into $H(X)$), but we present the calculation in full since that reference assumes the so-called ``side conditions'' in (\ref{ipsides}) while 
our proof does not use these conditions.

\begin{proof}

$Q'$ is given by (\ref{qprimis}), which we write using (\ref{morphrsi}) as:
\be
Q'{}^a=  R^{-1}
\left[ Q^b  \frac{\partial  r^a} {\partial z^b} \right]\,.
\ee
Now we use (\ref{Qderrra}) to write this as
\be
Q'{}^a=  R^{-1}
\left[ Q^a+    h^a_c C^c_b W^b \right]\,,
\ee
which, using
(\ref{Qsplit}), (\ref{dsta}), becomes
\be
\label{asds}
Q'{}^a=  R^{-1}
\left[ C^a_c z^c +W^a+    h^a_c C^c_b W^b \right]\,.
\ee

Now,  $R^{-1}$ maps $z^a\to s^a(z)$ so that
\begin{align}
R^{-1} ( C^a_c z^c )
&=
C^a_c s^c (z) \\
&=C^a_c (z^c + h^c_b W^b(s(z))
)
 \\
 &= C^a_c z^c + C^a_c h^c_b R^{-1}(W^b)\,, 
\end{align}
where we have used (\ref{srecur}).
Then using this in (\ref{asds}) we have
\be
Q'{}^a= C^a_c z^c + R^{-1}(W^b) (\delta^a_b +h^a_c C^c_b + C^a_c h^c_b)\,,
\ee
which can be written as
\be
\label{RQR3}
Q'{}^a= C^a_c z^c +  (\delta^a_b +h^a_c C^c_b + C^a_c h^c_b) W^b(s(z))\,.
\ee
In the last parenthesis we recognise the index notation for the linear map ${\bf 1}+h\partial+\partial h=P$. 
The result can then be written as 
\be
\label{RQR4}
Q'= \partial ^\star+ R^{-1}(W^b) P_b^a \partial_a\,, 
\ee
and this indeed satisfies  \eqref{tangencycondition_derivation}.
In fact 
the two summands $A_1,A_2$ on the right hand side of  (\ref{RQR4}) separately satisfy $P^\star A_iP^\star=P^\star A_i$
on account of $\pd P = P \pd$ (equation \eqref{commutationnn}) and $P^2 = P$.
\end{proof}

The upshot is the formula for the transported \lf-structure $\bar Q$ on $\bar X$ and for  the morphism $E^\star:\bS^\star\to \bar \bS^\star$:
\be
\label{homotopytransferredstructure_derivationpicture}
\bar Q=\bar \partial^\star+\bar W\,,\quad \bar W=E^\star(W^b) p_b^{\bar a}\frac{\partial}{\partial \bar z^{\bar a}}\,,\quad E^\star(z^a)=\iota^\star(z^a)+ h^a_b E^\star(W^b)\,.
\ee
This does not require the use of side conditions. However, using them, we can immediately write the corresponding expression in the coalgebra picture:
\be
\bar D=\bar\pd +\bar B\,,\quad\bar B=pB E\,.
\ee
Using \eqref{app:pEeq1} we find $\bar B$ is the unique coderivation whose dual is $\bar W$. The formula for $E:\bS(\bar X)\to \bS(X)$ (whose dual is $E^\star$) requires more notation, and appears in appendix \ref{app:equality}.

\paragraph{Example: homotopy transfer from a differential graded Lie algebra.}
As in section \ref{section:coderivation} we illustrate homotopy transfer in the simple case where the original \lf-algebra  has vanishing ternary and higher brackets. In this case 
$X$ has the structure of a \emph{differential graded Lie algebra} (dgLa). Since the higher brackets vanish,
\be
Q^a= C^a_b z^b + \frac{1}{2} C^a_{b_1b_2}z^{b_1}z^{b_2}\,,\quad W^a=\frac{1}{2} C^a_{b_1b_2}z^{b_1}z^{b_2}\,,
\ee
where $C^a_b$ and $C^a_{b_1b_2}$ are the structure constants of the differential and the binary bracket on the dgLa respectively.

The \lf-algebra on $\bar X$ resulting from homotopy transfer will have brackets of all orders. We will determine the binary and ternary brackets explicitly here. In other words we determine $\bar Q(\bar z^{\bar a})=\bar Q^{\bar a}$  to cubic polynomial order in $\bar z^{\bar a}\in \bar X^\star$. Since $\iota^\star z^a=\iota^a_{\bar a}\bar z^{\bar a}$ we can instead just count polynomial order in $\iota^\star z^a$ instead, which is notationally convenient.

We will need $E^\star z^a$ to second order. Calculating directly from the recursive definition \eqref{homotopytransferredstructure_derivationpicture}
\be
E^\star z^a=\iota^\star z^a+ h^a_b E^\star\left(\frac{1}{2} C^b_{c_1c_2}z^{c_1} z^{c_2}\right)=\iota^\star z^a + \frac{1}{2} h^a_b C^b_{c_1c_2} (\iota^\star z^{c_1}) (\iota^\star z^{c_2})+\dots
\ee
This suffices to determine the nonlinear part of $\bar Q$, $\bar W$, to cubic order. From \eqref{homotopytransferredstructure_derivationpicture} we write $\bar W^{\bar a}=E^\star W^a p^\star(\bar z^{\bar a})=p^{\bar a}_a E^\star W^a$ whence
\begin{align}
\bar W^{\bar a}=\frac{1}{2} p^{\bar a}_aC^a_{b_1b_2} (\iota^\star z^{b_1}) (\iota^\star z^{b_2}) 
+\frac{1}{2}p_a^{\bar a} C^a_{b c_3}h^b_d C^d_{c_1c_2}(\iota^\star z^{c_1}) (\iota^\star z^{c_2}) (\iota^\star z^{c_3}) +\dots
\end{align}
This defines the structure constants of the ternary bracket on $\bar X$. We read off
\be
\bar b_3(\bar x_1,\bar x_2,\bar x_3)= p\big[h\big[\iota \bar x_1,\iota \bar x_2\big],\iota\bar x_3  \big] +\text{symmetrisations}\,.
\ee
This agrees with the homotopy transported bracket (\ref{transported3bracket}) determined in sec.~2.

\subsection{Side conditions and interpretation}

To motivate our explicit form for $R$, we further impose what are called \emph{side conditions} in the mathematical literature:
\be
\label{sideconditions}
h^2=0\,,\quad h\iota=0\,,\quad ph=0\,.
\ee
They imply that the three parts in the decomposition of the identity
\begin{equation}\label{eq:identitydecomposition}
{\bf 1} = P - \pd h - h \pd
\end{equation}
are complementary projectors, i.e.~${\rm p}_1 = P$, ${\rm p}_2 = - \pd h$ and ${\rm p}_3 = - h \pd$ satisfy $({\rm p}_i)^2 = {\rm p}_i$ and ${\rm p}_i {\rm p}_j = 0$ if $i\neq j$.

Anticipating the later discussion, we will view homotopy transfer from $X$ to $\bar X\cong PX$ as the construction of a tree-level effective action, where the subspace corresponding to $(1-P)X$ has been ``integrated out''. This subspace further splits into two parts, $\pd h X$ and $h \pd X$. Recall from the introduction that elements of $X_0$ can be interpreted as fields (see equation \eqref{Hextendedvector}), and elements of $X_{1}$ as gauge parameters. Then, fields $\Psi$ in the first part $\pd h X_0$ are pure gauge (they take the form of a gauge transformation, $\Psi = \pd \Lambda$ with $\Lambda = h\Psi \in X_1$). The second part $h\pd X_0$ is the one that should be integrated over: at tree-level, this means solving their equations of motion and substituting the solution back into the action.

We now go to the dual picture and use the  decomposition ${\bf 1} = P - \pd h - h \pd$, written in matrix form 
$\delta^a_b =P^a_b -C^a_c h^c_b -h^a_c C^c_b$. This splits the dual basis elements $z^a$ as
\begin{equation}
z^a = P^a_b z^b - C^a_c h^c_b z^b -h^a_c C^c_b z^b\, .
\end{equation}
Then, the discussion of the previous paragraph is realized as follows: we look for an $E^\star:\bS^\star \to \bar{\bS}^\star$ satisfying the three conditions
\begin{enumerate}
\item $E^\star(P^a_b z^b) = \iota^a_{\bar{a}} \bar{z}^{\bar{a}}$: fields in $PX$ remain the same.
\item $E^\star(-C^a_b h^b_c z^c) = 0$: the pure gauge part of $(1-P)X$ is set to zero.
\item The remaining fields (those in $h\pd X$) should be expressed in terms of the $\bar{z}^{\bar{a}}$ in such a way that their equations of motion are solved. For this reason, we introduce a cyclic structure so we can consider the BV master action $\Theta$: this condition then reads
\be \label{eq:eomhCz}
E^\star\left(\frac{\pd \Theta}{\pd \tilde{z}^a}\right) = 0
\ee where $\tilde{z}^a = -h^a_b C^b_c z^c$.
\end{enumerate}
Strictly speaking, the intuition explained above is only applicable at degree zero but  these equations are written for \emph{all} basis elements $z^a$. However, it turns out that these equations can be motivated  by a more careful path-integral treatment, as we will show in \cite{second}.

Using the side conditions, it is easily seen that the ansatz
\be
\label{homotopytransferansatz_derivation}
E^\star(z^a)=\iota^\star(z^a)+h^a_b f^b\,, \nonumber
\ee
with arbitrary $f$, satisfies the first two requirements automatically. To understand the third requirement, which involves the cyclic inner product $\kappa$, we further assume the compatibility conditions
\begin{equation}\label{eq:ortho}
\kappa(Px_1,(1-P)x_2)=0\quad \text{and} \quad \kappa(hx_1,hx_2)=0 \quad \forall x_1,x_2\in X\, .
\end{equation}
These have two useful consequences. First, we have \be \kappa(Px_1,hx_2)=0\qquad \forall x_1,x_2\in X\, . \ee
This is because the third side condition, $ph=0$, implies that $\im(h) \subseteq (1-P)X$. Second, one has the symmetry property
\be \label{eq:symkappah}
\kappa(h x_1,x_2)=(-1)^{(x_1+1)(x_2+1)}\kappa(hx_2,x_1)\qquad \forall x_1,x_2\in X\,. 
\ee
This is proven as follows:
\begin{align}
\kappa(h x_1,x_2) &= \kappa(h x_1,(P - \pd h - h\pd) x_2) = \kappa(h x_1, -\pd h x_2) \\
&= (-1)^{(x_1+1)(x_2+1)} \kappa(h x_2, -\pd hx_1) = (-1)^{(x_1+1)(x_2+1)} \kappa( hx_2, x_1)\, ,
\end{align}
using the usual decomposition of the identity, the orthogonality properties of $\kappa$ assumed above, the cyclicity of $\kappa$, and the reasoning of the first line again.

Using the assumptions above, condition \eqref{eq:eomhCz} can be rewritten as
 \be
 \label{EOMconditionOrSourceJequalzero}
 E^\star (h^a_b Q^b)=0\,.
 \ee
Indeed, we have
\begin{align}
\frac{\pd \Theta}{\pd \tilde{z}^a} &= - h^c_b C^b_a \frac{\pd \Theta}{\pd z^c} = - h^c_b C^b_a \kappa_{cd} Q^d = \pm h^c_d C^b_a \kappa_{cb} Q^d \\
&= \pm \kappa_{ab} C^b_c h^c_d Q^d
\end{align}
(keeping track of signs is unnecessary here). This already shows that \eqref{eq:eomhCz} is implied by \eqref{EOMconditionOrSourceJequalzero}. To show the converse, multiply by $h_f^e \kappa^{fa}$ and use the relation $h\pd h = - h$, which follows from the decomposition \eqref{eq:identitydecomposition} by acting with $h$ and using the side conditions.

To conclude this section, we show how this requirement of ``tree-level integrating out'', in the form \eqref{EOMconditionOrSourceJequalzero}, suggests formula \eqref{homotopytransferredstructure_derivationpicture} for $E^\star$. Using the split $Q = \pd^\star + W$, equation \eqref{EOMconditionOrSourceJequalzero} becomes
 \be
 h^a_b C^b_c\iota^\star(z^c)+ h^a_b (E^\star W^b+ C^b_c h^c_d f^d)=0
 \ee
 which upon use of the $\iota\bar\partial=\partial \iota$ formula \eqref{iotadelcomm}, the side conditions \eqref{sideconditions} and $h\pd h = -h$ reduces to
 \be
 h^a_b(E^\star W^b - f^b)=0\,.
 \ee
This motivates our choice $f^a = E^\star W^a$, leading to \eqref{Rmorphism} and \eqref{homotopytransferredstructure_derivationpicture}.

\subsection{Homotopy transfer for cyclic \lf-algebras}
Assuming the original \lf-algebra on $X$ has a (nondegenerate) cyclic inner product $\kappa$, when does the resulting \lf-algebra on $\bar X$ have a cyclic inner product?

We will use the geometric picture of \lf-algebras. Since a cyclic inner product can be interpreted as a symplectic form $\kappa$, and the \lf-structure as $Q=(\Theta,\,\cdot\,)$ for a hamiltonian $\Theta$, we can simply pull back, 
\be
\label{barTheta}
\bar \Theta\equiv E^\star \Theta\,,\quad \bar \kappa \equiv E^\star \kappa\,, 
\ee
to find a candidate cyclic \lf-structure. It turns out that the side conditions \eqref{sideconditions} and the orthogonality properties \eqref{eq:ortho} will prove sufficient to make $(\bar X,\bar Q,\bar \kappa)$ a cyclic \lf-algebra. First, these conditions ensure $E^\star \kappa$ is a \emph{constant} symplectic form which is in fact the restriction of $\kappa$ to $\im(P)\cong \bar X$:
\begin{align}
E^\star \kappa&=\frac{1}{2}d(E^\star z^a)\kappa_{ab} d(E^\star z^b)\\
&=\frac{1}{2}d(\iota^\star z^a)\kappa_{ab} d(\iota^\star z^b)+ d(h^a_c E^\star(W^c))\kappa_{ab}d(\iota^\star z^b)\\
&\quad\;\;  +\frac{1}{2}d(h^a_c E^\star(W^c))\kappa_{ab} d(h^b_d E^\star(W^d))\\
&=\frac{1}{2}\iota^a_{\bar a}\iota^b_{\bar b}d\bar z^{\bar a}\kappa_{ab} d\bar z^{\bar b}\,.
\end{align}
We thus identify
\be
\bar \kappa_{\bar a \bar b}\equiv \kappa_{ab} \iota^a_{\bar a}\iota^b_{\bar b}\,, 
\ee
which is nondegenerate: $P$ is a $\kappa$-orthogonal projector, so the inverse is
\be
\bar \kappa^{\bar a \bar b}=\kappa^{ab}p_a^{\bar a}p_b^{\bar b}\,.
\ee

We now show the transferred structure $\bar Q$ of \eqref{homotopytransferredstructure_derivationpicture} agrees with the hamiltonian vector field
\be
(\bar \Theta,\,\cdot\,)=(\bar\Theta,\bar z^{\bar a})\partial_{\bar a}\,, 
\ee
where the antibracket is the one associated to $\bar \kappa$.

\begin{proof}
By direct calculation one finds 
\begin{align}
(\bar \Theta,\bar z^{\bar a})&=(E^\star \Theta,\bar z^{\bar a})=\partial_{\bar b}(E^\star\Theta)\bar \kappa^{\bar b \bar a}= \frac{\partial E^\star z^c}{\partial \bar z^{\bar b}}\left(E^\star \frac{\partial \Theta}{\partial z^c}\right)\kappa^{\bar b \bar a}\\
&=\left(\iota^c_{\bar b} + h^c_d \frac{\partial(E^\star W^d)}{\partial \bar z^{\bar b}}\right)\left(E^\star \frac{\partial \Theta}{\partial z^c}\right)\kappa^{\bar b \bar a}\,.
\end{align}
Since $Q(z^a)=(\Theta,z^a)=\partial_b \Theta\kappa^{ba}$ the first term reads
\be
\iota^c_{\bar b} \left(E^\star \frac{\partial \Theta}{\partial z^c}\right)\kappa^{\bar b \bar a}=p^{\bar a}_b E^\star(Q^b)=p^{\bar a}_b E^\star W^b + p^{\bar a}_b E^\star\left(C^b_cz^c\right)\,, 
\ee
where we split $Q=W+\partial^\star$ on the right-hand side. We recognise $p^{\bar a}_b E^\star W^b=\bar W^{\bar a}$, while
\be
 p^{\bar a}_b E^\star\left(C^b_cz^c\right)= p^{\bar a}_b C^b_c \iota^c_{\bar d}\bar z^{\bar d}+ p^{\bar a}_b C^b_c  h^c_d E^\star W^d=\bar{\partial}^\star(\bar z^{\bar a})+0\,;
 \ee
 the second term here vanishes due to $p\pd = \bar{\pd} p$ and the side condition $ph=0$. Therefore
 \be
 (\bar \Theta,\bar z^{\bar a})=\bar{\partial}^\star(\bar z^{\bar a})+\bar W^{\bar a}+  h^c_d \,\frac{\partial(E^\star W^d)}{\partial \bar z^{\bar b}} \left(E^\star \frac{\partial \Theta}{\partial z^c}\right)\kappa^{\bar b \bar a}\,.
 \ee
It remains to prove the last term is zero. This is implied by
 \be
 \label{EOMconditionOrSourceJequalzero_cyclic}
 h^a_b \left(E^\star \frac{\partial \Theta}{\partial z^a}\right)=0\,,
 \ee
which is equivalent to \eqref{EOMconditionOrSourceJequalzero}, as is easily seen using $Q^a=\partial_b \Theta\kappa^{ba}$ and the matrix version of \eqref{eq:symkappah}.
 We have therefore proven that $\bar Q$ of \eqref{homotopytransferredstructure_derivationpicture} is the hamiltonian vector field for $\bar\Theta=E^\star \Theta$ with cyclic inner product $\bar \kappa$.
\end{proof}

 There is the same physical interpretation as in the last section: identifying $\Theta$ with the BV master action, \eqref{EOMconditionOrSourceJequalzero_cyclic} is an abstract algebraic analogue of solving the equations of motion of
\be
 \Theta(z^a)=\Theta\left((P^a_b - C^a_c h^c_b - h^a_c C^c_b)z^b\right)
 \ee
 perturbatively for $h^a_b C^b_c z^c$ in terms of $P^a_b z^b$ while setting $C^a_b h^b_c z^c$ to zero; for the case of homotopy transfer into $H(X)$, the recursive formula \eqref{homotopytransferredstructure_derivationpicture} for $E^\star$ has been recognised in the physics context as the Berends-Giele recursion or perturbiner expansion for tree-level amplitudes \cite{Macrelli:2019afx,Lopez-Arcos:2019hvg}.

\section{Field Theories and  $L_\infty$-algebras }
\label{sec:fieldtheoriesandlfalgebras}

\subsection{$L_\infty$-algebras and the Batalin-Vilkovisky formalism }

In the BV formalism \cite{Batalin:1981jr,Batalin:1984jr}, the  field space  is an infinite-dimensional supermanifold $M$, graded according to ghost number.
 It is equipped with an odd symplectic structure $\omega$ defining Poisson brackets which are called \emph{anti-brackets}, and a homological vector field $Q$. This is a hamiltonian  vector field  whose hamiltonian function $\Theta$ is called the \emph{master action}. We provide a short review of the formalism in this section (see also \cite{Gomis:1994he,Henneaux:1992ig,Barnich:2000zw} for more extensive pedagogical references). While many formulas look similar to those of the previous section, we want to point out an important conceptual difference: here, we have a vector field on a supermanifold, which in general could be curved and topologically non-trivial\footnote{For example, this happens in Einstein gravity because of the condition $\det (g_{\mu\nu})\neq 0$. This introduces non-trivial topology in field space, even if the space-time manifold itself is homeomorphic to $\mathbb{R}^n$ \cite{Barnich:1995ap}.}, while in the previous section we were dealing with a derivation on a linear vector space.
  
This formalism arises in field theory as follows.
A bosonic field theory  is formulated in terms of a \emph{minimal set} of fields 
$\phi^i$
consisting of the commuting classical fields (with ghost number zero), the anti-commuting ghosts (with ghost number one) corresponding to the gauge parameters, and, if the theory is reducible, higher generation ghosts-for-ghosts (with higher ghost number).
The fields with even ghost number are commuting and those with odd ghost number are anti-commuting.\footnote{ The generalisation to include fermionic classical fields is straightforward, leading to a double grading with respect to both ghost number and fermion number, but we will not consider this here.} 
 The index $i$ includes both discrete indices and  continuous spacetime coordinates, and  summation over $i$ includes integration over spacetime.
 For each field $\phi^i$ in the minimal set, a corresponding \emph{anti-field} $\starred {\phi} _i$ is introduced.
For a field $\phi^i$ of ghost number $|\phi^i|=n$, the  corresponding anti-field  $\starred {\phi}_i$ has ghost number $|\starred {\phi}_i|=-(n+1)$ and has
  the opposite statistics. 
  To this minimal sector, one can add a non-minimal sector which has trivial BRST cohomology but plays a role in gauge fixing. This includes anti-ghosts and Nakinishi-Lautrup auxiliary fields, and again for each field in the non minimal sector there is a corresponding anti-field.

The fields and anti-fields provide local coordinates $\Phi^a = (\phi^i, \starred {\phi}_i)$
   for the BV supermanifold $M$.
   These are Darboux coordinates in which the odd symplectic form is
   \begin{equation}\label{eq:omegadef}
\omega = \frac{1}{2} d \Phi^a \wedge \omega_{ab} \,d \Phi^b= (-1)^i d\phi^i \wedge d\starred {\phi}_i\qquad ( (-1)^i\equiv (-1)^{|\phi^i|})
\end{equation}
with
\begin{equation}\label{eq:omegasign}
 \omega_{ab} = (-1)^{(a+1)(b+1)} \omega_{ba}\, .
\end{equation}
(Note that, for objects such as $d\Phi^a$ that carry multiple degrees, in this case form degree and ghost number, we use the convention where the degrees are added to determine signs. This explains how \eqref{eq:omegasign} follows from \eqref{eq:omegadef}. This is sometimes known as ``Bernstein's rule'', and matches the conventions of \cite[appendix A]{Arvanitakis:2018cyo} which we follow.) The matrix $\omega_{ab}$ is non-degenerate with inverse $\omega^{ab}$,
giving anti-brackets 
\begin{equation}
\label{antibracketdefBV}
(f, g) = (-1)^a \frac{\partial_r f}{\partial \Phi^a} \omega^{ab} \frac{\partial g}{\partial \Phi^b} \, ,
\end{equation}
where $\partial_r$ denotes the right derivative
\be
\frac{\partial_r f}{\partial \Phi^a}\equiv(-1)^{a(f+1)}\frac{\partial f}{\partial \Phi^a}
\ee
(recall that unlabeled derivatives are left derivatives).
In Darboux coordinates, the form \eqref{eq:omegadef} reproduces the usual antibracket formula
\begin{equation}
    (f, g) = \frac{\partial_r f}{\partial \phi^i} \frac{\partial g}{\partial \starred{\phi}_i} - \frac{\partial_r f}{\partial \starred{\phi}_i} \frac{\partial g}{\partial \phi^i}\, .
\end{equation}

The homological vector field $Q$ is a Hamiltonian vector field
with hamiltonian function given by the master action $\Theta$, $\iota _Q \omega = d \Theta$ so that with 
$Q = Q^a \partial _a$
\begin{equation}
Q_a \equiv \omega_{ab} Q^b =\partial _a \Theta\,. 
\end{equation}
The differential $Q$ can be written in terms of the antibracket as
\begin{equation}
Q = ( \Theta\, , \, \cdot)\, ,
\end{equation}
and the fact that $Q^2 = 0$ is equivalent to the \emph{classical master equation}
\begin{equation}
\label{master}
(\Theta, \Theta) = 0\, .
\end{equation}

We will expand around a solution of the classical equations of motion at which $Q$ vanishes. It will be convenient to choose coordinates such  that this solution is at $\Phi = 0$. In a neighbourhood of $\Phi=0$, the vector field $Q$ will be assumed to have a Taylor expansion
\begin{equation}
\label{Qisphi}
    Q(\Phi) = \sum_{n=1}^\infty \frac{1}{n!} \, C^a_{b_1\dots b_n} \Phi^{b_1} \, \cdots \Phi^{b_n} \,\frac{\partial}{\partial \Phi^a}\, ,
\end{equation}
for some constant coefficients $C^a_{b_1\dots b_n}$.
The index-down structure constants $C_{a_1 \dots a_n} \equiv \omega _{a_1 b}C^b_{a_2\dots a_n}$ then have the  symmetry
\begin{equation}
C_{a_1 a_2\dots a_n}=(-1)^{a_1 a_2}C_{a_2a_1\dots a_n}\, 
\end{equation}
and  the master action is
\begin{equation}
    \Theta = \sum_{n=2}^\infty \frac{1}{n!} \, C_{b_1\dots b_n}  \Phi^{b_1} \, \cdots \Phi^{b_n}\, .
\end{equation}
Conversely, given a master action $\Theta$, the coefficients  $C_{b_1\dots b_n}$ are given by
\begin{equation}
 C_{ b_1\dots b_n} = 
\frac{ \partial ^n \Theta}
{ \partial  \Phi^{b_n}\dots \partial  \Phi^{ b_1}} \,  \Bigg| _{\Phi=0}\, .
\end{equation}

The condition  $Q^2 = 0$ implies that the coefficients $C^a_{b_1\dots b_n}$ satisfy the generalised Jacobi identities that imply that they are the structure constants for an $L_\infty$-algebra.
Introducing a graded vector space $X$ of the same dimension as $M$ (isomorphic to the tangent space $T_0M$ at $\Phi=0$) with basis vectors $T_a$, 
 the $L_\infty$ brackets defined by the  structure constants $C^a_{c_1\dots c_n}$ are
\begin{equation}
    b_n(T_{c_1}, \dots,  T_{c_n}) = C^a_{c_1\dots c_n} T_a\, .
\end{equation}
Here if $\Phi^a $ has ghost number $|a|$, $T_a$ has the opposite ghost number $-|a|$, and ghost number agrees with the degree of section 3 (which is minus the degree of section 2), $degree =(ghost~number)$. 
In this way, the BV manifold defines an $L_\infty$-algebra on $X$.
This can be used to define brackets of fields $\Phi \equiv \Phi^aT_a$ by
\begin{equation}
\label{Bnis}
    B_n( \Phi_1 \dots \Phi_n)= 
    \Phi_1 ^{c_1} \dots \Phi_n ^{c_n}
   b_n (T_{c_1}, \dots,  T_{c_n}) = \Phi_1 ^{c_1} \dots \Phi_n ^{c_n}C^a_{c_1\dots c_n} T_a\, .
\end{equation}
The symplectic form $\omega$ on $M$ 
gives a cyclic structure to this $L_\infty$ algebra: in Darboux coordinates, they are identical,
\begin{equation}\label{eq:signkappaomega}
    \kappa = (-1)^{\phi^i} d\phi^i \wedge d\starred{\phi}_i =\omega\,.
\end{equation}
As discussed in the previous section, this defines a degree $-1$ graded antisymmetric inner product $\kappa : X\times X \rightarrow \mathbb{R}$ satisfying \eqref{eq:kappacyclic} so that the $L_\infty$ algebra is \emph{cyclic}.

Using the split $Q=\partial^\star + W$ into a linear piece $\partial^\star $ and a non-linear one, we have now
 \be 
 \label{dstaph}
 \partial^\star =C^a_b \Phi^b \partial_a\,,
 \ee
where  $ \partial_a= \partial/ \partial \Phi^a$,  together with $W=W^a\partial_a$, where
 \be
 W^a(z)= \sum_{n=2}^\infty \frac{1}{n!} \, C^a_{b_1\dots b_n} \Phi^{b_1} \, \cdots \Phi^{b_n} \, .
 \ee

\subsection{Relation of BV to the Algebraic Formulation}\label{sec:BVrelation}

As explained in the previous sections, for such an $L_\infty$-algebra on $X$ we can introduce a coderivation $D$, dual basis vectors $z^a$ and a  derivation $Q=D^{\star}$ given by
\begin{equation}
\label{Qisa}
    Q(z) = \sum_{n=1}^\infty \frac{1}{n!} \, C^a_{b_1\dots b_n} z^{b_1} \, \cdots z^{b_n} \,\frac{\partial}{\partial z^a}\, .
\end{equation}
Here $z^a$ has the same ghost number (degree) as $\Phi^a$.
Clearly $Q(z)$ is obtained from $Q(\Phi)$ by the simple replacement $\Phi ^a\to z^a$ so their algebraic properties are isomorphic.
However, $Q (\Phi)$ is a vector field on the BV manifold $M$ while $Q(z)$ is a derivation on ${\bf S}^*$.

The BV manifold $M$  is locally isomorphic to a (topological) super-vector space $V$. In the finite dimensional case with $m$ commuting fields of even ghost number and $n$ anti-commuting ones of odd ghost number, this  will be $V=\mathbb{R} ^{m|n}$. An open set $U$ in an open cover of $M$ is mapped to an open set in $V$ by a diffeomorphism
$\psi: U\to V$, with a point  $p \in U$ mapped to 
$\psi (p)= \Phi^a (p) e_a$
where the fields $\Phi^a$ are coordinates for $V$ and $e_a$ is a basis for $V$, where each $e_a$ has ghost number zero. Thus here the ghost number is carried by the coordinate $ \Phi^a $ and not the basis elements $e_a$, which is in contrast to our treatment of the graded vector space $X$ where the basis elements $T_a$ carry the degree and the coordinates are real numbers with degree zero.
The diffeomorphism maps a function ${\bf f}(p)$ on $M$ to a function $f(\Phi)$ on $V$, with  $f(\Phi)={\bf f}(\psi^{-1} (\Phi))$.

The vector space $V$ is graded by ghost number.
We denote the space spanned by the $\Phi^a$ of ghost number $n$ by $V_n$.
In the BV formalism with non-degenerate $\kappa $, $V_0$ is the space of classical fields $\varphi$ of ghost number 0  and has the same dimension as the space
 $V_{-1}$  of anti-fields of the classical fields $ \starred{\varphi}$  which have ghost number $-1$,
  $V_{1}$ is the space of (first generation) ghosts $c_1$ of ghost number  1 and has the same dimension as 
 $V_{-2}$ which is the space of antifields of (first generation) ghosts $\starred{c}_1$  of ghost number -2. For reducible gauge theories,
  $V_{r} $ is the space of $r$'th generation ghosts $c_r$ of ghost number  $r$ and has the same dimension as 
 $V_{-r-1}$  which is the the space of antifields for the  $r$'th generation ghosts $\starred{c}_r$ of ghost number $-r-1$.
 The antifields for the ghosts are neccessary for the non-degeneracy of $\kappa $, for the existence of an antibracket structure, and also for the crucial physical property $H^0(Q) = \{\text{observables}\}$ of the BRST cohomology (see for example chapter 17 of \cite{Henneaux:1992ig}).
The structure is indicated by the following diagram, which is to be compared with (\ref{Hextendedvector}).
\be\label{Qextendedvector}
  \begin{split}
  \cdots \  \xlongrightarrow{\partial ^*} \; 
  & V_{-2} \xlongrightarrow{\partial ^*} V_{-1} \xlongrightarrow{\partial ^*} 
V_{0} \xlongrightarrow{\partial ^*}  V_{1} \xlongrightarrow{\partial ^*}  \  \cdots \,,  \\
   &\,  \starred{c}_1 \ \   \qquad \  \starred{\varphi}\ \  \  \qquad  \varphi 
 \qquad \   \   c_1  \end{split}
 \ee
Here  each arrow indicates the action of $Q$ or its linear part $\partial^*$.
Note that $Q$ transforms $\starred{\varphi}$ into the equations of motion for $\varphi$.

Recall that ${\bf S}^{\star}$ consists of 
 \emph{formal power series} in $z^a$.
There is a natural map $\Lambda$
 from the space of functions on $V$ 
 that are defined by formal power series in $\Phi^a$
 to ${\bf S}^{\star}$.
Consider a function $f(\Phi)$  on $V$ which is 
given by
\begin{equation}
f=\sum _n a_{b_1\dots b_n}  \Phi^{b_1} \, \cdots \Phi^{b_n}
\end{equation}
for some constants $a_{b_1\dots b_n}$.
Then $\Lambda$ takes this to the power series in $z$ defined by the same coefficients,
\begin{equation}
\Lambda: \sum _n a_{b_1\dots b_n}  \Phi^{b_1} \, \cdots \Phi^{b_n} \mapsto \sum_n a_{b_1\dots b_n}  z^{b_1} \, \cdots z^{b_n}\, .
\end{equation}
Acting on such functions, 
\begin{equation}
Q(z) = \Lambda  Q(\Phi) \Lambda ^{-1}\,. 
\end{equation}
The map $\Lambda$, along with the sign \eqref{eq:signkappaomega} translating between $\omega$ and $\kappa$, takes the BV structures discussed above, the master action, the antibrackets and the master equation, to the  corresponding structures on ${\bf S}^{\star}$ discussed in subsection \ref{section:cyclicLinftyBRSTBV}. In particular, the antibrackets \eqref{antibracketdef} and \eqref{antibracketdefBV} agree.

\subsection{The Effective Action and Gauge Fixing }

In the quantum theory, the functional integral is  taken over a Lagrangian submanifold $\Sigma \subset M$, specified in terms of the \emph{gauge fermion}, which is
 a function $\Psi (\phi^i)$ of ghost number $-1$, denoted  by
 \begin{equation}\label{eq:antifieldgaugefixed}
\starred{\phi_i}= \frac{ \partial \Psi} { \partial \phi ^i}\, .
\end{equation}
The functional integral is then of the form
\begin{equation}
\int _\Sigma d\Phi^a \exp{ \frac i \hbar \Theta (\Phi)}
\end{equation}
and the master action is required to satisfy the quantum master equation
\begin{equation}
\label{masterq}
(\Theta, \Theta) = 2i \hbar \Delta  \Theta\, , \qquad \Delta \equiv  \frac{\partial _r  }{\partial \phi^i}  \frac{\partial _l }{\partial \starred{\phi }_i} \, 
\end{equation}
instead of the
 classical master equation  (\ref{master}). The functional integral can be rewritten as 
\begin{equation}
\int  d\phi^i \exp{ \frac i \hbar S_\Psi (\phi)}
\end{equation}
where
\begin{equation}
S_\Psi (\phi)=  \Theta (\phi,  \starred{\phi } )
 \Bigg| _{\starred{\phi_i}= \frac{ \partial \Psi} { \partial \phi ^i}}\, .
 \end{equation}
 
 In theories without gauge symmetry, one typically takes the gauge fermion to be zero, so that $\starred{ \phi}=0$ on $\Sigma$. 
The case with non-zero gauge fermion can be obtained from this by a canonical transformation.
\begin{equation}
\label{canonhomotopydual}
{\phi'}^i=\phi^i\,,\qquad \starred\phi ' _i=\starred\phi_i+ \frac
{\partial \Psi} { \partial \phi ^i}
\end{equation}
In theories with gauge symmetry, a non-minimal sector must be added and the gauge fermion chosen in such a way that the functional integral is well-defined, which requires $S_\Psi$ to satisfy a non-degeneracy condition (see e.g.~\cite{Henneaux:1992ig}).
 An important result is that physical quantities in the quantum theory are independent of  the choice of gauge fixing fermion, provided it is non-degenerate; in particular, they are unchanged by small deformations of $\Psi$.

The effective action can be defined by introducing sources and taking a Legendre transform with respect to the sources.
We shall be interested in the classical effective action defined by taking the classical $\hbar \to 0$ limit.

We  now discuss integrating out a subset of the fields $\phi$. We  divide the fields $\phi^i$ into two sets, $\phi^i=(\bar \phi^{\bar i}, \chi ^u)$ with corresponding antifields $ \starred{\phi }_i=( \starred{\bar \phi}_{\bar i},  \starred{\chi } _u)$ (so that $\Phi ^a =( \bar \Phi ^{\bar a}, X^\alpha)$).
Integrating out the fields $\chi$ gives an effective action depending just on $\bar \phi$. The leading contribution to the integral over $\chi$ is given by the saddle points that extremise the action $S_\Psi(\bar \phi,\chi)$, and this is sufficient for the tree level effective action. We will  justify this and expand on this point in the sequel \cite{second}.

For example, consider the case in which
\begin{equation}
S_\Psi(\bar \phi,\chi)= S_1(\bar \phi) + \frac 1 2 K_{uv} \chi^u\chi^v - \chi^u J_u(\bar \phi)
\end{equation}
for some function  $J_u(\bar \phi)$. Then the $\chi$ field equation gives
\begin{equation}
\chi^u=K^{uv}J_v(\bar \phi)
\end{equation}
and substituting this back in gives the effective action
\begin{equation}
S_{\rm{eff}}(\bar \phi)= S_1(\bar \phi) - \frac 1 2 J_u(\bar \phi) K^{uv}J_v(\bar \phi)
\end{equation}
depending only on the $\bar{\phi}$. Note the appearance of $K^{uv}$, which is the propagator for the fields $\chi^u$ that have been integrated out (this is a generic feature, as we will see in the examples below and more generally in \cite{second}).


\section{Examples}
\subsection{Homotopy as projected propagator}
\label{section:homotopyasprojectedpropagator}

In the  examples we shall consider here, we shall see that the homotopy is realized as the projected propagator. This comes as the result of a very simple observation: suppose we have a map $G : X \to X$ of degree $+1$ satisfying
\begin{equation}\label{eq:homotopyG}
{\bf 1} = \pd G + G \pd
\end{equation}
that commutes with the projector,
\begin{equation}\label{eq:GP=PG}
G P = P G\, .
\end{equation}
Then, the map
\begin{equation}\label{eq:hfromG}
h = - G({\bf 1}-P) = - ({\bf 1}-P) G
\end{equation}
satisfies the required homotopy relation
\begin{align}
\pd h + h \pd &= - \pd G ({\bf 1}-P) - ({\bf 1}-P) G \pd  \\
&= - (\pd G +  G\pd ) ({\bf 1}-P) \\
&= P - {\bf 1}\,, 
\end{align}
where we recalled that $P$ and $\partial$ commute. 
Thus, this $h$ can be used to perform homotopy transfer. 

In  field theories  without gauge symmetry (as in  some of the examples considered below), $G$ is essntially given by the propagator (inverse of the quadratic part of the action). When a gauge symmetry is present, one must first gauge-fix it in order to have solutions to \eqref{eq:homotopyG}, and is of course also what must be done to have a well-defined path integral.

Note that equation \eqref{eq:homotopyG} means that the homology of $\pd$ vanishes, $H(X)=0$. Indeed, \eqref{eq:homotopyG} states that $(-G)$ is a homotopy between the projector $P = 0$ onto the zero subspace $\{ 0 \} \subset X$ and the identity operator ${\bf 1}$. According to the discussion of section \ref{sec:homologyhomotopy}, $\{ 0 \}$ and $X$ therefore have isomorphic homologies, which shows $H(X) = 0$. 

Finally, consider the case where $H(X)\neq 0$. Then there exists no $G$ satisfying \eqref{eq:homotopyG}. Instead, the best one can do is
\be
1-\Pi=\pd G+G\pd
\ee
with $\Pi$ a projector onto $H(X)$. In other words $(-G)$ is a homotopy onto $H(X)$. If $P'$ is a projector commuting with $\Pi$, $G$ and $\pd$, we define the map
\be
h=-G(1-P')
\ee
which satisfies the homotopy relation
\be
P=1+h\pd+\pd h
\ee
with
\be
\label{PeqPprimeplusoneminusPprimeEtc}
P=P' + (1-P')\Pi\,.
\ee
We see $P=P'$ if $P'\Pi=P\Pi=\Pi$. This means we can only do homotopy transfer to a subspace $PX$ containing the entirety of $H(X)$.

\subsection{0D scalar example}

We consider vector variables $\phi ^i \in \mathbb{R}^n$ ($i=1,\dots n$) and the action
\begin{equation}\label{eq:action0D}
    S[\phi] = \frac{1}{2} A_{ij} \phi^i \phi^j + \sum_{k=2}^\infty \frac{1}{k!} A_{i_1 \dots i_{k}} \phi^{i_1} \dots \phi^{i_k}
\end{equation}
with an invertible, symmetric kinetic matrix $A_{ij}$ and symmetric higher interaction coefficients $A_{i_1 \dots i_k}$. This is some zero-dimensional field theory in which everything can be made very explicit; in the following, we explain the associated $L_\infty$ structure (both in the algebra and coalgebra pictures) and check that the homotopy transfer theorem gives the expected result.

\subsubsection{Associated $L_\infty$ algebra}
\label{section:0dparticlelinftyalgebra}

\paragraph{BV picture.} 

In addition to the fields $\phi ^i$ of ghost number $0$, we introduce
$n$ antifields $\starred{\phi}_i$ of ghost number $-1$, so that $(\phi^i, \starred{\phi}_i)\in \mathbb{R}^{n|n}$.
The symplectic form is  by
\begin{equation}
\kappa _i{}^j= \delta _i{}^j, \qquad \kappa_{ij}=\kappa^{ij}=0
\end{equation}
 so that the antibracket induced   is the usual
\begin{equation}
(f, g) = \frac{\pd_{r} f}{\pd \phi^i} \frac{\pd g}{\pd \starred{\phi}_i} - \frac{\pd_{r} f}{\pd \starred{\phi}_i} \frac{\pd g}{\pd \phi^i}\, .
\end{equation}
The  BV master action $\Theta$ is then
\begin{equation}\label{eq:theta0D}
\Theta = \sum_{k=2}^\infty \frac{1}{k!} A_{j_1 \dots j_k} \phi^{j_1} \dots \phi^{j_k}\, ,
\end{equation}
and the homological vector field is
 \begin{equation}\label{eq:Q0D}
Q = \sum_{k=1}^\infty \frac{1}{k!} A_{i j_1 \dots j_k} \phi^{j_1} \dots \phi^{j_k} \frac{\pd}{\pd \starred{\phi}_i}\, ,
\end{equation}
and indeed we have $Q = (\Theta, \,\cdot\,)$.

\paragraph{Coalgebra picture.} Since we have no gauge symmetry, there are only two spaces, $X_{0} = \mathbb{R}^n$ and $X_{-1} = (\mathbb{R}^n)^*$. The $L_\infty$ algebra structure on $X=X_0 \oplus X_{-1}$ can be read off from the
BV formalism or from the
 equations of motion
\begin{equation}\label{eq:eom0D}
\frac{\pd S}{\pd \phi^i} = A_{ij} \phi^j + \sum_{k=2}^\infty \frac{1}{k!} A_{i j_1 \dots j_{k}} \phi^{j_1} \dots \phi^{j_k} = 0 
\end{equation}
by comparing with (\ref{EoM}). 
Introducing basis elements $T_i$ of $X_0$ and basis elements $\starred{T}^{i}$ of $X_{-1}$, 
one has non-zero brackets given on the basis elements by
\begin{equation}\label{generalNbrackets}
    b_n( T_{i_1} , \dots , T_{i_n} ) = A_{i_1 \dots i_n i_{n+1} } \starred{T}^{ i_{n+1}}\, .
\end{equation}

For an element $v= v^iT_i\in X_0$ with coordinates $v^i\in \mathbb{R}^n$,
\begin{equation}
    b_n(v_1, \dots, v_n)= A_{j\,i_1 \dots i_n} v_1^{i_1} \dots v_n^{i_n} \starred{T}^{j}\, .
\end{equation}
In particular, the $1$-bracket $\pd \equiv b_{1}$ is given by 
\begin{equation}
   \pd v = A_{ij} v^i \starred{T}^{j} \, .
\end{equation}

Since the only non-zero brackets have all their arguments in $X_0$ and their image in $X_{-1}$, and since all the $L_\infty$ relations are quadratic in the $b_n$, they are 
automatically satisfied. This is also a cyclic $L_\infty$ algebra with inner product given by the canonical pairing between a vector space ($X_{0} = \mathbb{R}^n$) and its dual. 
With respect to a basis, 
\begin{equation}
\kappa( T_i, \starred{T}^{j} ) = \delta_i^j\, .
\end{equation}
The cyclicity property of $\kappa( v_1, b_n(v_2, \dots v_{n+1}))$ follows from the symmetry of the $A_{i_1 \dots i_k}$.
With this inner product and the general brackets (\ref{generalNbrackets}) it follows that the action (\ref{eq:action0D}) indeed takes the universal $L_{\infty}$ form 
given in the introduction.

\paragraph{Algebra picture.} We write the dual basis as $z^a=(\zeta ^i, \starred{\zeta}_i)$ (those are the $z^a$'s of section \ref{sec:dualpicture}), with pairing $\langle \zeta^i, T_j\rangle = \delta^i_j = \langle\starred{\zeta}_j, \starred{T}^{i}\rangle$. After the degree flip of section 3, they are of degree  $0$ and $-1$, respectively. The derivation $Q(z)$ is then
\begin{equation}
Q = \sum_{k=1}^\infty \frac{1}{k!} A_{i j_1 \dots j_k} \zeta^{j_1} \dots \zeta^{j_k} \frac{\pd}{\pd \starred{\zeta}_i}\, ,
\end{equation}
which squares to zero automatically, and the associated  $\Theta$ is
\begin{equation}
\Theta = \sum_{k=2}^\infty \frac{1}{k!} A_{j_1 \dots j_k} \zeta^{j_1} \dots \zeta^{j_k}\, .
\end{equation}
This is  obtained from  the BV structure by  the replacement $\phi \to \zeta$.

\subsubsection{Homotopy transfer}

Consider now a projector $P = \iota p: X \rightarrow X$ onto a subspace $PX = \iota \bar{X} \cong \bar{X}$. Then, in order to apply the homotopy transfer theorem, the first task is to find a homotopy map $h$ of degree $+1$ such that
\begin{equation}
    P = {\bf 1} + \pd h + h \pd\, .
\end{equation}
Since $h$ is of degree $+1$, its only non-zero component is $h_{-1}:X_{-1} \rightarrow X_0$ and the situation is as follows:
\begin{equation}
\begin{tikzcd}
0 \arrow[r,"\pd_1"] & X_0 \arrow[r,"\pd_0"] \arrow[d,"P_0"'] & X_{-1} \arrow[ld,"h_{-1}"'] \arrow[d,"P_{-1}"] \arrow[r,"\pd_{-1}"] & 0 \\
0 \arrow[r,"\pd_1"'] & \iota\bar{X}_0 \arrow[r,"\pd_0"'] & \iota\bar{X}_{-1} \arrow[r,"\pd_{-1}"'] & 0
\end{tikzcd}
\end{equation}
(This diagram is not commutative.)

In matrix form, the homotopy condition on $X_0$ reads 
\begin{equation}
 (P_0)\indices{^i_j} = \delta^i_j + (h_{-1})^{ik} A_{kj} \, ,
\end{equation}
and on $X_{-1}$ we have
\begin{equation}
    (P_{-1})\indices{_j^i} = \delta^i_j + A_{jk} (h_{-1})^{ki}\, .
\end{equation}
Abstractly, we can think of the matrices here as maps 
$P_0: \mathbb{R}^{n}\rightarrow \mathbb{R}^n$, $P_{-1}:(\mathbb{R}^{n})^*\rightarrow (\mathbb{R}^n)^*$ 
and $A^{-1}:(\mathbb{R}^{n})^*\rightarrow \mathbb{R}^n$. 
There is a homotopy map $h_{-1}$ satisfying the above conditions provided 
we have  $AP_0A^{-1}=P_{-1}$ as maps $(\mathbb{R}^{n})^*\rightarrow (\mathbb{R}^{n})^*$. 
The solution for $h$ is then 
 \begin{equation}
  h=-({\bf 1}-P_0)A^{-1}=-A^{-1}({\bf 1}-P_{-1})\,. 
 \end{equation}
If, furthermore, $P_{-1}$ is the dual map to $P_0\equiv P$, i.e.~$P_{-1}\equiv P^*$, which  means 
that the matrix forms of $P_{0}$ and $P_{-1}$ can be written as  $(P_0)\indices{^i_j} = (P_{-1})\indices{_j^i} \equiv P^i_j$, then 
the matrix form of $h$ reads 
\begin{equation}
    (h_{-1})^{ij} = - ({\bf 1} - P)^i_k (A^{-1})^{kj} = -A^{ik}({\bf 1}-P)^j_k \, .
\end{equation}

Let us make these formulas more concrete.
We split the fields $\phi^i $ into  $\phi^i=( \phi^{\bar i}, \chi ^u)$
where $ \phi^{\bar i}$ are the fields we keep and $\chi ^u$ are the fields we integrate out.
Thus, the projector acting on fields  takes the form $P^{\bar{i}}_{\bar{j}} = \delta^{\bar{i}}_{\bar{j}}$  with all 
other components being zero.
We further assume that the kinetic matrix has a block-diagonal form in this basis, $A_{\bar{i} u} = 0$. Then the above conditions are satisfied, and the solution reads
\begin{equation}
    h^{uv} = - (A^{-1})^{uv}\; , \quad \text{other components} = 0\, .
\end{equation}
This is an example of the  important result  derived in subsection \ref{section:homotopyasprojectedpropagator}:  with a minus sign, the homotopy $h$ is given by the propagator of fields that have been integrated out (the propagator being given as usual by the inverse of the quadratic part of the action). Note that this homotopy already satisfies the side conditions: $h\iota = ph=0$ since $h$ only acts in the complement of $\bar{X}$, and $h^2 = 0$ for degree reasons.

We now show that tree-level integrating out is realised by  homotopy transfer. We start for simplicity from an action with at most cubic terms,
\begin{equation}
    S[\phi^{\bar{i}}, \chi^u] = \frac{1}{2} A_{ij} \phi^i \phi^j + \frac{\lambda}{3!} A_{ijk} \phi^i \phi^j \phi^k\, .
\end{equation}
Integrating out at tree level means solving the equations of motion for the $\chi^u$ variables in term of the $\phi^{\bar{i}}$, and plugging back to get an action depending on the $\phi^{\bar{i}}$ alone. The solution for the $\chi^u$ at order $\lambda$ is $\chi^u = - \frac{1}{2}\lambda (A^{-1})^{uv} A_{v{\bar{i}}{\bar{j}}} \phi^{\bar{i}} \phi^{\bar{j}}$, and this gives
\begin{equation}
    \bar{S}[\phi^{\bar{i}}] = \frac{1}{2} A_{{\bar{i}}{\bar{j}}} \phi^{\bar{i}} \phi^{\bar{j}} + \frac{\lambda}{3!} A_{{\bar{i}}{\bar{j}}{\bar{k}}} \phi^{\bar{i}} \phi^{\bar{j}} \phi^{\bar{k}} - \frac{\lambda^2}{8} A_{{\bar{i}}{\bar{j}} u} (A^{-1})^{uv} A_{v {\bar{k}} {\bar{l}}} \phi^{\bar{i}} \phi^{\bar{j}} \phi^{\bar{k}} \phi^{\bar{l}} + O(\lambda^3)
\end{equation}
up to quartic order in the fields. 
This means that the corresponding $L_{\infty}$-algebra carries the non-trivial 3-bracket 
\begin{equation}
\bar{b}_3(\phi_1, \phi_2, \phi_3)_{\bar{i}} = -3\lambda^2 A_{u{\bar{i}}({\bar{j}}} (A^{-1})^{uv} A_{{\bar{k}} {\bar{l}})v} \phi_1^{\bar{j}} \phi_2^{\bar{k}} \phi_3^{\bar{l}}\, .
\end{equation}This formula, which is obtained here through the field theory computation, should be reproduced by formula \eqref{transported3bracket}. In the case where all the $x_i$'s are at degree zero, there are no signs and the formula reduces to
\begin{align}
    \bar{b}_3(x_1,x_2,x_3) = p\left( [h([x_1,x_2]),x_3]+[h([x_1,x_3]),x_2]+[h([x_2,x_3]),x_1]\right)\, .
\end{align}
The first term is computed as
\begin{align}
[h([x_1,x_2]),x_3]_{\bar{l}} &=
 \lambda A_{{\bar{l}} ij} h([x_1,x_2])^i (x_3)^j = \lambda A_{{\bar{l}} ij} h^{ik} [x_1,x_2]_k (x_3)^j \nonumber\\
&= \lambda^2 A_{{\bar{l}} ij} h^{ik} A_{kmn} (x_1)^m (x_2)^n (x_3)^j \nonumber \\
&= - \lambda^2 A_{{\bar{l}}{\bar{k}} u} (A^{-1})^{uv} A_{v {\bar{i}}{\bar{j}}} (x_1)^{\bar{i}} (x_2)^{\bar{j}} (x_3)^{\bar{k}}\, ,
\end{align}
where in the last step we used the the form of $h$ found above and the fact that the $x_i$ are in the projected subspace (so they only carry barred indices). The other two terms reproduce the factor three and the symmetrization. Note that this 3-bracket, which corresponds to the quartic term in the action $\bar{S}$, can be pictured diagrammatically as
\begin{equation}
\begin{tikzpicture}[scale=0.75]
\draw (0,0)--(1,1); \draw (0,0)--(-1,-1); \draw (0,0)--(1,-1); \draw (0,0)--(-1,1);
\filldraw (0,0) circle (5pt); 
\node at (1.5,0) {=};
\begin{scope}[shift={(3,0)}]
\draw (1,1) -- (0,0.5); \draw (0,-0.5)--(-1,-1); \draw (0,-0.5)--(1,-1); \draw (0,0.5)--(-1,1); \draw[dashed] (0,-0.5)--(0,0.5);
\end{scope}
\node at (4.5,0) {+};
\begin{scope}[shift={(6,0)},rotate=90]
\draw (1,1) -- (0,0.5); \draw (0,-0.5)--(-1,-1); \draw (0,-0.5)--(1,-1); \draw (0,0.5)--(-1,1); \draw[dashed] (0,-0.5)--(0,0.5);
\end{scope}
\node at (7.5,0) {+};
\begin{scope}[shift={(9,0)}]
\draw (0,-0.5)--(-1,-1); \draw (0,0.5)--(-1,1); \draw[dashed] (0,-0.5)--(0,0.5); \draw (0,-0.5)--(1,1); \draw (0,0.5)--(1,-1);
\end{scope}
\end{tikzpicture}
\end{equation}
i.e. it has the expected structure of two cubic terms joined by a propagator. This is consistent with the field theory picture of tree-level integrating out.

\subsection{Scalar field theory}

For a theory of $N$ scalar fields $\phi ^\alpha(x)$ in $D$-dimensional spacetime, we could use the notation $\phi^i$ where $i$ is a composite index consisting of the discrete index $\alpha =1,\dots, N$ and the continuous spacetime coordinate $x^\mu$, so that summation over $i$ represents integration over space-time as well as summation over $\alpha$.
Then a general action can be written in the form \eqref{eq:action0D},
and the results of the last section immediately carry over to the infinite dimensonal field theory case. In particular, it is clear that the construction of the homotopy in terms of the propagator goes through for any field theory without gauge symmetry; we will see an example of this in the next section.

\subsection{Scalar field and high-energy modes}\label{sec:scalarfieldandhighenergymodes}

In this section, we consider a single scalar field $\phi(x)$ in $D$-dimensional Euclidean space and the projector $P = \Pi_\Lambda$ onto low-energy modes given by
\begin{equation}\label{eq:homotopycutoff}
\Pi_\Lambda \phi (x) = \int_{p^2 < \Lambda^2} \tilde{\phi}(p) \,e^{-ipx} \,d^D\!p\, ,
\end{equation}
where $\Lambda$ is some arbitrary energy scale and $\tilde{\phi}$ is the Fourier transform of $\phi$. As in the previous example, we will show that we can take the propagator of high-energy modes ($p^2 > \Lambda^2$), with a minus sign, as the homotopy.

The theory has a  field  $\phi(x)$ of ghost number $0$  and  an antifield $\starred \phi (x)$ of ghost number $-1$.
The master action is the same as the classical action,
\begin{equation}
\Theta =  \int d^D \!x \,  \left( \frac 1 2 \phi (-\Delta + m^2) \phi + V(\phi) \right)
\end{equation}
where the kinetic term is given by the (Euclidean) Klein-Gordon operator and we have included a scalar potential $V$.
The BV-BRST operator is
\begin{equation}
Q= \int d^D x \, [(-\Delta + m^2) \phi +V'](x) \frac {\partial} {\partial \starred \phi (x)} 
\end{equation}
with linear part
\begin{equation}
\partial^*= \int d^D x \, [(-\Delta + m^2) \phi](x) \frac {\partial} {\partial \starred \phi (x)} 
\end{equation}

In the coalgebra formulation, we then have two spaces $X_0, X_{-1}$ at degree $0$ and $-1$. 
These spaces are infinite-dimensional, but we will not need to specify the exact space of functions we are dealing with because the differential and homotopy entering the homotopy transfer construction are both linear. The variable $x$ just plays the role of a continuous index. 
The differential (1-bracket) $\partial$ of $\phi\in X_0$ is given by the (Euclidean) Klein-Gordon operator,
\begin{equation}
(\partial \phi)(x) = (-\Delta + m^2) \phi(x)\, .
\end{equation}
Higher brackets depend on the self-interactions of $\phi$ and will not be needed in what follows.

Now, following section \ref{section:homotopyasprojectedpropagator}, we take (minus) the propagator of high-energy modes as the homotopy operator $h: X_{-1} \rightarrow X_0$: this reads
\begin{equation}
(h\starred \phi)(x) =  - \int_{p^2 > \Lambda^2} \frac{e^{-ipx}}{p^2 + m^2} \tilde{\starred\phi}(p) \,d^D\!p  \, .
\end{equation}
We now check the homotopy relation $\Pi_\Lambda = {\bf 1} + \pd h + h \pd$: acting on $X_0$, it is
\begin{align}
(\Pi_\Lambda \phi)(x) &= \phi(x) + 0 + (h\pd\phi)(x) \\
&= \phi(x) - \int_{p^2 > \Lambda^2} \frac{e^{-ipx}}{p^2 + m^2} \widetilde{(\partial \phi)}(p) \,d^D\!p \\
&= \phi(x) - \int_{p^2 > \Lambda^2} \frac{e^{-ipx}}{p^2 + m^2} (p^2 + m^2) \tilde{\phi}(p) \,d^D\!p \\
&= \int \tilde{\phi}(p) \,e^{-ipx} \,d^D\!p\, - \int_{p^2 > \Lambda^2} \tilde{\phi}(p) \,e^{-ipx} \,d^D\!p\, \\
&= \int_{p^2 < \Lambda^2} \tilde{\phi}(p) \,e^{-ipx} \,d^D\!p\, ,
\end{align}
and on $X_{-1}$, we have
\begin{align}
(\Pi_\Lambda \starred \phi)(x) &= \starred\phi(x) + (\pd h \starred\phi)(x) + 0 \\
&= \phi(x) + (-\Delta + m^2) (h\starred\phi)(x) \\
&= \phi(x) - (-\Delta + m^2) \int_{p^2 > \Lambda^2} \frac{e^{-ipx}}{p^2 + m^2} \tilde{\phi}(p)  \,d^D\!p \\
&= \phi(x) - \int_{p^2 > \Lambda^2} (p^2 + m^2) \frac{e^{-ipx}}{p^2 + m^2} \tilde{\starred\phi}(p) \,d^D\!p\\
&= \int_{p^2 < \Lambda^2} \tilde{\starred\phi}(p) \,e^{-ipx} \,d^D\!p\, .
\end{align}
Thus, the homotopy relation is satisfied and one can indeed view (tree-level) integrating out of high-energy modes as homotopy transfer: the formulas of section 2 will produce all desired vertices, and this corresponds to the expansion in tree diagrams using the propagator of high-energy modes in the internal legs.

Note, however, that this computation
\begin{enumerate}
\item is insensitive to the value of $m$, in particular we could have taken $\Lambda < m$;
\item would work in the same way if we ``integrate out'' the low-energy modes with momenta satisfying $p^2 < \Lambda^2$.
\end{enumerate}
Although this is unorthodox from a field-theory point of view, it is interesting that the $L_\infty$ algebra structure can nevertheless be transferred consistently to the resulting theories.

\subsection{Yang-Mills theory}

We now turn to Yang-Mills, a theory with gauge symmetry and therefore a bigger complex $X = X_1 \oplus X_0 \oplus X_{-1} \oplus X_{-2}$, where the new spaces $X_1$ and $X_{-2}$ correspond to ghosts (gauge parameters) and their antifield conjugates. In this case, the homotopy will consist of three maps $h_0$, $h_{-1}$, $h_{-2}$, as in the diagram below.
\begin{equation}
\begin{tikzcd}[column sep=3.5em,row sep=3em]
0 \arrow[r,"\pd_2"] & X_{1} \arrow[r,"\pd_1"]  \arrow[d,"P_1"'] & X_0 \arrow[ld,"h_0"'] \arrow[r,"\pd_0"] \arrow[d,"P_0"'] & X_{-1} \arrow[ld,"h_{-1}"'] \arrow[d,"P_{-1}"'] \arrow[r,"\pd_{-1}"] & X_{-2} \arrow[r,"\pd_{-2}"] \arrow[ld,"h_{-2}"'] \arrow[d,"P_{-2}"] & 0 \\
0 \arrow[r,"\pd_{2}"'] & \iota\bar{X}_1 \arrow[r,"\pd_{1}"'] & \iota\bar{X}_0 \arrow[r,"\pd_{0}"'] & \iota\bar{X}_{-1} \arrow[r,"\pd_{-1}"'] & \iota\bar{X}_{-2}\arrow[r,"\pd_{-2}"'] & 0
\end{tikzcd}
\end{equation}
Note that there is no difference between Yang-Mills and a sum of free Maxwell actions for the problem of finding $h$, as only the $1$-bracket $\partial$ (given by the quadratic part of the action) appears.

As indicated in section \ref{section:homotopyasprojectedpropagator}, we will first go to the gauge-fixed action and construct a ``propagator'' $G$ satisfying \eqref{eq:homotopyG}. Then, given a projector $P$ commuting with $G$, the homotopy is simply given by projecting $G$ according to  \eqref{eq:hfromG}.

\subsubsection{Gauge-fixed action and propagator}

The minimal sector for Yang-Mills consists of the gauge field $A_\mu^a$, the ghost $C^a$, and their antifields, with BV master action 
\begin{equation}\label{eq:SBVYMmin}
\Theta^\text{min.}_\text{BV}[A_\mu^a, C^a, \starred A ^{\mu}_a, \starred C_a] = \int d^4x \left( -\frac{1}{4} \delta_{ab}\cF^a_{\mu\nu} \cF^{b\mu\nu} + \starred A^{\mu}_a (D_\mu C)^a + \frac{1}{2} \starred C_a f^a_{bc} C^b C^c \right)\,. 
\end{equation}
The field strength and covariant derivatives are
\begin{equation}
    \cF^a_{\mu\nu} = \pd_\mu A^a_\nu - \pd_\nu A^a_\mu + g f^a_{bc} A^b_\mu A^c_\nu\, ,\quad (D_\mu C)^a = \pd_\mu C^a + g f^a_{bc} A^b_\mu C^c \, ,
\end{equation}
and gauge indices are contracted with $\delta_{ab}$. 
We then follow the usual procedure:
\begin{enumerate}
\item we enlarge the space of fields and antifields to include  $(\bar{C}^a, B^a)$ where $\bar C$ is the anti-ghost and $B$ is an auxiliary field, together with the antifield conjugate $(\starred{\bar{C}}_a, \starred B_a)$;
\item add the term $\int \,d^4x \starred{\bar{C}}_a B^a$ to the master action; and
\item introduce a gauge-fixing fermion
\begin{equation}
 \Psi = \int \!d^4\!x \,\bar{C}_a \left( -\frac{\xi}{2} B^a + \pd^\mu \!A^a_\mu  \right)\, , 
\end{equation}
and shift the antifields by the derivative of the  gauge-fixing fermion,
\begin{equation}
\starred \phi_i \mapsto \starred\phi_i+\frac{\delta \Psi}{\delta \phi^i}\, .
\end{equation}
\end{enumerate}
  The complete non-minimal set of fields is then
$\{\phi^i\} = \{ A^a_\mu , C^a, \bar{C}^a, B^a \}$, of ghost degrees $0$, $1$, $-1$ and $0$ respectively while the antifields are $\{ \starred\phi_i \} = \{ \starred{A}^{\mu}_a , \starred{C}_a, \starred{\bar{C}}_a, \starred{B}_a \}$ which  carry ghost number $-1$, $-2$, $0$ and $-1$ respectively (so $\text{gh}(\starred\phi) = - \text{gh}(\phi) -1$).
After the shift, the master action becomes
\begin{equation}\label{eq:SBVYM}
\Theta_\text{BV}[\phi,\starred \phi] = S_\Psi[\phi] + \int d^4x \left( \starred A^{\mu}_a (D_\mu C)^a + \frac{1}{2} \starred C_a f^a_{bc} C^b C^c  +\starred {\bar{C}}_a B^a\right)\, ,
\end{equation}
where 
\begin{equation}
S_\Psi [\phi] = \int d^4x \left( -\frac{1}{4} \cF^a_{\mu\nu} \cF^{a\mu\nu} - \pd_\mu \bar{C}^a (D^\mu C)^a - \frac{\xi}{2} B^a B^a + B^a \pd^\mu \! A^a_\mu \right)\, .
\end{equation}
The action $S_\Psi$ is
the Fadeev-Popov gauge-fixed action, simply obtained from the shifted master action by  setting the antifields to zero.
(Equivalently, it is obtained from the unshifted master action
using \eqref{eq:antifieldgaugefixed}.)
The field $B^a$ is an auxiliary field; eliminating it from $S_\Psi$ produces the usual gauge-fixing term $+(\pd^\mu A^a_\mu)^2/(2\xi)$.

The quadratic  part of $\Theta$  gives the linear part $\partial^*$ of the BRST differential $Q = (\Theta_\text{BV},\, \cdot \,) = \pd^* + W$: it increases the ghost number by one and is given by
\begin{align}
\pd^* A^a_\mu &= \partial_\mu C^a \\
\pd^* C^a &= 0\\
\pd^* \bar{C}^a &= - B^a\\
\pd^* B^a &= 0
\end{align}
on the fields and
\begin{align}
\pd^* A^{*\mu}_a &= - \pd_\nu (\pd^\mu A^{\nu}_a - \pd^\nu A^{\mu}_a) - \pd^\mu B_a \\
\pd^* C^{*}_a &= \Box \bar{C}_a - \pd_\mu A^{*\mu}_a \\
\pd^* \bar{C}^{*}_a &= - \Box C_a\\
\pd^* B^{*}_a &= \bar{C}^*_a - \xi B_a + \pd_\mu A^\mu_a
\end{align}
on the antifields. The nilpotency of $\pd^*$ encodes the linearized gauge invariance of the equations of motion as well as the linearized Noether identities.

These formulas can be dualized to get the differential $\partial$ on the vector space $X$. 
Since there are now several fields/antifields of the same degree, it is useful (if a bit cumbersome) to refine the notation for the basis elements $T_a$ of $X$.
For each field $\Phi ^a$\footnote{Note we are here temporarily reverting to the notion of earlier sections where $a$ labels all fields; we will shortly revert to using the index $a$ as a gauge index. The meaning of the indices should be clear from the context.} there is a corresponding basis element $z^a$ of $X^*$  and this is  dual to a basis element $T_a$ of $X$, which we will write  $T_x[\phi]_a$\footnote{We  can take these to be delta functions centred on $x$, with possible extra Grassmann variables to account for the degree, and multiplied by a basis vector pointing in the direction $a$.}.
A general vector in $X$ is $v^aT_a$ where the components are all real numbers of degree zero.
The dual pairing is
\begin{equation}\label{eq:pairingYM}
\langle T_x[z]_a, z^b(y) \rangle = \delta^b_a\, \delta^4(x- y)\, .
\end{equation}
The basis elements for the various subspaces $X_k$ at degree $k$ are given by
\begin{equation}
\begin{array}{cccc}
X_1 & X_0 & X_{-1} & X_{-2} \vspace{0.1cm}\\
\{T_x[C]_a\} & \{T_x[A]^\mu_a\, , T_x[B]_a\, , T_x[\starred{\bar{C}}]^a \} & \{T_x[\bar{C}]_a\, , T_x[\starred A]^a_\mu\, , T_x[\starred B]^a \}& \{T_x[\starred C]^a \}
\end{array}
\end{equation}
so that, for example, a generic element of $X_0$ can be written as
\begin{equation}
a + b + \bar{c}^* = \int\left( a_\mu^a(x) \,T_x[A]^\mu_a + b^a(x) \,T_x[B]_a + \starred{\bar{c}}_a(x) \,T_x[\starred{\bar{C}}]^a \right) d^4\!x\, .
\end{equation}
We will use in the following this notation with lowercase letters where $a = \int a_\mu^a(x) T_x[A]^\mu_a d^4\!x$, etc. 
Thus the components $v^a$  of a vector $v^aT_a\in X$ are then  $a_\mu^a(x), b^a(x), \starred{\bar{c}}_a(x), \dots $and are all of degree zero.
For example, $C^a $ has ghost number $1$, $\{T_x[C]_a\} $ has ghost number $-1$ and $c^a $ has ghost number $0$. 
With this notation, we can compute the differential $\pd$ acting on the various subspaces using $\langle \pd x, z \rangle = \langle x, \pd^* z \rangle$.
Note that the space $X_k$ is paired with the space of fields at ghost number $k$: therefore, $\partial$ decreases the degree by one, as it should (since $\partial^*$, as $Q$, \emph{increases} the ghost number by one). 

The result is, omitting the arguments $x$ and the integrals for brevity,
\begin{align}
\pd a &= - \pd_\nu \left( \pd^\mu a^\nu_a - \pd^\nu a^\mu_a \right) \,T[\starred A]_\mu^a + \pd_\mu a^\mu_a \,T[\starred B]^a & &\in X_{-1}\\
\pd c &= \pd_\mu c^a\, T[A]^\mu_a - \Box c_a \,T[\starred{\bar{C}}]^a & &\in X_{0}\\
\pd \bar{c} &= \Box \bar{c}_a \,T[\starred C]^a & &\in X_{-2}\\
\pd b &= - b^a \,T[\bar{C}]_a - \pd^\mu b_a \,T[\starred A]^a_\mu - \xi b_a \,T[\starred B]^a & &\in X_{-1}
\end{align}
and
\begin{align}
\pd \starred a &= -\pd_\mu \starred a^{\mu}_a \,T[\starred C]^a & &\in X_{-2}\\
\pd \starred c &= 0 & &\in X_{-3} = \{ 0 \} \\
\pd \starred{ \bar{c}} &=\starred{ \bar{c}}_a \,T[\starred B]^a & &\in X_{-1}\\
\pd \starred b &= 0 & &\in X_{-2}\, .
\end{align}

We can then look for a ``propagator'' in the sense of the previous section, i.e. a map $G$ of degree $-1$ such that
\begin{equation}
{\bf 1} = \pd G + G \pd\, .
\end{equation}
A solution of this is given by
\begin{align}
G \starred c &= \int \!d^4\!x\,d^4\!p\, e^{-ipx} \left( - \frac{1}{p^2}\right) \tilde{\starred c}^{a}(p) \,T_x[\bar{C}]_a \\
G \starred a &= \int \!d^4\!x\,d^4\!p\, e^{-ipx} \left( -\frac{1}{p^2}\left( \eta_{\mu\nu} - (1+\xi) \frac{p_\mu p_\nu}{p^2} \right) \tilde{\starred a}^{a \nu}(p) \,T_x[A]^\mu_a - \frac{i p_\mu}{p^2} \tilde{\starred a}^{a\mu}(p) \,T_x[B]_a \right) \\
G \starred b&= \int \!d^4\!x\,d^4\!p\, e^{-ipx} \frac{ip_\mu}{p^2} \tilde{\starred b}^{a}(p) \,T_x[A]^\mu_a \\
G \starred{\bar{c}} &= \int \!d^4\!x\,d^4\!p\, e^{-ipx} \frac{1}{p^2} \tilde{\starred {\bar{c}}}^{a}(p) \,T_x[C]_a\, ,
\end{align}
where tildes denote the Fourier transform.

\subsubsection{Homotopy transfer}

Given a projector $P$ that commutes with $G$, we can define a map $h$ from the propagator $G$ of the previous section by formula \eqref{eq:hfromG}. This $h$ will satisfy the homotopy relation automatically, and one can therefore perform the homotopy transfer to get a $L_\infty$ structure on $\bar{X} \simeq PX$.

Here are two examples of such projectors.
\begin{enumerate}
\item A projector that just removes some of the fields. Writing $\bar{a}$ for the indices in the image of $P$, the projectors $P_k$ (with $k = 1, 0, -1, -2$) take then all the same matrix form
\begin{equation}
P^{\bar{a}}_{\bar{b}} = \delta^{\bar{a}}_{\bar{b}}\, , \quad \text{other components } = 0\, ,
\end{equation}
with no action on the space-time index $x$. This commutes with $G$ because $G$ is diagonal in the gauge indices.
\item As in the scalar example, we can go to Euclidean signature and consider the projector $P = \Pi_\Lambda$ onto low-energy modes, acting as
\begin{equation}
\Pi_\Lambda \Phi^a(x) = \int_{p^2 < \Lambda^2} \tilde{\Phi}^a(p) \,e^{-ipx} \,d^4\!p
\end{equation}
on any field or antifield $\Phi^a$. This commutes with $G$ because $G$ is local in momentum space.
\end{enumerate}
Although the resulting theories are in general highly non-local (in position space), $L_\infty$ and BRST-BV algebraic structures still exist for them and follow from the original ones by homotopy transfer.

\section{Discussion}

We have investigated the physical content of the algebraic procedure of \emph{homotopy transfer} as applied to the formulation of classical field theories in terms of \lf-algebras. 
Given an \lf-algebra over a vector space $X$ along with a homotopy $h:X\to X$ and projective map onto a quasi-isomorphic chain complex $\bar X$, homotopy transfer provides an \lf-algebra over $\bar X$ along with a quasi-isomorphism of \lf-algebras between $X$ and $\bar X$. The main new technical results are the explicit homotopy transfer formulae \eqref{BFormula} and \eqref{homotopytransferredstructure_derivationpicture}, realised by the pair of \lf-algebra morphisms given by $F:\bS(X)\to \bS(\bar X)$ in \eqref{Fformula} and $E:\bS(\bar X)\to \bS(X)$ 
(the dual of (\ref{homotopytransferredstructure_derivationpicture})). These are compatible with a cyclic inner product, if present.\footnote{More precisely, $E^\star$ was shown to be compatible with the cyclic structures in the sense of \eqref{barTheta}, and $F$ was shown to produce the same \lf-algebra $E$ does. This is subject to dualisation caveats in the infinite dimensional case. However, $E$ can be written in the coalgebra picture directly as \eqref{Erecursionalgebrapicture}, and then compatibility with the cyclic structure can be  checked.} {The most similar results to ours can be found in Huebschmann's work \cite{huebschmann2011lie,huebschmann}, where morphisms $F$ and $E$ in both directions are given and shown to be \lf-morphisms for ``connected'' \lf-algebras (concentrated in purely positive or negative degree).}

We have attempted to give a self-contained and pedagogical treatment, including issues which are  not easily accessible in the mathematical literature on homotopy transfer, including a discussion of homotopy transfer in the dual picture as well as the relationship of the latter with the BV formalism. 
In the following we highlight  our main technical results and their physical interpretations: 

\paragraph{The homotopy transfer formulae and their applications.} There is a relative dearth of homotopy transfer formulae for \lf-algebras as opposed to $A_\infty$-algebras, presumably due to the difficulty of lifting $h$ to the symmetric algebra $\bS(X)$; as discussed in \cite{manetti}, this precludes a straightforward application of the homological perturbation lemma. We overcame this difficulty in appendix \ref{app:lift}, following \cite{berglund}. (For homotopy transfer of \lf-algebras see also \cite{huebschmann,huebschmann1999formal}.) Our explicit expressions for $F:\bS(X)\to \bS(\bar X)$ and $E:\bS(\bar X)\to \bS(X)$ have the added advantage of being \emph{strict} partial inverses of each other (by Appendix \ref{app:equality}):
\be 
F E={\bf 1}_{\bS(\bar X)}\,.
\ee
In general, the right-hand side here could be any invertible morphism of coalgebras $\bS(\bar X)\to\bS(\bar X)$; we thus obtained the simplest possible result.

A special case is homotopy transfer onto $H(X)$, which establishes the existence of a \emph{minimal} $L_{\infty}$ structure on the homology of $X$ (i.e.~with $\bar\pd=0$), also known as the \emph{minimal model}. This homotopy transfer is known to calculate the \lf-algebra encoding S-matrix elements from an input \lf-algebra describing off-shell data, for any quantum field theory on Minkowski spacetime \cite{Arvanitakis:2019ald} (see also \cite{Jurco:2019yfd}, which employed loop \lf-algebras instead). At tree level in either approach the starting point is the \lf-algebra associated to the classical theory as discussed in  this paper. Definition \eqref{homotopytransferredstructure_derivationpicture} for $E:\bS(\bar X)\to \bS(X)$ leads to a recursion relation for the associated multilinear maps $e_n:S^n(\bar X)\to X$,  which may be immediately evaluated, as the coefficients can be read off from the classical Lagrangian. In fact, it was recently argued \cite{Macrelli:2019afx} that for Yang-Mills theory this recursion for the minimal model ($\bar X\equiv H(X)$)  yields  the Berends-Giele off-shell recursion relations  \cite{BerendsGiele} for tree-level amplitudes. (Observe that in this case $e_n$ takes precisely $n$ on-shell gluons as input and produces a single off-shell gluon as output, exactly like the Berends-Giele currents do.)

What about $F$? $F$ leads to what might be called ``dual (off-shell) recursion'', again for tree amplitudes. The recursion relation is \eqref{eq:bigFrecursion}:
\be
F =  p + F Bh\;. 
\ee
This is compact notation for a recursive definition of the multilinear maps $f_n:S^n(X)\to \bar X$ associated to $F$. For $\bar X=H(X)$ the new $L_{\infty}$ structure (specified by a coderivation $\bar D$) will describe tree-level S-matrix elements, and can be calculated given knowledge of $F$ (i.e.~of the collection of maps $\{f_n\}$) and the original \lf-algebra structure (specified by $D$):
\be
\bar D=FD\iota\,.
\ee
For Yang-Mills theory, $f_n$ would take precisely $n$ off-shell gluons as input and produce a single on-shell gluon as output. This is the exact opposite behaviour with respect to the Berends-Giele currents --- hence reasonably referred to as ``dual recursion''. 

\paragraph{The physical interpretation of homotopy transfer.}
We illustrated how homotopy transfer produces an \lf-algebra describing effective dynamics. More precisely, if we introduce a projector $P$ and a split
\be
X=PX\oplus(1-P)X\,,
\ee
homotopy transfer onto $\bar X\cong PX$ produces the \lf-algebra derived from the tree-level effective action for the $\bar X$ degrees of freedom, calculated by a ``tree-level path integral over $(1-P)X$''. Following the usual lore where a tree-level path integral amounts to solving the classical equations of motion, the morphism $E^\star$ is seen to provide this ``solution''. This idea has recently been discussed in string field theory \cite{Masuda:2020tfa,Erbin:2020eyc,Koyama:2020qfb}, motivated by a construction of Ashoke Sen \cite{Sen:2016qap}; however, as seems to be the general trend with ideas from string field theory expressed in the language of homotopy algebra, we demonstrated in a few examples that this effective field theory interpretation is valid beyond string field theory. In this interpretation, the Sen construction gives a convenient ansatz \eqref{eq:hfromG} for the construction of a homotopy $h$.

There is a restriction on admissible projectors $P$: that $\pd$ restricted to $(1-P)X$ has trivial homology, so $X$ and $\bar X\cong PX$ have isomorphic homologies. Physically, this means that homotopy transfer is not allowed to ``forget'' any on-shell states and their scattering amplitudes: the two \lf-algebras are quasi-isomorphic, and therefore have isomorphic minimal models \cite{doubek2007deformation}, and thus equivalent S-matrix elements. What, then, of Wilsonian low-energy effective field theory, in which we want to ``forget'' scattering processes involving high-mass modes? In subsection \ref{sec:scalarfieldandhighenergymodes} on scalar fields, where we constructed a homotopy \eqref{eq:homotopycutoff} realising a sharp cutoff in momentum space ($p^2<\Lambda^2$), the scalar field was assumed to be Euclidean, which evades the issue because there is no scattering in Euclidean space, so the corresponding \lf-algebra has $H(X)=0$. This suffices for effective Lagrangians.

We could work in Minkowski space instead. The space $X$ now consists of scalar fields with rapid falloff at infinity (the ``off-shell'' fields, see section \ref{sec:infinited}), scalar fields solving the Klein-Gordon equation (the ``on-shell'' fields), and their antifields. The homology $H(X)$ is naturally identified with the on-shell fields \cite{Arvanitakis:2019ald,Macrelli:2019afx}. We can  consider a sharp momentum cutoff again (at the price of Lorentz invariance) and a homotopy similar to \eqref{eq:homotopycutoff} realising it. Formula \eqref{PeqPprimeplusoneminusPprimeEtc} then gives the space $\bar X$ of ``effective'' degrees of freedom: off-shell fields with momenta below the cutoff, \emph{plus all} on-shell fields, including ones with momenta above the cutoff{\footnote{There is a similar observation in \cite[section 3.1]{Erbin:2020eyc}.}}. We have thus retained more information than expected, in the form of on-shell fields with $p^2>\Lambda^2$ and their scattering amplitudes. In fact all quasi-isomorphisms of \lf-algebras retain all on-shell states (i.e.~the homology $H(X)$), so this is not a deficiency of our particular formulae. One conceivable resolution is that in the limit where the cutoff is much smaller than the masses of on-shell fields, the \lf-algebra over $\bar X$ splits into a direct sum of the low-energy effective \lf-algebra $\bar X_{\rm IR}$ plus the minimal \lf-algebra encoding amplitudes of high-mass fields (that were ``integrated out''), perhaps due to locality arguments. Then we can compose the obvious morphisms $\bar X_{\rm IR}\to \bar X\to X$ to find a morphism $\bar X_{\rm IR}\to X$ which is \emph{not} a quasi-isomorphism, and realises Wilsonian low-energy effective field theory. A proper understanding here will necessarily involve subtleties of infinite-dimensionality in the \lf-algebra formulation of field theories.

\paragraph{Beyond (conventional) (tree-level) effective field theory.} The homotopy transfer construction clearly accommodates a more general notion of effective field theory than is usually considered. 
For instance, double field theory could never be obtained as a Wilsonian low-energy effective field theory from closed string field theory, because there is no duality frame in which all ``effective'' degrees of freedom are simultaneously of low mass. However, it can be obtained via homotopy transfer, employing the approach due to Sen. 
The details will appear in the forthcoming work \cite{second}.

The most glaring omission in the current paper is the treatment of effective field theory beyond the tree-level approximation. In the mathematical literature progress has been made using the notion of \emph{loop} \lf-algebra, which is a generalisation involving an infinite collection of brackets at each loop order, and is closely related to the quantum master equation in the BV formalism. We will discuss this approach in the forthcoming paper also, and introduce a complementary approach involving the Zinn--Justin 1PI \lf-algebra of \cite{Arvanitakis:2019ald}, which is suggestive of a non-perturbative formulation. Along the way, we will prove the assertion that $E^\star$ is the leading-order contribution to the purely-algebraic ``path'' integral that defines a procedure analogous to homotopy transfer for loop \lf-algebras.

The last claim suggests, again, the question: what about $F$? (Given that it is the dual of $E$ that admits a path integral interpretation.) From the algebraic perspective neither $F$ nor $E$ is privileged over the other; both are quasi-isomorphisms, and knowledge of one suffices to construct the new \lf-algebra structure. We are thus led to the idea that perhaps we should elevate the notion of quasi-isomorphism of \lf-algebras to an equivalence of the underlying field theories. For field theories on Minkowski space at least this is perfectly reasonable, since quasi-isomorphisms preserve minimal models (hence S-matrices). 
Then one might think, somewhat more provocatively, of quasi-isomorphisms as the mathematical implementation of dualities in field theory.

\section*{Acknowledgements}

We would like to thank Chris Blair, Ezra Getzler, Ashoke Sen, Alexandre Sevrin, Dennis Sullivan, Dan Thompson, and Barton Zwiebach for helpful discussions and correspondence. 

\noindent{We would also like to thank Alberto Cattaneo, Harold Erbin, Branislav Jur\v{c}o, Carlo Maccaferri,    Pavel Mnev, Christian S\"amann, Martin Schnabl, Jim Stasheff, Jakub Vosmera, and Martin Wolf for comments on the
first version of this paper and for bringing some references to our
attention.
}

\noindent The work of A.S.A. is supported in part by the “FWO-Vlaanderen” through the project G006119N and by the Vrije Universiteit Brussel through the Strategic Research Program “High-Energy Physics”. V.L. is supported by European Union's Horizon 2020 Research Council grant 724659 MassiveCosmo ERC-2016-COG.
The work of O.H.~is supported by the ERC Consolidator Grant ``Symmetries and Cosmology".

\appendix

\section*{Appendix}

\section{Lift of homotopy map }
\label{app:lift}

\subsection{Proof of homotopy relation}

In this appendix we prove that the lift of the homotopy map $h$ to the full symmetric algebra ${\bf S}(X)$ given in the main text still satisfies the 
homotopy relation (\ref{iotaPRELLL}). 
The first step is to rewrite the recursive definition
 \be\label{strangeLeibniz}
 \begin{split}
  h(x_1\wedge \ldots\wedge x_n) \equiv \frac{1}{n!}\sum_{\sigma\in S_n}&\epsilon(\sigma;x)\Big(
  h(x_{\sigma(1)}\wedge\ldots\wedge x_{\sigma(n-1)})\wedge  x_{\sigma(n)}\\
  &\;\;\;\quad +(-1)^{x_{\sigma(1)}+\cdots x_{\sigma(n-1)}}\iota p(x_{\sigma(1)}\wedge\ldots \wedge x_{\sigma(n-1)})\wedge   h x_{\sigma(n)}\Big) 
  \end{split} 
 \ee
as
\begin{equation}\label{leibnizlessterms}
\begin{split}
  h(x_1\wedge \ldots\wedge x_n) = \frac{1}{n}\sum_{k=1}^n&(-1)^{x_k(x_{k+1}+\cdots+x_n)}\Big[
  h(x_{1}\wedge\ldots\wedge \hat{x}_k \wedge\ldots\wedge x_{n})\wedge  x_k\\
  &\;\;\;\quad +(-1)^{x_1+\cdots +\hat{x}_k + \cdots + x_{n}}\iota p(x_{1}\wedge\ldots\wedge \hat{x}_k \wedge\ldots\wedge x_{n})\wedge   h x_k\Big]\;, 
  \end{split} 
\end{equation}
where the hat means that the element is omitted from the product or sum.
This can be done because in \eqref{strangeLeibniz} the permutations over the $n-1$ first arguments always give the same terms and add up: what matters for the Koszul sign is which of the $x_k$'s is outside. What remains is a sum over $k$, with coefficient $(n-1)!/n!=1/n$. This rewriting allows for a better control of the signs and the permutations. To simplify a bit the notation, in the following we will also use $P=\iota p$ and omit the wedge products.

The proof proceeds by induction, assuming that the relation holds for $n$.  Evaluating $\pd h(x_1 \ldots x_{n+1})$ is straightforward:
\begin{align}
\pd h(x_1 \ldots x_n x_{n+1}) &= \frac{1}{n+1}\sum_{k=1}^{n+1} (-1)^{x_k(x_{k+1}+\cdots+x_{n+1})}\pd \Big[
  h(x_{1}\ldots\hat{x}_k \ldots x_{n+1}) x_k\\
  &\qquad\qquad\qquad +(-1)^{x_1+\cdots +\hat{x}_k + \cdots + x_{n+1}}P(x_{1}\ldots \hat{x}_k \ldots x_{n+1}) h x_k\Big]\\
&= \frac{1}{n+1}\sum_{k=1}^{n+1} (-1)^{x_k(x_{k+1}+\cdots+x_{n+1})} \Big[
  \pd h(x_{1}\ldots\hat{x}_k \ldots x_{n+1}) x_k\\
  &\qquad\qquad\qquad+(-1)^{x_1+\cdots +\hat{x}_k + \cdots + x_{n}+1}h(x_{1}\ldots\hat{x}_k \ldots x_{n+1}) \pd x_k \label{line6}\\
  &\qquad\qquad\qquad +(-1)^{x_1+\cdots +\hat{x}_k + \cdots + x_{n+1}}\pd P(x_{1}\ldots \hat{x}_k \ldots x_{n+1})  h x_k \label{line7}\\
  &\qquad\qquad\qquad + P(x_{1}\ldots \hat{x}_k \ldots x_{n+1}) \pd h x_k \Big]\;. 
\end{align}
Evaluating $h\pd (x_1 \ldots x_{n+1})$ is trickier. First, we write explicitly
\begin{equation}\label{delproduct}
\pd (x_1 \ldots x_{n+1}) = \sum_{j=1}^{n+1} (-1)^{x_1 + \dots + x_{j-1}} \, x_1 \ldots \pd x_j \ldots x_{n+1}\, .
\end{equation}
Then, when using formula \eqref{leibnizlessterms}, the case $k = j$ is special (since now $\pd x_k$ is moved instead of $x_k$, which gives different signs), so we split the sum:
\begin{align}
h\pd (x_1 \ldots x_{n+1}) &= \sum_{j=1}^{n+1} (-1)^{x_1 + \dots + x_{j-1}} \, h(x_1 \ldots \pd x_j \ldots x_{n+1}) \\
&= \frac{1}{n+1} \sum_{j=1}^{n+1} (-1)^{x_1 + \dots + x_{j-1}} \times \\
&\qquad \times \Bigg\{ \sum_{k=1}^{j-1} (-1)^{x_k(x_{k+1} + \cdots + x_{n+1}-1)} \Big[ h(x_1 \ldots \hat{x}_k \ldots \pd x_j \ldots x_{n+1}) x_k \label{line12}\\
&\qquad\qquad\qquad\qquad + (-1)^{x_1 + \cdots + \hat{x}_k + \cdots + x_{n+1}-1} P(x_1 \ldots \hat{x}_k \ldots \pd x_j \ldots x_{n+1}) h x_k \Big] \label{line13}\\
&\qquad\qquad + (-1)^{(x_j-1)(x_{j+1} + \cdots + x_{n+1})} \Big[ h(x_1 \ldots \hat{x}_j \ldots x_{n+1}) \pd x_j \label{line14}\\
&\qquad\qquad\qquad\qquad + (-1)^{x_1 + \cdots + \hat{x}_j + \cdots + x_{n+1}} P(x_1 \ldots \hat{x}_j \ldots x_{n+1}) h \pd x_j \Big] \label{line15}\\
&\qquad\quad + \sum_{k=j+1}^{n+1} (-1)^{x_k(x_{k+1} + \cdots + x_{n+1})} \Big[ h(x_1 \ldots \pd x_j \ldots \hat{x}_k \ldots x_{n+1}) x_k \label{line16}\\
&\qquad\qquad\qquad\qquad + (-1)^{x_1 + \cdots + \hat{x}_k + \cdots + x_{n+1}-1} P(x_1 \ldots \pd x_j \ldots \hat{x}_k \ldots x_{n+1}) h x_k \Big] \Bigg\} \, .\label{line17}
\end{align}
Lines \eqref{line12} and \eqref{line16} combine nicely to give back a term of the form $h\pd(x_1 \ldots \hat{x}_k \ldots x_{n+1}) x_k$. To see this, we have to swap the sums, using
\begin{equation}
\sum_{j=1}^{n+1} \sum_{k=1}^{j-1} f(j,k) = \sum_{\substack{j,k = 1 \\ k< j}}^{n+1} f(j,k) = \sum_{\substack{k,j = 1 \\ j > k}}^{n+1} f(j,k) = \sum_{k=1}^{n+1} \sum_{j=k+1}^{n+1} f(j,k)
\end{equation}
and the same reasoning for the other switch. This gives an outer $k$ sum, and the inside $j$ sum reproduces $\pd(x_1 \ldots \hat{x}_k \ldots x_{n+1})$ using formula \eqref{delproduct}:
\begin{align}
\eqref{line12} + \eqref{line16} &= \frac{1}{n+1} \sum_{k=1}^{n+1} (-1)^{x_k(x_{k+1} + \cdots + x_{n+1})} \times \\
&\qquad \times \Big[ \sum_{j=k+1}^{n+1}(-1)^{x_k} (-1)^{x_1 + \cdots + x_{j-1}} h(x_1 \ldots \hat{x}_k \ldots \pd x_j \ldots x_{n+1}) x_k \\
&\qquad\qquad + \sum_{j=1}^{k-1} (-1)^{x_1 + \cdots + x_{j-1}} h(x_1 \ldots \pd x_j \ldots \hat{x}_k \ldots x_{n+1}) x_k \Big] \\
&= \frac{1}{n+1} \sum_{k=1}^{n+1} (-1)^{x_k(x_{k+1} + \cdots + x_{n+1})} h \pd(x_1 \ldots \hat{x}_k \ldots x_{n+1}) x_k  \label{line22}
\end{align}
where we used $(-1)^{x_k} (-1)^{x_1 + \cdots + x_{j-1}} = (-1)^{x_1 + \cdots + \hat{x}_k + \cdots + x_{j-1}}$. Using the same reasoning, lines \eqref{line13} and \eqref{line17} combine into
\begin{equation}\label{line23}
\eqref{line13} + \eqref{line17} = \frac{1}{n+1} \sum_{k=1}^{n+1} (-1)^{x_k(x_{k+1} + \cdots + x_{n+1})} (-1)^{x_1 + \cdots + \hat{x}_k + \cdots + x_{n+1} - 1} P \pd (x_1 \ldots \hat{x}_k \ldots x_{n+1}) x_k \, .
\end{equation}

We can finally compute $(\pd h + h \pd)(x_1 \dots x_{n+1})$. In this, \eqref{line6} exactly cancels \eqref{line14}, \eqref{line7} cancels \eqref{line23} owing to $\pd P = P \pd$, and the signs of the other terms work out to give
\begin{align}
&(\pd h + h \pd)(x_1 \dots x_{n+1}) = \frac{1}{n+1} \sum_{k=1}^{n+1} (-1)^{x_k(x_{k+1}+\cdots+x_{n+1})} \times \\
&\qquad \times \Big[ \underbrace{(\pd h + h\pd)}_{(P-{\bf 1)}}(x_{1}\ldots\hat{x}_k \ldots x_{n+1}) x_k + P(x_{1}\ldots\hat{x}_k \ldots x_{n+1})\underbrace{(\pd h + h\pd)}_{(P-{\bf 1})} x_k \Big] \\
&= \frac{1}{n+1} \sum_{k=1}^{n+1} (-1)^{x_k(x_{k+1}+\cdots+x_{n+1})} \Big[ -(x_{1}\ldots\hat{x}_k \ldots x_{n+1}) x_k + P(x_{1}\ldots\hat{x}_k \ldots x_{n+1})P x_k \Big] \\
&= (P-{\bf 1}) (x_1 \ldots x_{n+1})\;, 
\end{align}
where we used the morphism property $P(x_{1}\ldots\hat{x}_k \ldots x_{n+1})P x_k = P(x_{1}\ldots\hat{x}_k \ldots x_{n+1} x_k)$ and also put $x_k$ back in its place, 
using $(-1)^{x_k(x_{k+1}+\cdots+x_{n+1})} x_{1}\ldots\hat{x}_k \ldots x_{n+1} x_k = x_1 \ldots x_{n+1}$.
This proves the homotopy relation acting on monomials of degree $n+1$, thereby completing the proof by induction.

\subsection{Side conditions}
We will now prove that under the side conditions the lift of $h$ implies the relations (\ref{module}).  
The trick is to rewrite $h$ as
\begin{equation}\label{eq:factorh}
h = h^d q = q h^d \, ,
\end{equation}
where $h^d$ is the lift of $h$ as a normal derivation and $q$ takes care of inserting the $\iota p$'s. More specifically, $h^d$ is defined by the Leibniz rule
\begin{align}
\label{eq:defhd}
h^d(x_1 \wedge \ldots \wedge x_n) &= h^d(x_1 \wedge \ldots \wedge x_{n-1}) \wedge x_n + (-1)^{x_1 + \cdots + x_{n-1}} x_1 \wedge \ldots \wedge x_{n-1} \wedge h(x_n) \\
&= \sum_{i=1}^{n} (-1)^{x_1 + \cdots + x_{i-1}} x_1 \wedge \ldots \wedge h(x_i) \wedge \ldots x_n\, ,
\end{align}
and $q$ is given by
\begin{align}\label{eq:qperm}
q(x_1 \wedge \ldots \wedge x_n) &= \frac{1}{n!} \sum_{\sigma\in S_n} \sum_{i=0}^{n-1} \frac{1}{n-i} \;\epsilon(\sigma ;x) \,\iota p (x_{\sigma(1)} \wedge \ldots \wedge x_{\sigma(i)}) \wedge x_{\sigma(i+1)} \wedge \ldots \wedge x_{\sigma(n)}\;. 
\end{align}
With these definitions, the proof of $h = h^d q$ is direct. To prove that $q$ and $h^d$ commute, we follow \cite{berglund} and rewrite $q$ in a different way: first, notice that permutations of the $i$ first arguments and the $n-i$ last ones always give the same term, so each individual term comes with a factor \begin{equation}
Q^n_i = \frac{i! (n-i)!}{n!(n-i)} = \frac{i!(n-i-1)!}{n!}\qquad (i<n)\; .
\end{equation}
Then, because $\iota p$ has no intrinsic degree, we get no sign as we reorder the terms to appear in the right order $x_1 \wedge x_2 \wedge \ldots \wedge x_n$, with some of the $x_i$'s carrying $\iota p$. This gives the formula
\begin{align}
q(x_1 \wedge \ldots \wedge x_n) &= \sum_{\epsilon \in \{ 0, 1\}^n} Q^n_{|\epsilon|}\; (\iota p )^{\epsilon_1}(x_1) \wedge (\iota p )^{\epsilon_2}(x_2) \wedge \ldots \wedge (\iota p )^{\epsilon_n}(x_n)\, ,
\end{align}
where: $\epsilon = (\epsilon_1, \epsilon_2 , \dots, \epsilon_n)$ is a sequence of zeroes and ones, with $\epsilon_i$ telling us if $x_i$ carries an $\iota p$ or not, and we sum over all such sequences; $|\epsilon| = \epsilon_1 + \dots + \epsilon_n$ is the total number of $\iota p$'s in that term (which is the index $i$ in \eqref{eq:qperm}); and we define $Q^n_n = 0$ to be consistent with \eqref{eq:qperm}. Now, because of the side conditions, we have $\iota p h = 0 = h\iota p$, so this way of writing $q$ makes it clear that $q h^d = h^d q$. Notice also that this nice factorization would not hold without the side conditions: without them, one would get extra terms of the form $h \iota p$ (in $h^d q$) or $\iota p h$ (in $q h^d$).

Now, since $h^d$ is linear, it is a coderivation as well as a derivation,
\begin{equation}
\Delta h^d = (h^d \otimes {\bf 1} + {\bf 1} \otimes h^d) \Delta \; .
\end{equation}
This is the crucial trick which helps below: it allows us to move all the $h$'s across $\Delta$, i.e.
\begin{equation}
\Delta h = (h^d \otimes {\bf 1} + {\bf 1} \otimes h^d) \Delta q\;, 
\end{equation}
without having to worry about insertions of $\iota p$'s. The $h^d$ part also satisfies the side conditions, 
\be
h^d h^d = 0\, ,\quad ph^d = 0\, ,\quad h^d\iota = 0\, ,
\ee
as does  the full lift $h=h^d q = q h^d$. These properties imply that the seven maps below vanish (``annihilation conditions''):
\begin{align}\label{eq:annihilation}
0 &= (h \otimes h) \Delta h = (h\otimes h) \Delta \iota = (h \otimes p) \Delta h = (h \otimes p) \Delta \iota \\
&= (p \otimes h) \Delta h = (p \otimes h) \Delta \iota = (p \otimes p) \Delta h\, .\nonumber
\end{align}
This is easily checked using the decompositions of $h$, the coderivation property of $h^d$, the morphism property of $\iota$ and the side conditions: for example,
\begin{align}
(h \otimes p) \Delta h &= (qh^d \otimes p) \Delta h^d q \\
&= (qh^d \otimes p) (h^d \otimes {\bf 1} + {\bf 1} \otimes h^d) \Delta q \\
&= (q h^d h^d \otimes p + q h^d \otimes p h^d) \Delta q \\
&= 0\, .
\end{align}

To finish the proof, we use the homotopy relation $\iota p = {\bf 1} + \partial h + h \partial$ 
to show that $h$ satisfies the three properties in \eqref{module}. This uses the annihilation conditions \eqref{eq:annihilation} (which themselves followed from the symmetrized $(\iota p, {\bf 1})$-Leibniz rule and the side conditions).
The first relation in (\ref{module}) is now proved as follows 
\begin{align}
(h\otimes {\bf 1} - {\bf 1} \otimes h) \Delta h &= \left[ h\otimes (\iota p - \partial h - h \partial) - (\iota p - \partial h - h \partial)\otimes h \right] \Delta h \\
&= ({\bf 1}\otimes \iota) (h\otimes p) \Delta h + ({\bf 1} \otimes \partial + \partial\otimes {\bf 1})(h\otimes h) \Delta h \nonumber\\
&\quad - ({\bf 1}\otimes \iota)(p\otimes h) \Delta h - (h\otimes h)({\bf 1} \otimes \partial + \partial\otimes {\bf 1}) \Delta h \\
&= - (h\otimes h) \Delta \partial h \\
&= - (h\otimes h) \Delta (\iota p - {\bf 1} - h \partial) \\
&= (h\otimes h) \Delta\, ,
\end{align}
where we used the homotopy relation, the coderivation property of $\partial$ and the annihilation conditions.

For the second relation in (\ref{module}) we compute 
\begin{align}
(p\otimes {\bf 1})\Delta h &= p\otimes (\iota p - \partial h - h \partial) \Delta h \\
&= ({\bf 1}\otimes \iota) (p\otimes p)\Delta h - ({\bf 1}\otimes \partial) (p\otimes h) \Delta h - (p\otimes h)({\bf 1}\otimes \partial) \Delta h \\
&= -(p\otimes h) ({\bf 1} \otimes \partial + \partial\otimes {\bf 1}) \Delta h + (p\otimes h) (\partial \otimes {\bf 1}) \Delta h
\end{align}
Using $p \partial = \bar{\partial} p$, the second term gives
\begin{align}
(p\otimes h) (\partial \otimes {\bf 1}) \Delta h = -(p\partial \otimes h) \Delta h = - (\bar{\partial} p \otimes h) \Delta h = - (\bar{\partial} \otimes  {\bf 1}) (p\otimes h) \Delta h = 0\, ,
\end{align}
and the first is 
\begin{align}
-(p\otimes h) ({\bf 1} \otimes \partial + \partial\otimes {\bf 1}) \Delta h &= -(p\otimes h) \Delta \partial h = -(p\otimes h) \Delta (\iota p - {\bf 1} - h \partial) \\
&= (p\otimes h) \Delta\, .
\end{align}
The final property in \eqref{module} is proved in exactly the  same way.

\section{Equality of homotopy-transferred \lf-algebra structures}
\label{app:equality}

We will prove that
\be \label{eq:FE=1}
F E={\bf 1}_{\bS(\bar X)}
\ee
where $F:\bS(X)\to \bS(\bar X)$ is the morphism of coalgebras \eqref{Fformula} and $E:\bS(\bar X)\to \bS(X)$ is the \emph{dual} to the morphism of algebras $E^\star: \bS(X)^\star\to \bS(\bar X)^\star$ defined in \eqref{homotopytransferredstructure_derivationpicture}, when both are constructed from the same collection of \lf-algebra brackets $(\partial,b_1,b_2\dots)$ and the same homotopy $h$ satisfying the side conditions \eqref{ipsides}. (Recall that $F$ is a morphism of coalgebras when the side conditions are satisfied.) This implies that, when both are defined (e.g.~for finite-dimensional \lf-algebras),
\be
\bar D=\bar Q^*\,,
\ee
where on the left-hand side $\bar D=\bar\partial +\bar B$ is the homotopy-transferred \lf-algebra structure of \eqref{barBBB}, while on the right $\bar Q=\bar \partial^\star+\bar W$ is the homotopy-transferred structure of \eqref{homotopytransferredstructure_derivationpicture}.

In the algebra picture it is clear that a derivation or a morphism of algebras is completely specified by its values on generators $\{z^a\}$ of $\bS(X)^\star$. Dually, a coderivation or a morphism $M:\bS (X)\to\bS(X)$ of coalgebras is completely specified by $\prone M:\bS(X)\to X$ \cite{Markl:1997bj}, where $\prone$ is the projection $\bS(X)\to X$:
\be
\prone x_1=x_1\,,\qquad \prone(x_1\wedge x_2\wedge \cdots \wedge x_n)=0\quad\forall n>1\,,\quad \forall x_1,x_2,\dots x_n\in X\,.
\ee
This is clear from the defining duality relation using the pairing \eqref{eq:pairingS} of $\bS$ and $\bS^\star$ with a choice of basis:
\be
\langle M^\star(z^a),T_{b_1}\wedge T_{b_2}\wedge \cdots T_{b_n}\rangle=\langle z^a, M(T_{b_1}\wedge T_{b_2}\wedge \cdots T_{b_n})\rangle\,;
\ee
on the right-hand side, $\langle z^a,\,\cdot\,\rangle$ projects onto 
$X$. We therefore specify $E:\bS(\bar X)\to \bS(X)$ in the coalgebra picture directly, as the unique morphism of coalgebras that satisfies the recursion relation
\be
\label{Erecursionalgebrapicture}
\prone  E=\prone \iota + h \prone B E\,.
\ee
Its dual is \eqref{homotopytransferredstructure_derivationpicture}. (This is identical to a morphism constructed in \cite{manetti}; side conditions are invoked therein to prove that $E$ is a morphism of \lf-algebras, which is unnecessary.)

Since $F=p+FBh$ it suffices to show
\be
\label{app:pEeq1}
pE={\bf 1}_{\bS(\bar X)}
\ee
and
\be
h E=0
\ee
where $h$ is the lifted homotopy \eqref{strangeLeibniz2} that is a map $\bS(X)\to \bS(X)$. The maps $p$ and $E$ are both coalgebra morphisms, so $pE$ is specified by $\bar{\pi}_1 pE$ (where $\bar \pi_1:\bS(\bar X)\to\bar X$ is the analogous projection involving $\bar X$). By \eqref{iotaPext} for $p$ we see $\bar \pi_1 p=p\prone$, so we can use \eqref{Erecursionalgebrapicture} and the side conditions to calculate directly
\be
\bar{\pi}_1 p E=p\prone E=p\prone\iota + p h \prone BE=\bar\pi_1\implies pE={\bf 1}_{\bS(\bar X)}\,.
\ee

It remains to show $hE=0$. By \eqref{eq:factorh} it suffices to show instead $h^d E=0$ where $h^d$ is the derivation \eqref{eq:defhd} on $\bS(X)$. It is easiest to prove
\be
h^d E(\bar x_1\wedge \bar x_2\wedge\cdots \wedge\bar x_n)=0
\ee
inductively in $n$. First, $h^d E(\bar x_1)=h^d \iota(\bar x_1)=0$ by the side conditions. For the inductive step, we assume
\be
\label{FequalsEinductiveassumption}
h^d E(\bar x_1\wedge \cdots \wedge \bar x_m)=0\qquad \forall m\leq n\,,\quad \forall \bar x_1,\bar x_2,\dots \bar x_m\in X\,.
\ee
and prove
\be \label{eq:hdE}
h^d E(\bar x_1\wedge \cdots \wedge \bar x_{n+1})=0\,.
\ee
Let us write $e_n : S^n(\bar{X}) \rightarrow X$ for the collection of maps defining the morphism $E$, as in equations \eqref{coalgebramorphismaction}. The inductive assumption is equivalent to
\begin{equation}
h \,e_m = 0 \qquad \forall m \leq n\, .
\end{equation}
Since $E$ is a morphism of coalgebras, we have an equation of the type
\begin{align} \label{eq:EminusPiE}
(E-\prone E)(\bar x_1\wedge \cdots \wedge \bar x_{n+1}) =&\sum_{\text{partitions}} \;\sum_{\sigma \in S_{n+1}} \;c(\bar x, \{a_i\}) \\
&\times\; e_{a_1}(\bar x_{\sigma(1)}\wedge\ldots \wedge\bar x_{\sigma (a_1)})\wedge \ldots \wedge e_{a_k}(\bar x_{\sigma(a_1 + \dots + a_{k-1})}\wedge\cdots \wedge\bar x_{\sigma (n+1)})\, .\nonumber
\end{align}
for some coefficients $c(\bar x,\{a_i\})$. Here, the first sum runs over all ways of splitting the integer $n+1$ into $k \geq 2$ parts of length $a_i > 0$,
\begin{equation}
n+1 = a_1 + a_2 + \dots + a_k \quad (k \geq 2\,, \;a_i > 0)\, .
\end{equation}
The crucial point here is that all the maps $e_{a_i}$ appearing on the right-hand side have $a_i < n+1$, which will allow us to use the inductive hypothesis. The existence of a formula like \eqref{eq:EminusPiE} is apparent from \eqref{coalgebramorphismaction}: removing $\pi_1 E$ amounts to removing the first term, and we remain only with wedge products of maps of lower order. (Nevertheless, a proof of \eqref{eq:EminusPiE} for all $n$ is given below.)

Now, the side conditions and \eqref{Erecursionalgebrapicture} immediately show $h^d\prone E=0 $. Then, since $h^d$ is a derivation, acting on \eqref{eq:EminusPiE} with $h^d$ and using the inductive assumption implies \eqref{eq:hdE}. This concludes the proof of \eqref{eq:FE=1}.

To prove \eqref{eq:EminusPiE} at all orders, we define a (rescaled) wedge product operation
\be
\Lambda: \bS(X) \otimes \bS(X) \rightarrow \bS(X)
\ee by
\begin{align}
\Lambda((x_1 \wedge \dots \wedge x_i) \otimes (x_{i+1} \wedge \dots \wedge x_{i+j}))= \frac{1}{2^{i+j}-2} \, x_1 \wedge \dots \wedge x_i \wedge x_{i+1} \wedge \dots \wedge x_{i+j}\, .
\end{align}
It satisfies the property
\begin{equation}
\Lambda \Delta = {\bf 1}_{\bS(X)} - \pi_1\, ,
\end{equation}
as can be checked by direct computation: on $x_1$ it is $0 = 0$, and on a higher-order monomial,
\begin{align}
\Lambda \Delta (x_1 \wedge \dots x_n) &= \Lambda \sum_{i=1}^{n-1} \sum_{\sigma \in (i,n-i)} \epsilon(\sigma; x)\, (x_{\sigma(1)}\wedge\ldots \wedge x_{\sigma(i)})\,\otimes \, 
  (x_{\sigma(i+1)}\wedge \ldots \wedge x_{\sigma(n)}) \\
  &= \frac{1}{2^n-2} \sum_{i=1}^{n-1} \sum_{\sigma \in (i,n-i)} \epsilon(\sigma; x)\, x_{\sigma(1)}\wedge\ldots \wedge x_{\sigma(i)} \wedge x_{\sigma(i+1)}\wedge \ldots \wedge x_{\sigma(n)} \\
  &= \frac{1}{2^n-2} \left( \sum_{i=1}^{n-1} \frac{n!}{i!(n-i)!}\right)x_1 \wedge \ldots \wedge x_n \\
  &= x_1 \wedge \ldots \wedge x_n\, .
\end{align}
Now, since $E$ is a morphism of coalgebras, it satisfies $\Delta E = (E\otimes E) \Delta$. This implies
\begin{equation}
E - \prone E = \Lambda (E \otimes E) \Delta\, ,
\end{equation}
and equation \eqref{eq:EminusPiE} follows by evaluating on $\bar x_1\wedge \cdots \wedge \bar x_{n+1}$.

\section{Dependence on the choice of homotopy}

In this appendix, we prove that the homotopy-transferred $L_\infty$ structures corresponding to different choices of $h$ are isomorphic; what matters is the choice of projector $P$.

So, we start with a map $h$ satisfying the homotopy relation \eqref{homotopyID} and also assume that it satisfies the side and orthogonality conditions \eqref{sideconditions} and \eqref{eq:ortho}; in particular,
\begin{align}
{\bf 1} - P &= \pd h + h \pd \label{eq:homotopyapp}\\
h^2 &= 0 \label{eq:h2=0app}\\
\kappa(hx, hy) &= 0\, .
\end{align}
The orthogonality condition means that $\im h$ is an isotropic subspace for the cyclic inner product $\kappa$, and so is $\im\pd$ on account of the cyclicity of $\kappa$ and $\pd^2 = 0$. They are in fact also Lagrangian inside $({\bf 1} - P) X$, and by \eqref{eq:homotopyapp} we find that
\be \label{eq:decomp(1-P)X}
({\bf 1} - P) X=(\im h)\oplus(\im \pd)
\ee
is the direct\footnote{The sum is direct because, if $x \in ({\bf 1} - P) X$ belongs to both images, then $x = ({\bf 1} - P) x = \pd hx + h \pd x = 0$ since $\pd^2 = 0 = h^2$. } sum of Lagrangian subspaces $\im h$ and $\im \pd$. We can introduce generators $T_i\in\im h$ and $\starred T^i\in \im \pd$ generalising the notation of \ref{section:0dparticlelinftyalgebra}, for which $\kappa$ takes the canonical form
\be
\kappa(T_i,\starred T^j)=\delta^j_i\,,
\ee
$\pd$ takes the form
\be
\partial \starred T^i=0\,,\quad \partial T_i=K_{ij} \starred T^j\,,
\ee
and $h$ takes the form
\be \label{eq:explicithapp}
h \starred T^i= K^{ji}T_j\,,\quad h T_i=0\,.
\ee
Condition \eqref{eq:homotopyapp} forces $K_{ij}$ to be invertible with inverse $K^{ji}$ (i.e. $K_{ij} K^{ik} = \delta^k_j$). Viewing these equations in another way, this means that $h$ is in fact completely determined by a choice of a Lagrangian complement $L$ to $\im \pd$ in the decomposition \eqref{eq:decomp(1-P)X} \cite{Arvanitakis:2019ald}: indeed, $\left.\pd\right|_L : L \to \im \pd$ must be invertible, and equations \eqref{eq:homotopyapp} -- \eqref{eq:h2=0app} then force $h$ to be the inverse of $\left.\pd\right|_L$ on $\im \pd$ and zero on $L = \im h$, as in equation \eqref{eq:explicithapp}.

We can reinterpret this via the BRST-BV interpretation of the dual picture of \lf-algebras given in sections \ref{section:cyclicLinftyBRSTBV} and \ref{sec:BVrelation}. We introduce dual coordinates $z^a=(\phi^i,\starred\phi_i)$, so $\langle \phi^i,T_j\rangle=\delta^i_j=\langle\starred\phi_i,\starred T^j\rangle$ are the non-vanishing duality pairings, and calculate the BV master action of \eqref{thetadefinition}:
\begin{align}
\Theta &= \frac{1}{2} \kappa_{ac}C^c_bz^a z^b+\mathcal O(z^3) \\
&= \text{(terms in the projected part)} + \frac{1}{2} K_{ij}\phi^i\phi^j+\mathcal O(z^3)\,.
\end{align}
We see therefore that a choice of $h$ corresponds to a splitting of the dual basis $z^a=(\phi^i,\starred\phi_i)$ of $(({\bf 1} - P)X)^\star$ which is such that the quadratic term in $\Theta|_{\starred\phi=0}$ is invertible. This corresponds directly to a \emph{proper} choice of ``fields'' $\phi$ versus ``antifields'' $\starred\phi$ in the BV path integral, meaning one where the quadratic term is non-degenerate when antifields are set to zero.

A proper choice of fields versus antifields is usually found by the introduction of a gauge-fixing fermion $\psi$. Typically the theory is given with an \emph{improper} choice of fields $\phi$ and antifields $\starred\phi$, which is then transformed into a proper one by the canonical transformation (with the antibracket \eqref{antibracketdef})
\be
\phi\to \phi\,,\qquad \starred\phi\to \starred\phi+ (\psi,\starred\phi)\,,
\ee
where $\psi$ is a function of fields $\phi$ alone. The partition function and gauge-invariant correlation functions are then shown to be invariant under deformations of $\psi$.

As we saw above, the homotopy $h$ is completely determined by its image $L = \im h$. Therefore, a deformation of $h$ (satisfying the homotopy, orthogonality and side conditions) can be encoded in a deformation of that subspace. These are parametrized by a matrix $\psi_{ij}$ as
\be
\label{deformationofhomotopy}
T'_i=T_i-\psi_{ij} \starred T^j\,,\qquad \starred {T'}^i=\starred T^i\,,
\ee
with the orthogonality condition $\kappa(T'_i,T'_j)=0$ implying\footnote{A more general deformation could be written down as $T''_i= A^j_i T_j-B_{ij} \starred T^j$, with $A$ invertible, but this is simply $T''_i = A^j_i T'_j$ (with $\psi = A^{-1} B$) and so parametrizes the same subspace $L$. The only relevant deformation is the part given in \eqref{deformationofhomotopy}.} (c.f.~\eqref{eq:omegadef} and \eqref{eq:signkappaomega})
\be
(-1)^i\psi_{ij}+(-1)^j \psi_{ji}=0
\,,
\ee
which is $\psi_{ij}=\psi_{ji}$ given $|i|=|j|+1\mod 2$. The deformed homotopy $h'$ is defined by
\be
h' \starred T^i= K^{ji}T'_j\,,\quad h T'_i=0\,.
\ee
It satisfies all the properties $h$ does if $\psi_{ij}=-\psi_{ji}$. In the dual basis, \eqref{deformationofhomotopy} gives
\be
\label{deformationofhomotopydual}
{\phi'}^i=\phi^i\,,\qquad \starred\phi ' _i=\starred\phi_i+ \psi_{ji}\phi^j\,,
\ee
which is generated by the gauge-fixing fermion
\be
\psi=\frac{1}{2}\psi_{ij}\phi^i\phi^j
\ee
via the antibracket \eqref{antibracketdef}. It follows from the BRST-BV formulation on $L_\infty$ algebras and the properties of the antibracket that a transformation of this form describes an isomorphism of \lf-algebras preserving the cyclic structure. We have therefore proved that, up to isomorphisms, a deformation of $h$ does not affect the transferred $L_\infty$ structure.

\bibliography{LINFTY}

\end{document}